\documentclass[11pt,preprint,trackchanges]{aastex63}

\usepackage{url}
\def\apjl{\rm ApJL}
\def\apjs{\rm ApJS}

\def\aap{\rm AAP}

\newcommand{\gtorder}{\mathrel{\raise.3ex\hbox{$>$}\mkern-14mu
            \lower0.6ex\hbox{$\sim$}}}
\newcommand{\ltorder}{\mathrel{\raise.3ex\hbox{$<$}\mkern-14mu
            \lower0.6ex\hbox{$\sim$}}}

\shorttitle{The Mass and Radius of PSR~J0030$+$0451}
\shortauthors{Miller, Lamb, Dittmann, et al.}

\begin{document}

\title{PSR J0030+0451 MASS AND RADIUS FROM {\it NICER} DATA AND IMPLICATIONS FOR THE PROPERTIES OF NEUTRON STAR MATTER
}

\correspondingauthor{M. C. Miller}
\email{miller@astro.umd.edu}

\author[0000-0002-2666-728X]{M.~C.~Miller}
\affiliation{Department of Astronomy and Joint Space-Science Institute, University of Maryland, College Park, MD 20742-2421 USA}

\author[0000-0002-3862-7402]{F.~K.~Lamb}
\affiliation{Center for Theoretical Astrophysics and Department of Physics, University of Illinois at Urbana-Champaign, 1110 West Green Street, Urbana, IL 61801-3080, USA}
\affiliation{Department of Astronomy, University of Illinois at Urbana-Champaign, 1002 West Green Street, Urbana, IL 61801-3074, USA}

\author[0000-0001-6157-6722]{A.~J.~Dittmann}
\affiliation{Department of Astronomy and Joint Space-Science Institute, University of Maryland, College Park, MD 20742-2421 USA}

\author[0000-0002-9870-2742]{S.~Bogdanov}
\affiliation{Columbia Astrophysics Laboratory, Columbia University, 550 West 120th Street, New York, NY 10027, USA}

\author{Z.~Arzoumanian}
\affiliation{X-Ray Astrophysics Laboratory, NASA Goddard Space Flight Center, Code 662, Greenbelt, MD 20771, USA}

\author{K.~C.~Gendreau}
\affiliation{X-Ray Astrophysics Laboratory, NASA Goddard Space Flight Center, Code 662, Greenbelt, MD 20771, USA}

\author[0000-0001-6119-859X]{A.~K.~Harding}
\affiliation{Astrophysics Science Division, NASA Goddard Space Flight Center, Greenbelt, MD 20771, USA}

\author[0000-0002-6089-6836]{W.~C.~G.~Ho}
\affiliation{Department of Physics and Astronomy, Haverford College, 370 Lancaster Avenue, Haverford, PA 19041, USA}
\affiliation{Mathematical Sciences, Physics and Astronomy, and STAG Research Centre, University of Southampton, Southampton SO17 1BJ, UK}

\author[0000-0002-5907-4552]{J.~M. Lattimer}
\affiliation{Department of Physics and Astronomy, Stony Brook University, Stony Brook, NY 11794-3800, USA}

\author[0000-0002-8961-939X]{R.~M.~Ludlam}
\affiliation{Cahill Center for Astronomy and Astrophysics, California Institute of Technology, Pasadena, CA 91125, USA\footnote{Einstein Fellow}}

\author[0000-0003-2386-1359]{S.~Mahmoodifar}
\affiliation{Department of Astronomy and Joint Space-Science Institute, University of Maryland, College Park, MD 20742-2421 USA}
\affiliation{Astrophysics Science Division, NASA Goddard Space Flight Center, Greenbelt, MD 20771, USA}

\author[0000-0003-4357-0575]{S.~M.~Morsink}
\affiliation{Department of Physics, University of Alberta, 4-183 CCIS, Edmonton, AB T6G 2E1, Canada}

\author[0000-0002-5297-5278]{P.~S.~Ray}
\affiliation{Space Science Division, U.S. Naval Research Laboratory, Washington, DC 20375, USA}

\author[0000-0001-7681-5845]{T.~E.~Strohmayer}
\affiliation{Astrophysics Science Division and Joint Space-Science Institute, NASA Goddard Space Flight Center, Greenbelt, MD 20771, USA}

\author[0000-0002-7376-3151]{K.~S.~Wood}
\affiliation{Praxis, resident at the U.S. Naval Research Laboratory, Washington, DC 20375, USA}

\author[0000-0003-1244-3100]{T.~Enoto}
\affiliation{The Hakubi Center for Advanced Research and Department of Astronomy, Kyoto University, Kyoto 606-8302, Japan}

\author[0000-0002-2731-9295]{R.~Foster}
\affiliation{MIT Kavli Institute for Astrophysics and Space Research, Massachusetts
Institute of Technology, 70 Vassar Street, Cambridge, MA, USA}

\author{T.~Okajima}
\affiliation{X-Ray Astrophysics Laboratory, NASA Goddard Space Flight Center, Code 662, Greenbelt, MD 20771, USA}

\author{G.~Prigozhin}
\affiliation{MIT Kavli Institute for Astrophysics and Space Research, Massachusetts
Institute of Technology, 70 Vassar Street, Cambridge, MA, USA}

\author{Y.~Soong}
\affiliation{X-Ray Astrophysics Laboratory, NASA Goddard Space Flight Center, Code 662, Greenbelt, MD 20771, USA}

\begin{abstract}

Neutron stars are not only of astrophysical interest, but are also of great interest to nuclear physicists because their attributes can be used to determine the properties of the dense matter in their cores. One of the most informative approaches for determining the equation of state (EoS) of this dense matter is to measure both a star's equatorial circumferential radius $R_e$ and its gravitational mass $M$. Here we report estimates of the mass and radius of the isolated 205.53~Hz millisecond pulsar PSR~J0030$+$0451 obtained using a Bayesian inference approach to analyze its energy-dependent thermal X-ray waveform, which was observed using the \textit{Neutron Star Interior Composition Explorer} (\textit{NICER}). This approach is thought to be less subject to systematic errors than other approaches for estimating neutron star radii. We explored a variety of emission patterns on the stellar surface. Our best-fit model has three oval, uniform-temperature emitting spots and provides an excellent description of the pulse waveform observed using \textit{NICER}. The radius and mass estimates given by this model are $R_e = 13.02^{+1.24}_{-1.06}$~km and $M = 1.44^{+0.15}_{-0.14}~M_\odot$ (68\%). The independent analysis reported in the companion paper by Riley et al. explores different emitting spot models, but finds spot shapes and locations and estimates of $R_e$ and $M$ that are consistent with those found in this work. We show that our measurements of $R_e$ and $M$ for PSR~J0030$+$0451 improve the astrophysical constraints on the EoS of cold, catalyzed matter above nuclear saturation density.

\end{abstract}


\section{INTRODUCTION}
\label{sec:introduction}

A key current goal of nuclear physics is to understand the properties of cold catalyzed matter above the saturation density of nuclear matter.  Matter at these densities cannot be studied in terrestrial laboratories. Hence observations of neutron stars---which contain large quantities of such matter---play a key role (see, e.g., \citealt{2007PhR...442..109L}). Over the last few years, the discovery of several high-mass neutron stars (\citealt{2010Natur.467.1081D,2013Sci...340..448A,2018ApJ...859...47A,2019arXiv190406759T}) and measurement of the binary tidal deformability during a neutron star merger \citep{2017PhRvL.119p1101A,2018PhRvL.121p1101A,2018PhRvL.121i1102D} have  advanced our knowledge of the properties of cold dense matter, but precise and reliable measurements of neutron star radii would significantly improve our understanding.

Various radius estimates have been made using models of the X-ray emission from quiescent neutron stars (see \citealt{2018MNRAS.476..421S} for a recent summary), from neutron stars during thermonuclear X-ray bursts (see \citealt{2010ApJ...722...33S}; \citealt{2016ApJ...820...28O}; and \citealt{2017A&A...608A..31N} for different perspectives), and from accretion-powered millisecond pulsars (see \citealt{2018A&A...618A.161S}), with inferred radii typically ranging from $\sim 10$~km to $\sim 14$~km, consistent with most theoretical predictions. However, these estimates are susceptible to significant systematic errors, in the sense that a model could provide a formally good fit to the data but yield a credible interval for the radius that strongly excludes the true value \citep{2013arXiv1312.0029M,2016EPJA...52...63M}.

In contrast, analyses of the soft X-ray pulse waveforms observed using the \textit{Neutron Star Interior Composition Explorer} (\textit{NICER}) are expected to be less susceptible to systematic errors. Analyses of synthetic waveforms carried out prior to the launch of \textit{NICER} showed that, for the cases considered, using model assumptions different from the true situation (e.g., different emission or beaming patterns, different spectra, or different surface temperature distributions) did not significantly bias parameter estimates, provided the fit was formally good \citep{2013ApJ...776...19L,2015ApJ...808...31M}. Simple pulse waveform models have previously been fit to the soft X-ray waveforms of rotation-powered pulsars observed using \textit{ROSAT} and the {\it Extreme Ultraviolet Explorer} (\textit{EUVE};  \citealt{1997ApJ...490L..91P,1998A&A...329..583Z}) and \textit{XMM-Newton} (see, e.g., \citealt{2007ApJ...670..668B,2008ApJ...689..407B,2013ApJ...762...96B}). These fits gave estimates for the radii of these pulsars that were consistent with the expected range of neutron-star radii, but the number of counts available was too small to obtain tight constraints. Nevertheless, these pioneering studies indicated that rotation-powered pulsars are promising targets for such measurements.

Here we present the results of an analysis by \textit{NICER} team members of the currently available data on the 205.53~Hz \citep{2018ApJ...859...47A} millisecond pulsar PSR~J0030$+$0451. The results of an independent analysis of these data by other \textit{NICER} team members are presented in the companion paper by \citet{2019ApJ...887L..21R}. The analysis presented there uses an independently developed code for computing pulse waveforms, a different approach to modeling the heated regions on the stellar surface, and different sampling methods than the ones used in the analysis presented here. The consistency of the surface temperature distributions and mass-radius posteriors reported there with those reported here suggests that the unavoidable uncertainties in models of the heated areas on the stellar surface have not greatly affected the results of either analysis. 
In Section~\ref{sec:observations} we describe how we obtained and processed the \textit{NICER} data on PSR~J0030$+$0451. 
In Section~\ref{sec:methods} we discuss our methods, including our modeling of the emission from the stellar surface, our approach to modeling the soft X-ray waveform, our parameter estimation and model evaluation methods, and our treatment of instrumental effects and bias. 
In Section~\ref{sec:fits-to-synthetic-data} we describe how we used analyses of synthetic waveforms to verify our parameter estimation codes and procedures, and to evaluate the consequences of our decision to analyze a subset of the \textit{NICER} data on PSR~J0030$+$0451.
In Section~\ref{sec:fits-to-nicer-data} we present and discuss our best estimates of the radius and mass of PSR~J0030$+$0451 using the currently available \textit{NICER} data on its pulse waveform.
In Section~\ref{sec:implications-for-EoS} we compute and discuss the constraints on two different parameterized models of the equation of state of neutron star matter implied by our measurements of the radius and mass of PSR~J0030$+$0451. We summarize our conclusions in Section~\ref{sec:conclusions}.

\section{OBSERVATIONS}
\label{sec:observations}

\subsection {Data Used}
\label{sec:data-used}

Here, we briefly summarize the \textit{NICER} X-ray Timing Instrument~(XTI) data set that was used for the parameter estimation analyses presented here. More detailed descriptions of the observations, data reduction, and event folding procedures, and a thorough examination of the properties of the final event data can be found in \citet{2019ApJ...887L..25B}. 

The \textit{NICER} XTI data on PSR~J0030$+$0451 were collected in a set of short observations (typically lasting a few hundred to $\sim$2000 s)  over the period from 2017 July 24 to 2018 December 9. Due to the exceptional rotational stability of PSR~J0030$+$0451 and the superb absolute timing accuracy of \textit{NICER}, neither the large time span nor the discontinuities in the exposure had any adverse effects, such as smearing due to long-term pulse-phase drifts, on the pulse profile.

\subsection {Filtering of the Data and Event Folding}
\label{sec:data-filtering}

For this analysis, we used products from the calibration database (CALDB) version 20181105 and gain solution version {\tt optmv7}. In addition to the standard pipeline processing event filtering that all \textit{NICER} data undergo, the PSR~J0030$+$0451 data set was subjected to screening criteria tailored to the inference analyses discussed later. 

We included only data accumulated when the source subtended an angle greater than $80^{\circ}$ from the Sun, in order to minimize the contamination from optical loading in the lowest-energy detector channels. All events from the XTI detector with DET\_ID~34 were excised because it frequently exhibits significantly elevated count rates relative to the other detectors (analogous to a ``hot pixel'' in charge-coupled device detectors). 

In the final filtering stage, we constructed a time series with 16~s time bins using the event data in channels 25--800 and then discarded all the events in those time bins that had an average count rate greater than $3\,{\rm counts}\,{\rm s}^{-1}$. The long-term average count rate during the \textit{NICER} observations of PSR~J0030$+$0451 was $\sim\,$0.7$\,{\rm counts}\,{\rm s}^{-1}$. This cut therefore excluded the event data in all the 16~s time bins that had count rates more than four times the long-term average count rate. The purpose of this cut was to improve the signal-to-noise ratio of the measured thermal emission from the pulsar by removing data contaminated by flares that were not filtered out by the standard pipeline processing.

The thermal emission pulse profile that was used to estimate the mass and radius of PSR~J0030$+$0451 was constructed by averaging the data from the roughly 400 million rotation periods observed by \textit{NICER} during more than 18 months of observations. We compared the profile measured during the first third of this observing time with the profile measured during the second third, and also compared the profile measured during the first half of this observing time with the profile measured during the second half. These comparisons showed no statistically significant differences between the compared profiles. We expect that any shorter-term flux variations were averaged out by using such a long data set.

To generate the final X-ray pulse profile as a function of pulsar rotational phase and energy, we used the ``photons'' plugin for the Tempo2 pulsar timing package \citep{2006MNRAS.369..655H} and the best publicly available radio timing solution, obtained by NANOGrav \citep{2018ApJS..235...37A}, to assign pulse phases to each event. The number of the ``pulse invariant'' (PI) detector channel provides an estimate of the photon energy: the photon energy in eV is approximately equal to $10\times$ the PI channel number in which the event occurs.

As discussed in \citet{2019ApJ...887L..25B} the final X-ray pulse profile was cross-verified by comparing it with an independently generated profile in which pulse phases were assigned to events using the PINT pulsar timing software. The pulse phases assigned to individual events by the two procedures differed by $\lesssim$1\,$\mu$s, and the differences between the two folded X-ray pulse profiles were negligible.  The final event list used in the analysis reported in this paper is available at https://doi.org/10.5281/zenodo.3524457.

\subsection {Calibration of the Instrument}
\label{sec:calibration}

To incorporate the performance of the \textit{NICER} XTI in our model, we used version 1.02 of the redistribution matrix file (RMF), which gives the probability that a photon in a particular energy range is registered by a particular detector channel, and version 1.02 of the ancillary response file (ARF), which provides the effective area of the telescope as a function of energy, as measured on axis.

Spectral calibration of the \textit{NICER} XTI has been carried out primarily using observations of the Crab pulsar and nebula. The response below channel~40 is more uncertain than the response at higher energies, because of current imperfections in the modeling of instrumental effects such as the detector trigger efficiency. Also, because at these low energies the Crab spectrum is strongly attenuated by the interstellar medium, only a relatively small number of counts are available to determine the spectrum at these energies. Fits to \textit{NICER} observations of the Crab pulsar and nebula in the energy channels 40--299 used in this study (see Section~\ref{sec:energy-channels}) show systematic residuals, but these are typically $\lesssim\,$5\% (\citealt{2018ApJ...858L...5L}). They are likely due to imperfect modeling of the microphysics of the concentrator optics and astrophysical features, such as oxygen edges and emission lines.

\section{METHODS}
\label{sec:methods}

In this section, we first describe our modeling of the soft X-ray emission from hot spots on the stellar surface and our approach to modeling the soft X-ray pulses produced by the rotation of these spots. We then present the specific pulse waveform models we consider in this work and our procedure for constructing and fitting these model waveforms to the \textit{NICER} pulse waveform data on PSR~J0030$+$0451. We end this section by discussing how we deal with the errors in the response of the \textit{NICER} instrument and the evidence that including data from channels below channel~40 would risk biasing our estimates of the mass and radius of PSR~J0030$+$0451. We defer to subsequent sections discussion of our use of synthetic pulse waveform data to verify our parameter estimation procedures, our use of \textit{NICER} pulse waveform data to estimate the mass and radius of PSR~J0030$+$0451, our assessment of the reliability of these estimates, and their implications for the EoS of cold dense matter.

\subsection{Modeling the Emission from the Stellar Surface}
\label{sec:modeling-the-emission}

The soft X-ray pulse waveform of PSR~J0030$+$0451 is thought to be produced by hot regions on its rotating surface created by inward moving particles created in electron-positron pair cascades interacting with the outermost layers of the neutron star \citep{1981ApJ...248.1099A, 2001ApJ...556..987H}. Models of global pulsar magnetospheres (see, e.g., \citealt{2014ApJ...793...97K, 2015ApJ...801L..19P, 2018ApJ...858...81B}) have delineated the current-closure patterns of pulsars, which show outward currents in the magnetic polar regions balanced by return currents over the outer parts of the polar caps that are connected to the current sheet. Some of the particles that make up the return currents (those that are accelerated in the current sheet) are responsible for the pulsar's $\gamma$-ray and nonthermal X-ray emission \citep{2018ApJ...857...44K,2018ApJ...855...94P}. The polar cap pair cascades that take place in both outgoing and return current regions \citep{2013MNRAS.429...20T} supply pairs for radio emission. Some of the particles produced in these pair cascades are accelerated downward and are expected to deposit energy deep in the neutron star's atmosphere (see \citealt{1974RvMP...46..815T} and \citealt{2007ApJ...670..668B}). This energy is expected to heat the atmosphere at large optical depths, producing thermal soft X-ray emission from the stellar surface \citep{2001ApJ...556..987H, 2002ApJ...568..862H}. The spectrum of this emission is expected to depend on the chemical composition of the star's atmosphere and, possibly, on the strength and orientation of the magnetic field at the locations of the hot spots \citep{2014PhyU...57..735P}.

The models of the thermal emission from these hot spots that we have used in the analysis of the soft X-ray waveform of PSR~J0030$+$0451 that we report here assume that the upper atmosphere of the pulsar is fully ionized pure hydrogen and that, for our purposes, the magnetic fields in the hot spots can be neglected. We now discuss these assumptions.

\textit{Composition of the upper atmosphere}.  Our model atmospheres were constructed assuming that the gas in the atmosphere is fully ionized, and hence that the dominant opacity is free-free absorption \citep{2001MNRAS.327.1081H}. While opacity tables for partially ionized matter exist \citep{1996ApJ...464..943I, 2005MNRAS.360..458B, 2016ApJ...817..116C}, they do not cover the full range of temperatures and energies needed for the present analysis. However, at the temperatures $\sim\,$$10^6$~K that are relevant for the present study, the fraction of hydrogen that is neutral is very low; comparison of the outward specific intensity given by our fully ionized hydrogen atmosphere model differs by at most 1--2\% at 0.5--1~keV from that for an atmosphere with an effective temperature of $10^6$~K given by a model constructed using the Opacity Project (OP) opacity table \citep{2005MNRAS.360..458B}, which allows for partial ionization (see \citealt{2019ApJ...887L..26B}).

Our assumption that the upper atmosphere is pure hydrogen is based primarily on the astrophysical abundance of hydrogen and calculations that show how the lightest element present in a neutron star atmosphere will float to the top within seconds. These calculations are extrapolations of the earlier calculations by, for example, \citet{1980ApJ...235..534A}. Given the astrophysical abundance of hydrogen and its tendency to float to the top, we expect it to completely dominate the composition of the upper atmosphere.

If the outer atmosphere of the pulsar were instead dominated by helium, the emission spectrum and beaming pattern in the energy range 0.5--1.0~keV would be expected to be very similar to those for a hydrogen atmosphere, but their normalizations would be slightly smaller than for hydrogen (see \citealt{2019ApJ...887L..26B}). In particular, hydrogen and helium atmospheres with an effective temperature of $10^6$~K (the approximate inferred temperature of the heated regions on the surface of PSR~J0030$+$0451) are predicted to produce specific intensities normal to the surface that differ from each other by $\lesssim\,$5\% in the 0.5--1~keV energy range (again see \citealt{2019ApJ...887L..26B}). Our current best fit of the zero-field NSX helium atmosphere \citep{2001MNRAS.327.1081H} to the  \textit{NICER} data on PSR~J0030$+$0451 appears disfavored relative to our current best fit of the zero-field NSX hydrogen atmosphere (\citealt{2001MNRAS.327.1081H}; see also \citealt{2009Natur.462...71H}). Specifically, our current fit using the NSX helium model yields a best-fit mass of $2.7\,M_\odot$, with little posterior probability below $2.0\,M_\odot$. Although more work would be needed to completely rule out a helium atmosphere for PSR~J0030$+$0451, this mass estimate is so large that we interpret it as disfavoring a helium atmosphere. 

Thus, while there is at present no direct spectroscopic confirmation that the upper atmosphere of PSR~J0030$+$0451 is pure hydrogen, the fact that the spectrum and beaming pattern predicted by hydrogen atmosphere models agree well with the  spectrum and beaming of the soft X-ray emission from PSR~J0030$+$0451 observed using \textit{NICER} and that the X-ray spectrum and beaming predicted by helium atmosphere models appear disfavored, supports our assumption that the upper atmosphere is hydrogen. Therefore, in computing model waveforms we used the local emission for a given surface gravity, effective temperature, angle from the surface normal, and photon energy that is predicted by the NSX code developed by \citet{2001MNRAS.327.1081H} (see also \citealt{2009Natur.462...71H}) for a fully ionized hydrogen atmosphere.

The 205.53~Hz rotation frequency of PSR~J0030$+$0451 \citep{2018ApJ...859...47A} is small enough that the only effects of the rotation that need to be taken into account in computing pulse waveforms are the special relativistic Doppler shift and aberration caused by the motion of the emitting surface and the rotationally induced oblateness of the star; the effect of frame-dragging and the effect on the exterior spacetime of the rotationally induced stellar mass quadrupole are negligible (see \citealt{2007ApJ...663.1244M} and \citealt{2019ApJ...887L..26B}). We therefore used the so-called oblate-star Schwarzschild-spacetime (OS) approximation \citep{2007ApJ...654..458C,2007ApJ...663.1244M,2014ApJ...791...78A} when tracing light rays from the stellar surface to the observer. Computations of pulse waveforms like those used here were verified by members of the \textit{NICER} Lightcurve Working Group, as described in \citet{2019ApJ...887L..26B}.

\textit{Strength and structure of the stellar magnetic field}.  We can estimate an approximate lower bound on the strength of the surface magnetic field of PSR~J0030$+$0451 by assuming that the field is a centered dipole and estimating the strength of the dipole moment using the best braking models currently available (see, e.g., \citealt{2006ApJ...643.1139C}). The period $P$ of PSR~J0030$+$0451 is 0.00487 seconds and its period derivative ${\dot P}$ is $1.02\times 10^{-20}$ \citep{2018ApJ...859...47A}. If we use Equation~(12) of \citet{2006ApJ...643.1139C} and, for the sake of argument, choose a magnetic field inclination $\theta$ of $\pi/2$ and the field configuration described by $\alpha=0$, then the inferred strength of the magnetic field at the stellar surface is $B\sim 4\times 10^8$~G. In principle, this represents an approximate lower bound on the strength of the total magnetic field in the atmospheres of the hot spots. For comparison, the electron cyclotron energy in a magnetic field of this strength is $\hbar\omega_c\approx 4$~eV, more than 50 times smaller than the lowest photon energies that can be observed using \textit{NICER} and $\sim\,$100 times smaller than the lowest energies we consider in our fits. 

We note, however, that the three hot spots in the ``southern'' rotational hemisphere (we denote the rotational hemisphere that includes the line of sight to the observer as the ``northern'' rotational hemisphere) that can explain the pulse waveform of PSR~J0030$+$0451 observed using \textit{NICER} suggest that the magnetic field of PSR~J0030$+$0451 is at least an offset dipole, and may have higher moments that dominate the field strength near the stellar surface. A complex field would be consistent with the structures of the magnetic fields of other millisecond rotation-powered pulsars inferred from their radio and $\gamma$-ray emission \citep{1991ApJ...366..261R, 1998ApJ...492..267R}, and the inferred magnetic field configurations of the accretion-powered millisecond pulsars that are thought to be the progenitors of the millisecond rotation-powered pulsars \citep{2009ApJ...706..417L,2009ApJ...705L..36L}.

If the magnetic field of PSR~J0030$+$0451 does have significant higher multipole moments, they may not be negligible at the light cylinder, which is closer to the stellar surface in millisecond pulsars than in pulsars with lower spin rates. Moreover, in some spin-down models (again see \citealt{2006ApJ...643.1139C}), the spin-down rate depends on the magnetic flux enclosed by the open-field-line region, which is defined by the boundary of the closed magnetosphere and may not extend all the way out to the light cylinder. The way this flux evolves with spin rate can be affected by the presence of multipolar field components, complicating the relation between the strength of the surface magnetic field and the spin-down rate.

Given this complexity, it is difficult to infer the strength and configuration of the magnetic field on the surface of PSR~J0030$+$0451. In current braking models, the braking torque scales with the strength of the dipole component of the magnetic field at some characteristic distance from the star. For a given strength of the dipole component at this distance, the strength of the magnetic field at the stellar surface scales as $(R_e/d)^3$, where $R_e$ is the circumferential radius of the star and $2d$ is the distance through the star from one magnetic pole to the other. Thus, if the large-scale magnetic field is essentially dipolar and the corresponding magnetic poles are a significant distance south of the rotational equator, the field strength at the surface of the star could be 5--10~$\times$ greater than it would be for a centered dipole field with the same strength a large distance from the stellar surface. In this case, the indicated strength of the magnetic fields at the hot spots could be as large as $B\sim 4\times 10^9$~G.

\textit{Effect of the magnetic field on the soft X-ray emission from the stellar surface}.  The model atmospheres we have used to compute the waveforms that we fitted to the \textit{NICER} waveform data assume that the effects of the stellar magnetic field on the structure and radiative properties of the atmosphere are negligible. This approximation is valid for magnetic field strengths less than $B_c = e^3m_e^2c/\hbar^3 = 2 \times 10^9$~G. The upper bound on the strength of the magnetic field that we estimated above by assuming that it is dipolar and using the best currently available braking models is an order of magnitude smaller than $B_c$, but our estimates of the possible field strengths at the hot spots are a factor of two larger than $B_c$.

If the magnetic field exceeds $B_c$, its effect is non-perturbative and dominant (see \citealt{2001RvMP...73..629L}): the radiation field becomes polarized, the beaming pattern becomes strongly anisotropic, and spectral features associated with the electron cyclotron resonance and bound species become important \citep{2014PhyU...57..735P}. The absence of any obvious spectral feature in the  spectra observed using \textit{XMM-Newton} (see, e.g., Bogdanov \& Grindlay 2009) and in the $\sim\,$$0.3$--3~keV spectra obtained using \textit{NICER} rules out $B = $ (3--30)$ \times 10^{10}$~G, assuming a redshift of 20\%. In addition, while an electron cyclotron line produced by a magnetic field $B \gg 10^{11}$~G could be hidden at $E > 3$~keV, such fields would produce pencil or fan beams that are unlikely to be able to produce an energy-dependent pulse profile that can match the profile observed using \textit{NICER}.

The changes in the spectrum and the beaming pattern that occur for magnetic fields $B \ga 10^{10}$~G are somewhat larger than the changes in the spectrum and beaming pattern that occur when going from a purely H model atmosphere to a purely He model atmosphere without any magnetic field. While spectra that assume such strong  fields might yield statistically acceptable fits to the \textit{NICER} observations of PSR~J0030$+$0451, it is likely, though not certain, that such fits would, as in the case of the He atmosphere models, yield mass and radius estimates that are so extreme that we would consider these models disfavored.

If the total strengths of the magnetic fields in the atmospheres of the hot spots of PSR~J0030$+$0451 are less than $B_c$, they are expected to have little effect on the spectrum and shape of its soft X-ray waveform. The agreement of our energy-resolved waveform models with the energy-resolved waveform data obtained using \textit{NICER} are consistent with this interpretation.

\subsection{Our Approach to Modeling the Soft X$-$ray Waveform}
\label{sec:waveform-modeling}

In modeling the soft X-ray pulse waveform of PSR~J0030$+$0451, we started with the simplest possible description of the heated region(s) on the stellar surface and then, as necessary, considered more complicated descriptions, guided at each step by the differences between the best-fit simpler pulse waveform models and the waveform observed using \textit{NICER}. Along the way, we verified our parameter estimation algorithms by fitting appropriate models to several synthetic waveforms that were generated using the models of the heated regions on the stellar surface that emerged from our fits to the observed waveform, and comparing the values of the parameters obtained from these fits with the values of these same parameters that were used to generate the synthetic waveforms.

Visual examination of the PSR~J0030$+$0451 waveform observed using \textit{NICER} shows evidence for two peaks, and waveform models constructed using a single hot spot fail to reproduce the qualitative features of the observed waveform. Consequently, we began our modeling of the observed waveform using models that had two uniform circular hot spots with possibly different locations, areas, and temperatures, but these models did not adequately describe the observed waveform.  In particular, the (relatively large) spot sizes that were required to reproduce the weak higher harmonic content of the observed waveform appeared to conflict with the (relatively small) spot areas that were required to reproduce the observed X-ray flux.

We therefore considered models with three and four different uniform-temperature circular hot spots, and finally models with two and three different uniform-temperature oval hot spots. In all cases we allowed the spots to overlap in any way that improved the fit; our procedure for treating overlapping spots is described below. This algorithm allowed the heated regions in the model to evolve toward a variety of different complex shapes as the fitting process proceeded, including multiple separate hot spots, multiple-temperature spots in which each spot has hotter and cooler regions, elongated or more complicated configurations produced by two or more partially or completely overlapping or adjacent circular or oval hot spots, crescent-shaped hot regions produced by cold oval spots partially or completely overlapping hot oval spots, and so on. These explorations found no evidence for different temperatures within the same hot spot and no evidence for more than three heated regions on the stellar surface. 

We found that a model with two different, non-overlapping, uniform oval spots is preferred over any models with two overlapping spots, adequately describes the observed waveform, and gives a much better fit to the data than a model with two circular spots. The larger east$-$west extent of the best-fitting oval spots allows them to reproduce the observed weak higher harmonic content of the waveform, while their smaller north$-$south extent allows them to have the relatively small areas required to reproduce the observed X-ray flux.

To assess whether a model with two different uniform-temperature oval spots is sufficient, we also considered configurations with three different uniform-temperature oval spots. We found that a model with three different, non-overlapping, uniform oval spots is preferred over any models in which the spots overlap, describes the observed waveform adequately, and gives a fit to the data that is slightly, but not significantly, better than the best-fit model with two non-overlapping, uniform oval spots. The two main spots in the best-fit model with three spots are very similar to the spots in the two-spot model, while the third spot has a much higher temperature but a much smaller area than either of the other two spots, is located close to the rotational pole on the far side of the star from the observer, and makes only a very small contribution to the waveform. 

The pulse waveforms and the mass and radius estimates given by the best-fit two- and three-spot models are statistically indistinguishable. We therefore concluded that both models adequately describe the observed waveform, that there is no evidence that a more complicated model is required, and that the mass and radius estimates given by these models are reliable. We slightly prefer the radius and mass estimates given by the model with three oval spots, only because the evidence for this model is slightly---although not significantly---greater than the evidence for the model with two oval spots.

In Section~\ref{sec:fits-to-synthetic-data}, we show and discuss the results of several of the analyses of synthetic waveform data that we performed to gain confidence in the accuracy and reliability of our parameter estimation procedure. Specifically, we show the results we obtained by fitting a model of the waveform produced by two circular spots to synthetic waveform data generated using this model, and by fitting a model of the waveform produced by two oval spots to synthetic waveform data generated using this more complicated model. In Section~\ref{sec:fits-to-nicer-data}, we describe in detail our systematic approach to modeling the pulse waveform of PSR~J0030$+$0451 and the evidence that led us to focus on the waveform model with three oval spots that we used to estimate the radius and mass of PSR~J0030$+$0451. But first we detail the various waveform models we constructed, the modeling procedure that led us to consider these models, and the algorithms we used to estimate the best-fit values of the parameters in these models. We also describe the procedures we used to evaluate and compare these models, the motivation for our selection of the \textit{NICER} data we analyzed, and our assessment of the effects of the errors in the \textit{NICER} response.

\subsection{Waveform Models}
\label{sec:waveform-models}

\begin{deluxetable*}{c|l|c}
    \tablecaption{Primary parameters of the pulse waveform models considered in this work.}
\tablewidth{0pt}
\tablehead{
      \colhead{Parameter} & \colhead{Definition} & \colhead{Assumed Prior}
    \label{tab:wf-primary-parameters}
}
\startdata
      \hline
      $GM/(c^2R_e)$ & Stellar compactness & $0.125-0.3125$ \\
      \hline
      $M$ & Gravitational mass & $1.0-2.4~M_\odot$ \\
      \hline
      $\theta_{\rm c1}$ & Spot 1 center & $0.1-3.1$ rad \\
      \hline
      $\Delta\theta_1$ & Spot 1 radius & $0-3$ rad \\
      \hline
      $f_1$ & Spot 1 oval ratio & $0.1-20$ \\
      \hline
      $kT_{\rm eff,1}$ & Spot 1 eff. temp. & $0.011-0.5$~keV\\
      \hline
      $\Delta\phi_2$ & Spot 2 longitude diff. & $0-1$ cycles \\
      \hline
      $\theta_{\rm c2}$ & Spot 2 center & $0.1-3.1$ rad \\
      \hline
      $\Delta\theta_2$ & Spot 2 radius & $0-3$ rad \\
      \hline
      $f_2$ & Spot 2 oval ratio & $0.1-20$ \\
      \hline
      $kT_{\rm eff,2}$ & Spot 2 eff. temp. & $0.011-0.5$~keV \\
      \hline
      $\Delta\phi_3$ & Spot 3 longitude diff. & $0-1$ cycles \\
      \hline
      $\theta_{\rm c3}$ & Spot 3 center & $0.1-3.1$ rad \\
      \hline
      $\Delta\theta_3$ & Spot 3 radius & $0-3$ rad \\
      \hline
      $f_3$ & Spot 3 oval ratio & $0.1-20$ \\
      \hline
      $kT_{\rm eff,3}$ & Spot 3 eff. temp. & $0.011-0.5$~keV \\
      \hline
      $\theta_{\rm obs}$ & Observer inclination & $0.1-\pi/2$ rad \\
      \hline
      $N_{\rm H}$ & H column density & $0-2.5\times 10^{20}$~cm$^{-2}$ \\
      \hline
      $D$ & Distance & $\exp[-(D-0.325~{\rm kpc})^2/2(0.017~{\rm kpc})^2]$ \\
      \hline 
\enddata
\tablecomments{We assumed the above priors, which are flat in the indicated range except for the distance $D$, and a rotation frequency of 205.53~Hz as seen by a distant observer (a)~when we analyzed the synthetic PSR~J0030$+$0451 waveform generated using two oval hot spots, which had parameter values (see Table~\ref{tab:synth2ovalchan40}) chosen so it would mimic the waveform observed by \textit{NICER}, and (b)~when we analyzed the actual PSR~J0030$+$0451 waveform observed by \textit{NICER}.
	When we analyzed the synthetic waveform that was generated using two circular hot spots, we used the priors listed above for most of the parameters, but different priors for four parameters.  This is because the waveform was constructed to mimic the waveform of PSR~J0437$-$4715 observed by \textit{NICER}, and in generating it we used values for these four parameters that are different from those listed in Table~\ref{tab:synth2circchan25}. In generating this synthetic waveform, we assumed a rotation frequency of 173.6~Hz and flat priors in the intervals listed for the following four parameters: $M$: $1.0 - 1.9~M_\odot$,  $\theta_{\rm obs}$: $0.6 - 0.9$~rad, $N_{\rm H}$: $0  - 2.5\times10^{20}~{\rm cm}^{-2}$,  $D$: $0.13 - 0.18$~kpc.
	See the text for further details about the assumed prior probability distributions for each parameter.}
\end{deluxetable*}

In the analyses we present here, we used pulse waveform models with the primary parameters listed and defined in Table~\ref{tab:wf-primary-parameters}. There are 12 primary parameters in the model that has two circular spots (this model assumes $f_1 = 1 = f_2$ and has no parameters for a third spot), 14 primary parameters in the model that has two oval spots, and 19 primary parameters in the model that has three oval spots. 

For the reasons discussed previously, all the waveform models we present in this report assume that the effective temperature of the emission from a given hot spot does not vary across the spot. The effective temperature listed in Table~\ref{tab:wf-primary-parameters} is the effective temperature that would be measured by a local observer on the stellar surface. 

When constructing an oval spot centered at a given colatitude and having a given latitudinal extent, we determined the longitudinal extent it would have if it were instead circular and then multiplied this extent by the colatitude-independent factor $f_1$ for spot~1, $f_2$ for spot~2, and $f_3$ for spot~3. The possible values of $f_1$, $f_2$, and $f_3$ are bounded from above by the requirement that no spot have a longitudinal extent greater than $2\pi$. 
 
We defined the northern hemisphere as the hemisphere containing the sightline to the observer. We defined longitude~0 as the longitude of the center of spot~1; after generating a trial waveform, we rotated its phase to give the best fit to the waveform data (see below).

\subsection{Our Waveform Modeling Procedure}
\label{sec:fitting-procedure} 

As noted earlier, when fitting model waveforms to the observed waveform, we allowed hot spots to overlap one another. This approach allowed the fitting process to create more complicated heated regions, such as multiple isolated spots, spots with hotter cores and cooler annuli, spots with cooler cores and hotter annuli, noncircular heated regions having two different temperatures, and so on, if these configurations were favored by the data. For the reasons described in Section~\ref{sec:waveform-modeling}, we discuss here only our modeling of waveforms with at least two hot spots and at most three.

In the pixels on the stellar surface where spots overlap, our fitting algorithm chooses the temperature of the lowest-numbered spot. For example, if spot~2 and spot~3 overlap, all the pixels in the overlap region are assumed to emit with the effective temperature currently assigned to spot 2. If spot~1, spot~2, and spot~3 all overlap, all pixels in the overlap region are assumed to emit with the effective temperature currently assigned to spot~1.

In addition to the primary parameters listed in Table~\ref{tab:wf-primary-parameters}, we introduced two types of ancillary parameters. The first set of ancillary parameters describes the counts in each energy channel that do not come from any of the hot spots. We refer to these as \textit{background} counts. These background counts allow for any unmodulated emission from the star that is not produced by any of the hot spots, any other unmodulated emission from the pulsar system, any emission within the field of view that does not come from the pulsar system, and the instrumental backgrounds. Another ancillary parameter describes the overall starting phase of the pulse waveform. In our fitting procedure, we chose to marginalize over these ancillary parameters, rather than fitting for them explicitly. Separately marginalizing the likelihood over the ancillary parameters is fully justified when, as here, the values of the primary parameters are uncorrelated with the values of the ancillary parameters.

We estimated the number of background counts in each energy channel and marginalized the likelihood with respect to the number of background counts in each channel as follows. We first generated a trial waveform by choosing a trial set of values for all the parameters in the waveform model and a trial starting phase for the waveform. Then, for each energy channel, we independently fit a Gaussian to the curve of the likelihood versus the number of background counts assumed present in that channel. By integrating over each Gaussian, we were able to estimate the number of background counts in each energy channel and marginalize the likelihood over the number of background counts in each energy channel, for each trial waveform. This procedure added one parameter to the waveform model for each energy channel we considered. As we discuss in detail below, for our final fits to the \textit{NICER} pulse waveform data we chose to use \textit{NICER} energy channels 40 through 299. This approach therefore added 260 parameters to the waveform model.

We performed a similar procedure to determine the best-fit overall phase of the waveform and marginalize the likelihood over this phase. Namely, for each trial offset of the overall phase (defined as the difference between the longitude of the center of the first spot relative to the longitude of the line of sight to the observer), we computed the likelihood over all energy channels. We then fit a Gaussian to the likelihood as a function of the overall phase, which allowed us to estimate the overall phase. We then marginalized the likelihood with respect to the overall phase by integrating the phase offset over the Gaussian curve, assuming a prior for the overall phase that was uniform from 0 to 1 cycles and zero outside this range. This added one more parameter to the waveform model.

Our procedure for estimating the number of background counts in each \textit{NICER} energy channel assumes that we have no independent knowledge of the number or spectrum of the background counts. In reality, we have some information about astrophysical and sky backgrounds and the backgrounds contributed by solar system, magnetospheric, terrestrial, and International Space Station sources. We have developed codes that can take this information into account, but many of these backgrounds are time-variable and are often not accurately known. The values of the waveform parameters we derived when we included models of the properties of these backgrounds differ little from the values we found when we used the procedure just described. We have therefore chosen to adopt the conservative approach of using this procedure, which does not rely on any models of the various backgrounds.

\subsection{Parameter Estimation and Model Evaluation}
\label{sec:parameter-estimation}

\textit {Parameter estimation and model comparison.}  To estimate the values of the parameters in our pulse waveform models and compare the evidence for different models using synthetic or \textit{NICER} data, we used standard Bayesian methods. For a given model ${\cal M}$ with parameters ${\vec\theta}$ that have a normalized prior probability density $q({\vec\theta})$, the normalized posterior probability density is given by Bayes' Theorem:
\begin{equation}
P({\cal M},{\vec\theta}|{\rm data})={{\cal L}({\rm data}|{\cal M},{\vec\theta})q({\vec\theta})\over{P({\cal M}|{\rm data})}}\; ,
\end{equation}
where ${\cal L}({\rm data}|{\cal M},{\vec\theta})$ is the likelihood of the data given the model with parameter values ${\vec\theta}$, and $P({\cal M}|{\rm data})\equiv \int_{\vec\theta} {\cal L}({\rm data}|{\cal M},{\vec\theta})q({\vec\theta})d{\vec\theta}$ is the evidence for model ${\cal M}$. 

Given a model ${\cal M}$, we can compute the joint posterior probability density distribution of the values of any subset of the parameters ${\vec\theta}$ by marginalizing the other parameters, i.e., by integrating $P({\cal M},{\vec\theta}\,|\,{\rm data})$ over the possible values of all parameters except those in the subset. For example, the 1D probability density distribution of the value of any single parameter can be computed by integrating $P({\cal M},{\vec\theta}\,|\,{\rm data})$ over the possible values of all the other parameters. 

The relative probability of two different models is given by their odds ratio, which is the ratio of their evidences (their Bayes factors) multiplied by the ratio of the prior probabilities of the two models.

\textit {Sampling methods.}  We used three statistical sampling methods to estimate the values of the parameters in our various waveform models and compute the evidence for these models: the publicly available nested sampler MultiNest (MN) \citep{2009MNRAS.398.1601F}, the parallel-tempering Markov chain Monte Carlo sampler (PT-emcee) that is included in the publicly available emcee package version~2.2.1 \citep{2013PASP..125..306F}, and a hybrid sampling algorithm  (see below) that we call MN+PT-emcee. 

The MN and PT-emcee sampling algorithms are complementary: nested samplers are optimized for computing the Bayesian evidence and are therefore well suited for computing the odds ratios of different models, whereas Markov chain Monte Carlo algorithms are optimized for parameter estimation. We carried out many tests of MN, PT-emcee, and MN+PT-emcee in the context of our waveform-fitting problem, using both synthetic and \textit{NICER} waveform data sets. 

For relatively simple waveform models and data sets, such as those with two uniform circular hot spots, we were able to adequately sample the full parameter space using either MN or PT-emcee in a reasonable amount of clock time. For these simpler models and data sets, the posterior probability density distributions given by these two algorithms appeared well converged and were in excellent agreement. 

For more complex models and data sets, such as our fits of models with two or three oval spots to synthetic data or to the \textit{NICER} data on PSR~J0030$+$0451, we found that, at least for the values of its sampling parameters that we explored, the MN algorithm abandoned regions of parameter space that contained solutions with log likelihoods comparable to the highest found in previous searches, did not give reproducible results, and---when we used it to analyze synthetic data---produced 1$\sigma$ and 2$\sigma$ credible regions that too often did not contain the values of the parameters that had been assumed in generating the synthetic data, even after we ran the code for a very long time. 

In contrast, even with poor initial seeds, the PT-emcee algorithm appeared able to sample well the parameter space of complicated models and data sets, gave reproducible results, and produced 1$\sigma$ and 2$\sigma$ credible regions that contained the values of the parameters that had been assumed in generating the synthetic data about as often as one would expect, i.e., about 68\% of the time for 1$\sigma$ credible regions and about 95\% of the time for 2$\sigma$ credible regions. We therefore adopted the PT-emcee algorithm for analyzing the more complex models and data sets we considered. However, with poor initial seeds, PT-emcee took large amounts of clock time to complete each parameter estimation.

In order to speed up parameter estimation using PT-emcee, we sought to choose high likelihood points as initial guesses. Having the MN nested sampling algorithm readily available, we typically used it to provide an initial guess. We tested this hybrid approach, which we refer to as MN+PT-emcee, extensively, and found that it was able to find correct solutions in much less clock time than was required if PT-emcee was not given a high likelihood initial guess. Consequently, we used MN+PT-emcee to estimate the values of the parameters in the models with two and three oval spots, and used the resulting samples to estimate the evidence for these models, given the data.

In our initial MultiNest surveys, we used 1000 active points (essentially the number of parameter combinations being actively updated), an efficiency of 0.01 (the target fraction of new parameter combinations that update the list of active points), and a tolerance of 0.1; for precise definitions of these parameters see \citet{2009MNRAS.398.1601F}. Based on our experience sampling the synthetic and \textit{NICER} waveform data, we limited the number of modes to five for models with two spots and 10 for models with three spots. In our PT-emcee analyses, we used 400 walkers when fitting models with two spots and 1000 walkers when fitting models with three spots, and the default temperature ladder.

\textit {Completeness of models.}  Although our parameter estimations and model comparisons are fully Bayesian, we also performed $\chi^2$ analyses to determine whether our models adequately describe the \textit{NICER} data. In computing $\chi^2$, we use the expression for model variance originally advocated by \citet{1900PHIL...50..302P}, namely,
\begin{equation}
\chi^2=\sum_i {(m_i-d_i)^2\over{m_i}}\; ,
\label{chi-squared}
\end{equation}
where $m_i$ is the number of counts in bin $i$ that are expected in the model, and $d_i$ is the observed number of counts in bin $i$, rather than the common data variance form in which the $m_i$ in the denominator is replaced by $d_i$.  The model variance form has the advantage that if the counts $d_i$ are drawn from a Poisson distribution with expected values $m_i$, then the expected value of $\chi^2$ is equal to the number of bins for any values of $m_i$. This is not true for the data variance form. We note that for our purposes the computation of $\chi^2$ is a one-way test: if the value of $\chi^2$ for a given number of degrees of freedom is large enough to be highly improbable then we are warned that our model of the spots and/or the instrument is likely to be incomplete, but if the value of $\chi^2$ is reasonably probable we cannot conclude that the model is correct.

\subsection {Treatment of Instrumental Uncertainties and Bias}
\label{sec:uncertainties-bias}

The method we used to make the final parameter estimates we report here takes into account the substantially larger uncertainties in the \textit{NICER} response below 0.4~keV (channel 40) and the evidence we found that including data from these channels might bias our results on PSR~J0030$+$045.

As we discuss in Section~\ref{sec:energy-channels}, we found that including data from channels below channel~40 biases waveform-based estimates of the distance to PSR~J0030$+$0451 toward values much larger than the independently known distance. The source of this bias is currently unknown. Since its origin is unknown, it could also bias waveform-based estimates of other parameters, including the radius and mass of the star. Given this risk, we judged it important to address this issue.

It might be possible to gain a better understanding of the origin of this distance bias if we also had reliable, independent estimates of the values of other stellar and waveform parameters, but for PSR~J0030$+$0451, we only have a reliable independent estimate of the distance. This provides too little information to identify the source(s) of this bias or to devise sophisticated corrections. We therefore considered only simple ways to address this bias. We also wanted to be careful not to do anything that might bias other, even more important model parameters.

As was mentioned in Section~\ref{sec:calibration}, the \textit{NICER} response below channel~40 is more uncertain than the response above channel~40. It was therefore natural to explore the effect of ignoring the data in channels below channel~40. As we show in detail in Section~\ref{sec:energy-channels}, discarding the data in these low channels eliminates the bias in our waveform-based estimates of the distance to PSR~J0030$+$045. It is our hope that discarding these data also reduces any potential biases in our estimates of the other parameters in the waveform model. Because this data selection procedure is so simple and does not touch the data that we do use, we think it is less likely to bias other waveform parameters than a more complex procedure.

In analyzing the \textit{NICER} data on PSR~J0030$+$0451, we found no detectable modulation in the energy channels above channel 299, and we therefore did not include data from these channels in our analyses of the \textit{NICER} data on PSR~J0030$+$0451.

One would expect that discarding the data in channels 25--39 would reduce the precisions of the parameter estimates. We investigated this possibility by comparing the precisions of the parameter estimates obtained by analyzing the data in channels 40--299 produced by a synthetic waveform similar to the PSR~J0030$+$0451 waveform observed by \textit{NICER} with the precisions obtained by analyzing the data in channels 25--299 produced by the same synthetic waveforms and found that discarding the data in channels 25--39 did not significantly reduce the precisions of parameter estimates (see Section~\ref{sec:analysis-2oval-synthetic-data}). We also compared the posterior probability distributions for the radius and mass of PSR~J0030$+$0451 obtained using the \textit{NICER} data in channels 40--299 with the probability distributions obtained using the data in channels 25--299 and found no evidence that the precisions of the parameter estimates were reduced by discarding the data in channels 25--39 (see Section~\ref{sec:energy-channels}). 

A remaining question is whether the fits of our waveform models to the \textit{NICER} waveform data on PSR~J0030$+$0451 are affected by the unavoidable systematic errors present in our model of the \textit{NICER} response in the energy channels 40--299 that we used. As we described in Section~\ref{sec:calibration}, the \textit{NICER} response in these channels was calibrated by comparing the results of \textit{NICER} observations of the Crab nebula, which is traditionally assumed to have a power-law spectrum, with the power-law spectrum of the nebula determined by observations made using other instruments. When \textit{NICER} observations of the Crab nebula spectrum are compared with the spectrum obtained by folding the power law determined by other instruments through the nominal \textit{NICER} response, the residuals in the energy channels we used in this study (channels 40--299) are $\lesssim\,$5\% \citep{2018ApJ...858L...5L}. As mentioned in Section~\ref{sec:calibration}, these residuals are likely due to imperfect modeling of the microphysics of the concentrator optics and astrophysical features, such as oxygen edges and emission lines.

We explored the likely effect of deviations of the nominal \textit{NICER} response from the true \textit{NICER} response by multiplying the nominal effective area curve of \textit{NICER} by the ratio between the count rate as a function of energy given by the assumed spectrum of the Crab nebula and the count rate as a function of energy measured by \textit{NICER} \citep{2018ApJ...858L...5L}. We then used this modified effective area to analyze both synthetic waveform data constructed to mimic the \textit{NICER} data on PSR~J0030$+$0451, and the actual \textit{NICER} waveform data on PSR~J0030$+$0451. 

We found that modifying the nominal effective area in this way had a negligible effect on the quality of our fits to synthetic waveform data and to the actual waveform data on PSR~J0030$+$0451. For example, the probability of the $\chi^2$ obtained by comparing the energy-resolved waveform given by the best-fit model with two oval spots to the \textit{NICER} data in energy channels assigned to 32 phase bins changed from 0.105 to 0.112 when we used the modified effective area curve instead of the nominal effective area curve. This indicates that the current systematic errors in the nominal effective area of \textit{NICER} in channels 40--299 have only a minor effect on our fits.

Given these results, we chose to ignore all events in the channels below channel~40 and use the current nominal \textit{NICER} effective area curve for channels 40--299, both when analyzing the actual \textit{NICER} data on PSR~J0030$+$0451 and when analyzing synthetic \textit{NICER} data.

\section {ANALYSIS OF SYNTHETIC PULSE WAVEFORM DATA}
\label{sec:fits-to-synthetic-data}

In order to verify that our parameter estimation procedures give correct posterior probability distributions for the range of pulse waveform models that we used to analyze the \textit{NICER} waveform data on PSR~J0030$+$0451, we applied these same procedures to synthetic waveform data sets that we created to mimic the \textit{NICER} data on PSR~J0030$+$0451, and checked that our procedures give appropriate parameter estimates and credible regions. 

As we explained in the previous section, we chose to discard the data on PSR~J0030$+$0451 in the \textit{NICER} energy channels below channel~40 and above channel 299. In order to verify that discarding the data in the channels below channel~40 would not bias our estimates of the waveform parameters, and to assess the effect of discarding these data on the precisions of our estimates, we first analyzed the data in channels 25--299 given by a particular synthetic waveform and then analyzed only the data in channels 40--299 given by the same synthetic waveform (we did not consider any channels below channel~25, because channel 25 was the lowest channel included in the cleaned \textit{NICER} data set). In each case, we analyzed the synthetic data using the MultiNest sampling algorithm and also using either the PT-emcee sampling algorithm or our hybrid MN+PT-emcee algorithm. 

By (1)~fitting models having two possibly different and overlapping circular spots to synthetic data generated assuming two different, non-overlapping circular spots and (2)~fitting models having two possibly different and overlapping oval spots to synthetic data generated assuming two different, non-overlapping oval spots, we were able to verify that our parameter estimation procedures yield appropriate posterior probability density distributions for these synthetic waveforms. Specifically, for these data sets the MultiNest and MN+PT-emcee sampling algorithms both produced posterior distributions for the waveform parameters that are fully consistent with the values of the parameters that were assumed when the synthetic waveforms were generated. We view the agreement of the results given by these two different sampling algorithms with each other and with the parameters assumed in generating the synthetic data as evidence that both algorithms give correct results when used to analyze synthetic waveforms like these, when the sampling is complete.  

By comparing the results we obtained by analyzing the synthetic waveform data in \textit{NICER} energy channels 40--299 with the results we obtained by analyzing the data in channels 25--299, we were also able to assess the effect on the precisions of our parameter estimates of using only the data in energy channels 40--299. We found that using this reduced data set had a negligible effect on the sizes of the credible regions.

We now provide the details of these tests.

\subsection {Analysis of Synthetic Waveforms Generated Using Two Uniform Circular Spots}
\label{sec:analysis-2circ-synthetic-data}

The first analysis of synthetic data we present here shows the results we obtained by fitting a 12-parameter waveform model with two possibly different and overlapping uniform circular spots to a synthetic waveform generated assuming two different, non-overlapping uniform circular spots, using our MN+PT-emcee sampling algorithm. Table~\ref{tab:synth2circchan25} lists the values of the waveform parameters that we assumed when we generated the synthetic waveform data. We note that the temperature we assumed for one of the two spots when we generated this synthetic waveform is substantially less than the temperatures of either of the two main spots we found when we analyzed the \textit{NICER} data on PSR~J0030$+$0451. We show our results for this synthetic waveform because it was the synthetic waveform that was selected---in advance of the analyses of the PSR~J0030$+$0451 waveform---to verify the parameter estimation procedures that would be used by the \textit{NICER} team to analyze pulse waveforms. The results obtained by others on the team when they analyzed this waveform are consistent with the results we obtained when we analyzed it \citep{2019ApJ...887L..26B}.

\begin{deluxetable*}{c|r|r|r|r|r|r}
    \tablecaption{Fits to Synthetic Data Generated Using Two Circular Spots.}
\tablewidth{0pt}
\tablehead{
      \colhead{Parameter} & \colhead{Assumed Value} & \colhead{Median} & \colhead{$-1\sigma$} & \colhead{$+1\sigma$} & \colhead{$-2\sigma$}  & \colhead{$+2\sigma$}
   \label{tab:synth2circchan25}
}
\startdata
      \hline
      $R_e ({\rm km})$ & 13.0 & 13.435 & 12.248 & 14.722 & 11.191 & 15.929  \\ 
      \hline
      $GM/(c^2R_e)$ & 0.1636  & 0.163 & 0.149 & 0.174 & 0.137 & 0.182  \\ 
      \hline
      $M (M_\odot)$ & 1.44 & 1.472 & 1.289 & 1.667 & 1.151 & 1.849  \\
      \hline
      $\theta_{\rm c1} ({\rm rad})$ & 0.6283 & 0.531 & 0.456 & 0.608 & 0.400 & 0.680  \\ 
      \hline
      $\Delta\theta_1 ({\rm rad})$ & 0.01 & 0.011 & 0.010 & 0.012 & 0.009 & 0.013  \\
      \hline
      $kT_{\rm eff,1} ({\rm keV})$ & 0.231 & 0.229 & 0.223 & 0.233 & 0.219 & 0.237  \\
      \hline
      $\theta_{\rm c2} ({\rm rad})$ & 2.077 & 2.096 & 1.959 & 2.260 & 1.871 & 2.401  \\
      \hline
      $\Delta\theta_2 ({\rm rad})$ & 0.33 & 0.358 & 0.322 & 0.397 & 0.288 & 0.446 \\
      \hline
      $kT_{\rm eff,2} ({\rm keV})$ & 0.0578 & 0.056 & 0.054 & 0.059 & 0.052 & 0.061  \\
      \hline
      $\Delta\phi_2 ({\rm cycles})$ & 0.5625 & 0.562 & 0.556 & 0.567 & 0.550 & 0.571  \\
      \hline
      $\theta_{\rm obs} ({\rm rad})$ & 0.733 & 0.774 & 0.662 & 0.894 & 0.583 & 0.996  \\
      \hline
      $N_{\rm H} (10^{20}~{\rm cm}^{-2})$ & 2.0 & 2.266 & 2.132 & 2.398 & 1.999 & 2.536 \\
      \hline
      $D ({\rm kpc})$ & 0.156 & 0.165 & 0.146 & 0.176 & 0.133 & 0.179 \\
      \hline 
\enddata
\tablecomments{Results obtained using our MN+PT-emcee sampling algorithm to fit a waveform model with two possibly different and overlapping uniform circular spots to the data in \textit{NICER} energy channels 25--299 from a synthetic waveform generated assuming two different, non-overlapping uniform circular spots.}
\end{deluxetable*}

\begin{deluxetable*}{c|c|c|c}
    \tablecaption{Verification of Fits to Synthetic Data Generated by Two Circular Spots}
\tablewidth{0pt}
\tablehead{
      \colhead{Energy Channels} & \colhead{$\pm 1\sigma$ (68.3\%)} & \colhead{$\pm 2\sigma$ (95.4\%)} & \colhead{$\pm 3 \sigma$ (99.7\%)}
    \label{tab:synth2circular}
}
\startdata
      \hline
      25--299 & 10 & 12 & 12\\
      \hline
      40--299 & 7 & 11 & 12\\
      \hline 
\enddata
\tablecomments{The number of values of the waveform parameters assumed in generating the synthetic waveform that are within the indicated 1D credible intervals derived from our fit of the 12-parameter waveform model with two possibly different and overlapping circular spots to the synthetic waveform generated assuming two different, non-overlapping circular spots, using the data in the indicated energy channels.}
\end{deluxetable*}

Table~\ref{tab:synth2circchan25} shows the results that we obtained by fitting the waveform model with two possibly different and overlapping uniform circular spots to synthetic waveform data generated assuming two different, non-overlapping uniform circular spots. The table lists the median value of each model parameter computed using its 1D posterior probability distribution and the boundaries of the $\pm1\,\sigma$ and $\pm2\,\sigma$ credible intervals for each parameter.

We found that when this waveform model was fit to the synthetic waveform data in \textit{NICER} energy channels 25--299, the values of~10 of the~12 parameters assumed in generating the synthetic waveform are within the relevant $\pm 1\,\sigma$ credible intervals for these parameters while the values of the remaining two parameters ($\theta_{c1}$ and $N_{\rm H}$) assumed in generating the synthetic waveform are within their $\pm 2\,\sigma$ credible intervals. 
We found that when this model was instead fit only to the synthetic data in \textit{NICER} energy channels~40--299, the assumed values of seven of the 12 parameters were within the relevant $\pm 1\,\sigma$ credible intervals, the assumed values of four others were within the relevant $\pm 2\,\sigma$ credible intervals, and the assumed value of the other parameter was within the relevant $\pm 3\,\sigma$ credible interval. 
Thus, for both  energy channel choices, the 1D posterior probability distributions derived for the waveform parameters are consistent with the values  of the parameters assumed when the synthetic waveform was generated. These results are summarized in Table~\ref{tab:synth2circular}.

\begin{figure*}[t!]
\begin{center}
  \resizebox{0.60\textwidth}{!}{\includegraphics{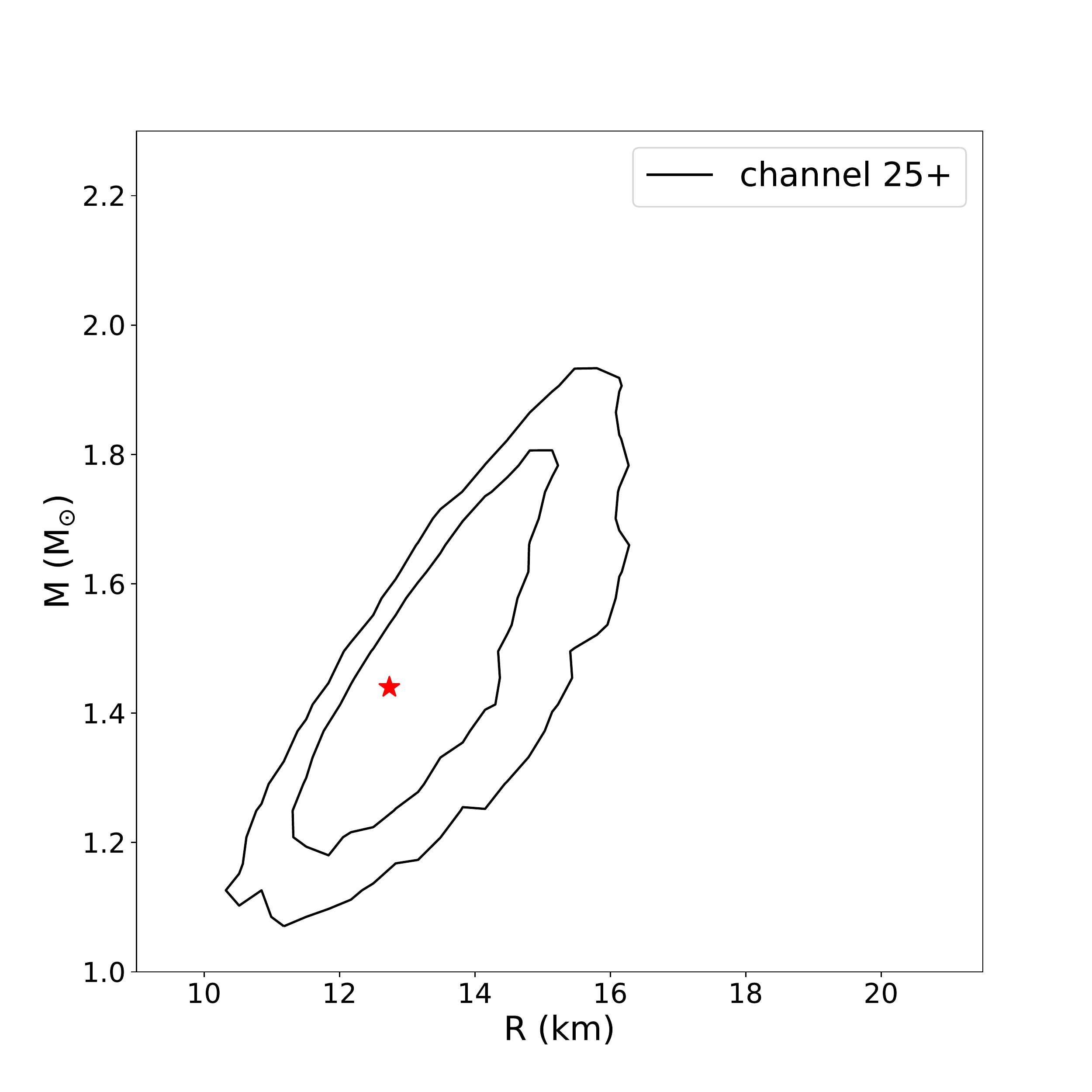}}
   \caption{68\% (inner) and 95\% (outer) contours of the joint probability density distribution of $M$ and $R_e$ obtained by fitting a model waveform with two possibly different uniform circular hot spots to a synthetic waveform that assumed two different uniform circular hot spots. This analysis used the synthetic waveform data in \textit{NICER} energy channels 25--299. The joint probability density of $M$ and $R_e$ obtained from this fit is consistent with the $M$ and $R_e$ values assumed when the synthetic data were generated, which are indicated by the star.}
\label{fig:synthcircmr}
\end{center}
\end{figure*}

Figure~\ref{fig:synthcircmr} shows that the joint probability density distribution for $M$ and $R_e$ that was obtained by fitting this model to the synthetic waveform data in \textit{NICER} energy channels 25--299 is consistent with the values of the stellar mass and radius assumed when the synthetic waveform was generated.

\begin{figure*}[ht!]
\gridline{\fig{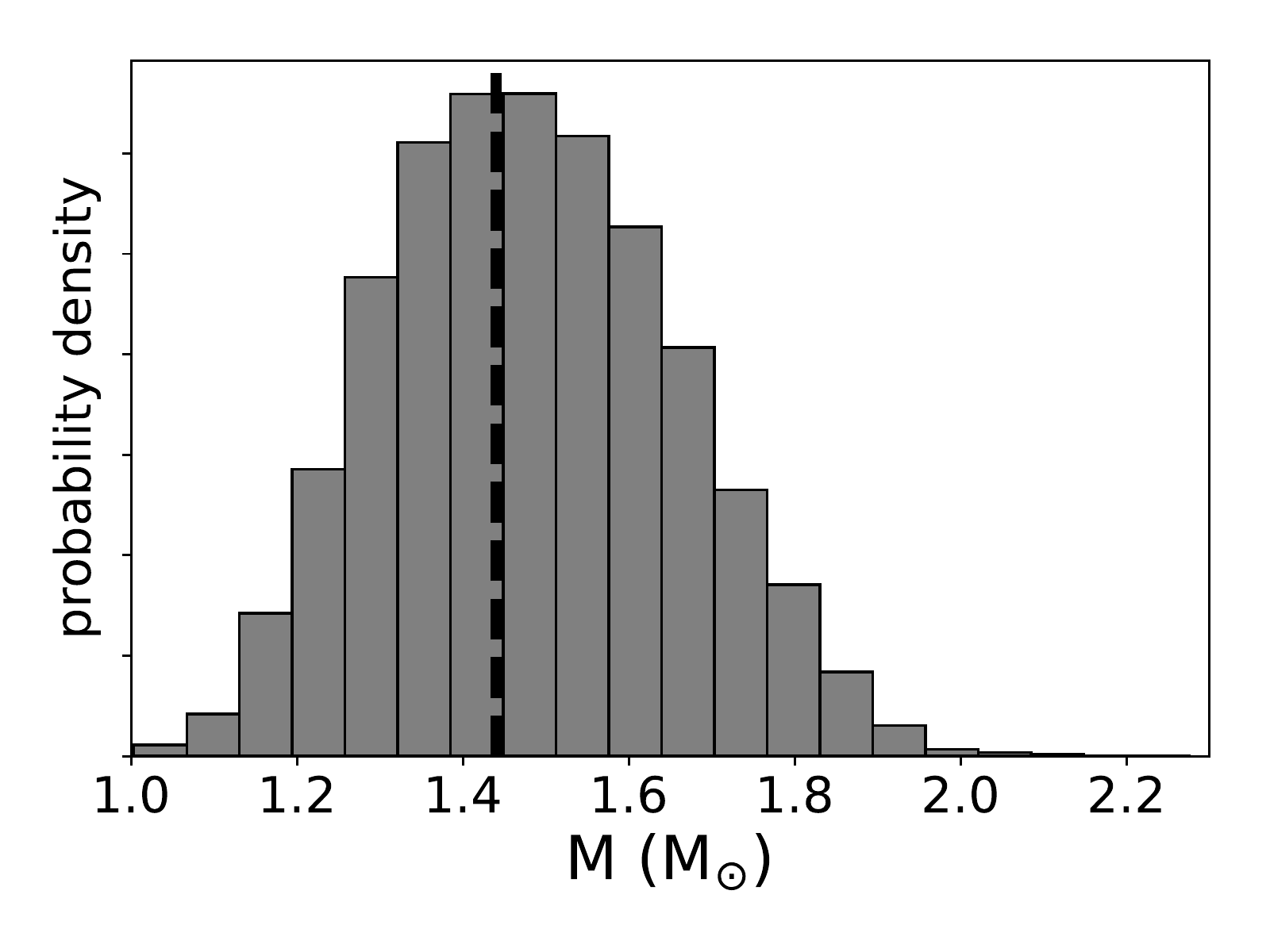}{0.3\textwidth}{(a)}
          \fig{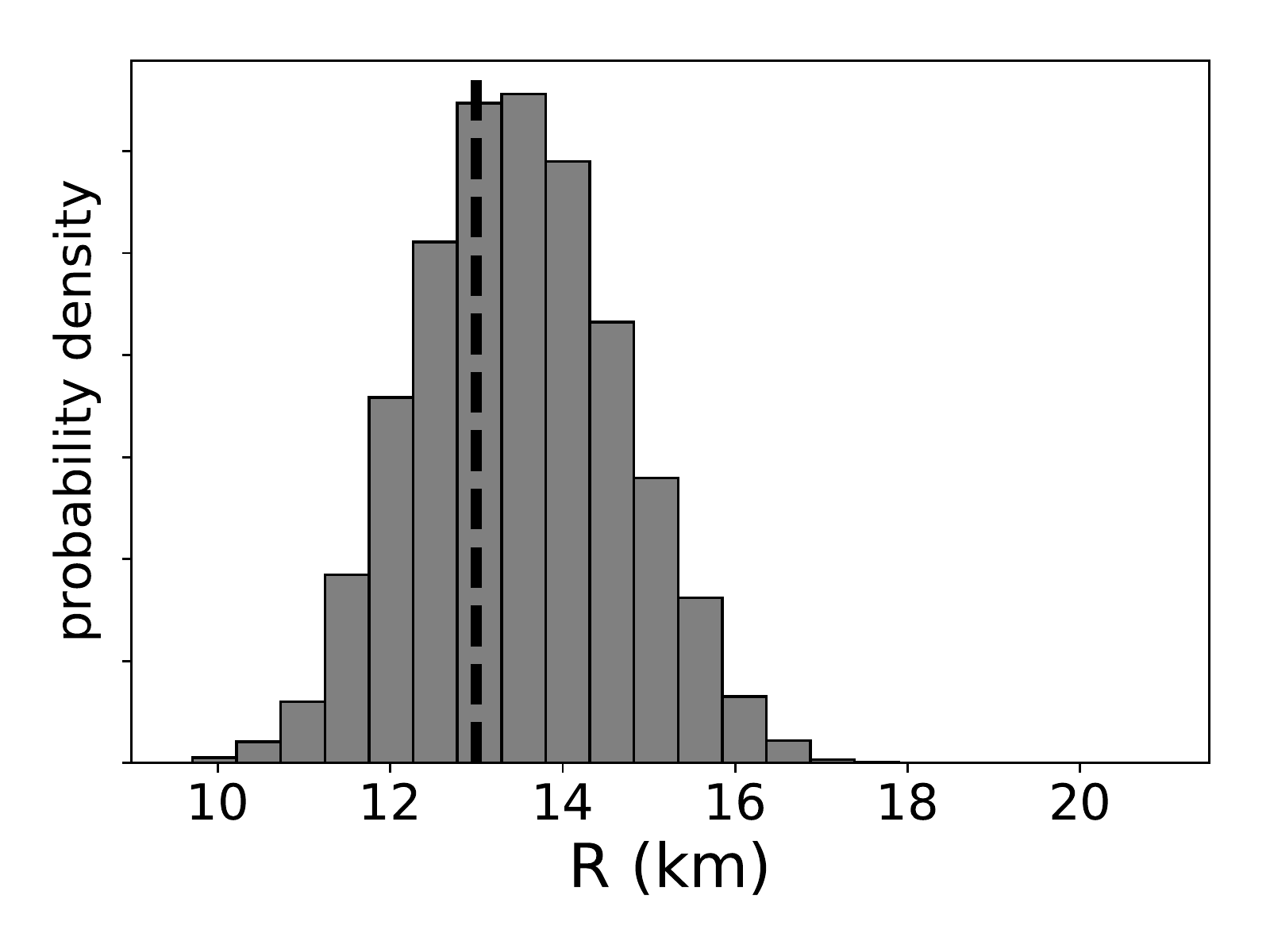}{0.3\textwidth}{(b)}
          \fig{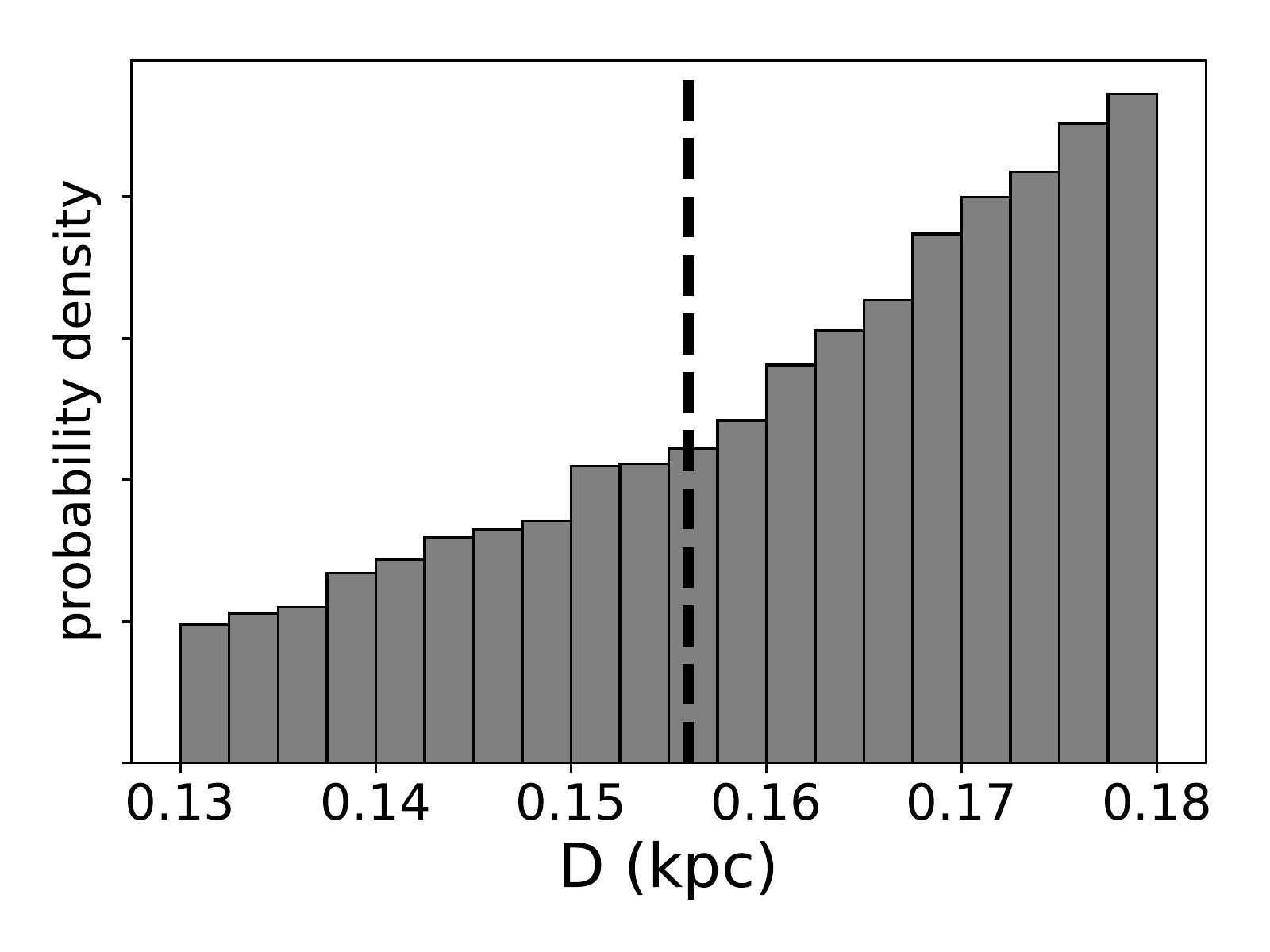}{0.3\textwidth}{(c)}}
\gridline{\fig{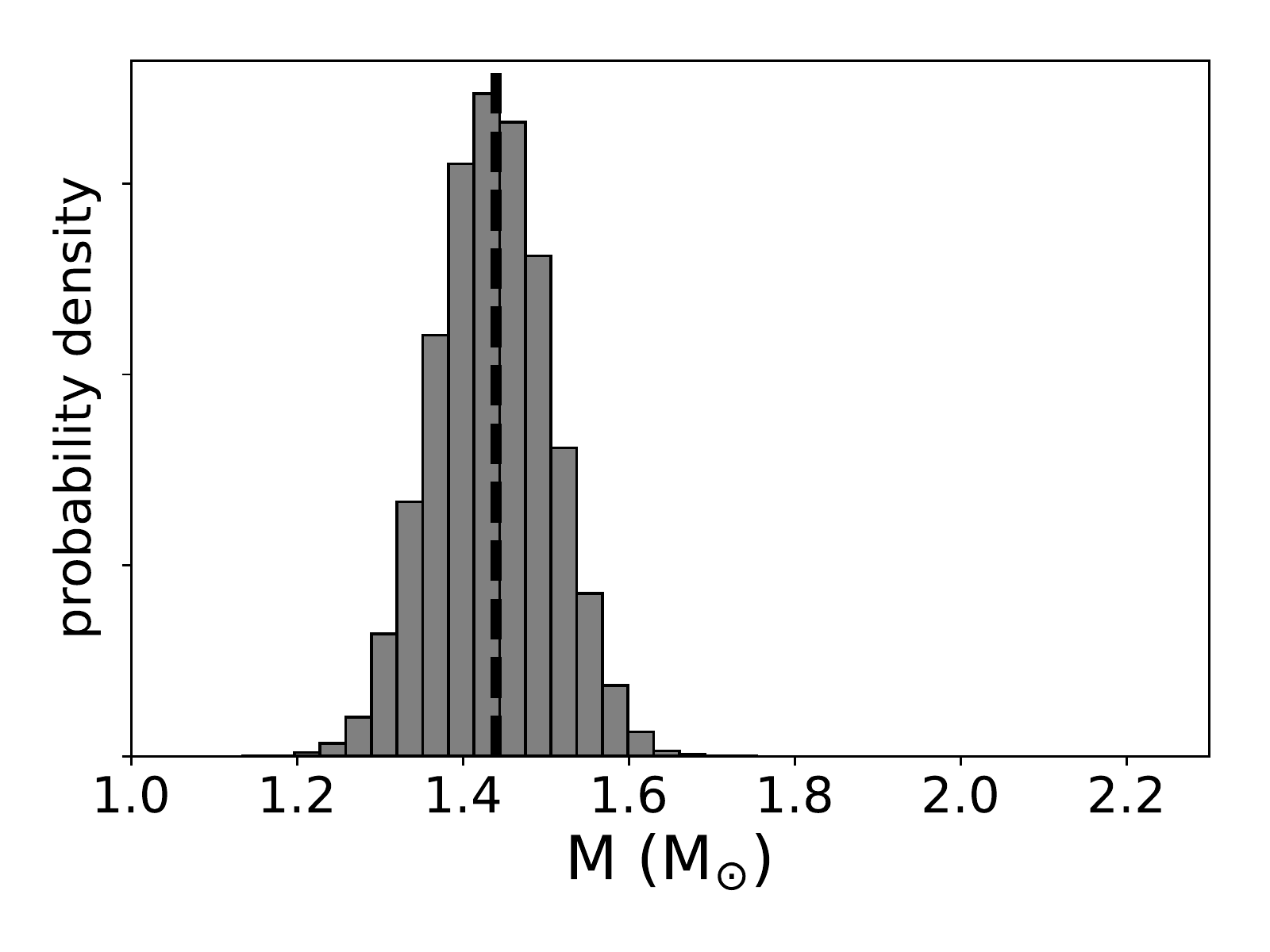}{0.3\textwidth}{(d)}
          \fig{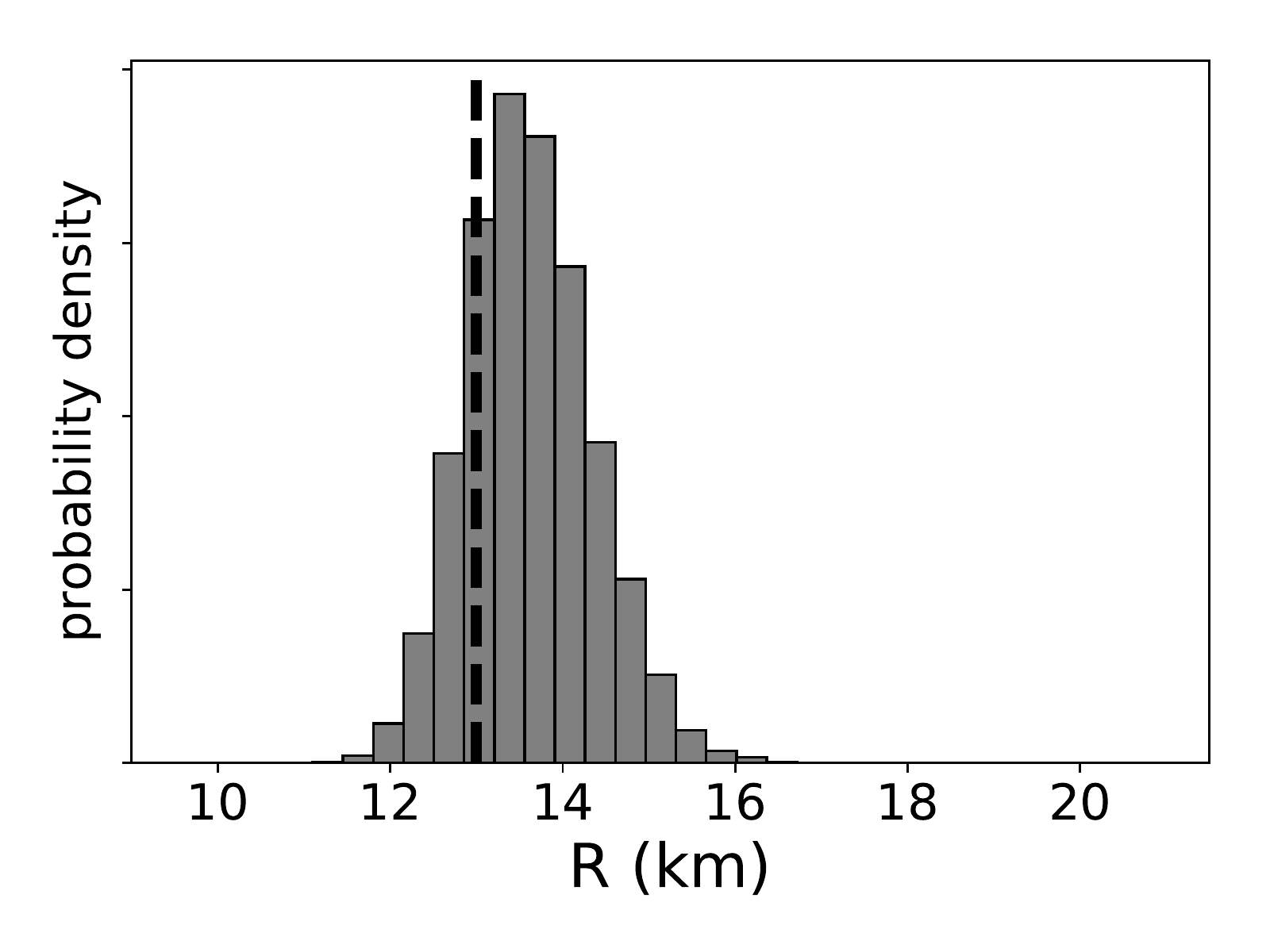}{0.3\textwidth}{(e)}
          \fig{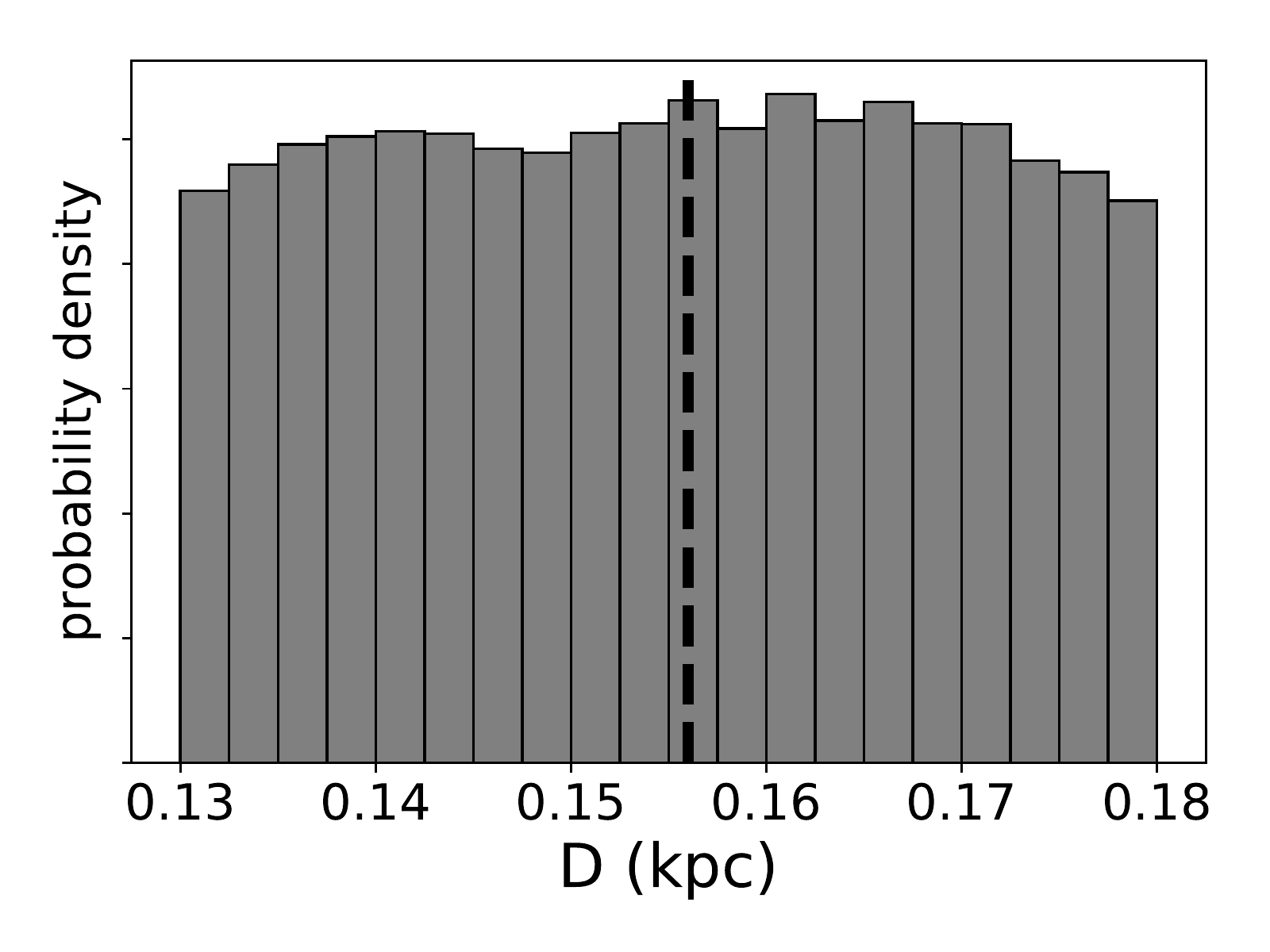}{0.3\textwidth}{(f)}}
\caption{Posterior probability density distributions for $M$, $R_e$, and $D$ obtained by fitting the waveform model with two possibly different uniform circular hot spots to synthetic waveform data generated using a model that assumed two different uniform circular hot spots. 
\textit{Top row:} results obtained by fitting the data in \textit{NICER} energy channels 25--299. 
\textit{Bottom row:} results obtained by fitting the data in \textit{NICER} energy channels 40--299.
The vertical dashed line in each panel shows the value of that waveform parameter that was assumed when generating the synthetic waveform data.}
\label{fig:synthcirc25vs40}
\end{figure*}

The panels in the top row of Figure~\ref{fig:synthcirc25vs40} show the 1D probability distributions of $M$, $R_e$, and $D$ obtained by fitting this model to the synthetic waveform data in \textit{NICER} energy channels~25--299, whereas the panels in the bottom row show the results obtained by fitting to the data in channels~40--299. This figure illustrates two noteworthy aspects of these distributions. 

First, these 1D probability distributions are consistent with each other and with the values of $M$, $R_e$, and $D$ that were assumed in generating the synthetic waveform. The widths of the distributions for $M$ and $R_e$ obtained using only the data in channels~40--299 underestimate the true uncertainties in the estimates of these parameters, because the temperature assumed for the lower-temperature spot when generating this synthetic waveform was sufficiently low that discarding the data in energy channels 25--39 narrowed these distributions. Our tests (see, e.g., Figure~\ref{fig:synthoval25vs40} below) have shown that---in contrast to the reduced widths of the $M$, $R_e$, and $D$ distributions obtained from this synthetic waveform when the data in channels 25--39 are discarded---discarding the data in these channels has no significant effect on the distributions of $M$, $R_e$, and $D$ obtained by analyzing the \textit{NICER} data on PSR~J0030$+$0451, which has higher-temperature hot spots.

Second, although the 1D probability distributions for $D$ obtained by fitting this model to data selected using both these energy-channel criteria are consistent with the value of $D$ assumed in generating the synthetic waveform data (the cumulative probability near the assumed value of $D$ is substantial for both channel selections and fits with wider priors show that the probability density eventually decreases with increasing distance), these results show that pulse waveform analysis does not determine $D$ very precisely. These analyses demonstrate the importance of using the precise and accurate independent measurement of $D$ that is available for PSR~J0030$+$0451 (see below) when analyzing the pulse waveform of this pulsar.

\subsection {Analysis of Synthetic Waveforms Generated Using Two Uniform Oval Spots}
\label{sec:analysis-2oval-synthetic-data}

The second analysis of synthetic data we present here shows the results we obtained by fitting a 14-parameter waveform model with two possibly different and overlapping uniform oval spots to a synthetic waveform generated assuming two different, non-overlapping uniform oval spots, using our MN+PT-emcee sampling algorithm. Table~\ref{tab:synth2ovalchan40} lists the values of the waveform parameters that we assumed when we generated the synthetic waveform data.

\begin{deluxetable*}{c|r|r|r|r|r|r}
    \tablecaption{Fits to Synthetic Data with Two Oval Spots.}
\tablewidth{0pt}
\tablehead{
      \colhead{Parameter} & \colhead{Assumed Value} & \colhead{Median} & \colhead{$-1\sigma$} & \colhead{$+1\sigma$} & \colhead{$-2\sigma$}  & \colhead{$+2\sigma$}
   \label{tab:synth2ovalchan40}
}
\startdata
      \hline
      $R_e ({\rm km})$ & 13.49 & 15.997 & 14.497 & 17.471 & 13.044 & 18.798  \\
      \hline
      $GM/(c^2R_e)$ & 0.1503 & 0.145 & 0.136 & 0.153 & 0.129 & 0.161  \\
      \hline
      $M (M_\odot)$ & 1.374 & 1.565 & 1.386 & 1.750 & 1.225 & 1.918  \\
      \hline
      $\theta_{\rm c1} ({\rm rad})$ & 2.251 & 2.247 & 2.181 & 2.319 & 2.113 & 2.393  \\
      \hline
      $\Delta\theta_1 ({\rm rad})$ & 0.026 & 0.024 & 0.021 & 0.028 & 0.018 & 0.033  \\ 
      \hline
      $f_1$ & 9.14 & 7.356 & 5.589 & 9.558 & 4.603 & 11.742  \\
      \hline
      $kT_{\rm eff,1} ({\rm keV})$ & 0.1151 & 0.114 & 0.111 & 0.117 & 0.109 & 0.119  \\
      \hline
      $\theta_{\rm c2} ({\rm rad})$ & 2.442 & 2.459 & 2.401 & 2.516 & 2.336 & 2.567  \\
      \hline
      $\Delta\theta_2 ({\rm rad})$ & 0.031 & 0.028 & 0.025 & 0.032 & 0.022 & 0.038  \\
      \hline
      $f_2$ & 16.23 & 17.465 & 14.672 & 19.288 & 11.941 & 19.907  \\
      \hline
      $kT_{\rm eff,2} ({\rm keV})$ & 0.1164 & 0.115 & 0.113 & 0.117 & 0.111 & 0.119  \\
      \hline
      $\Delta\phi_2 ({\rm cycles})$ & 0.457 & 0.455 & 0.452 & 0.457 & 0.450 & 0.459  \\
      \hline
      $\theta_{\rm obs} ({\rm rad})$ & 0.939 & 0.931 &  0.848 &  1.003 &  0.766 &  1.071 \\
      \hline
      $N_{\rm H} (10^{20}~{\rm cm}^{-2})$ & 0.084 & 0.163 &  0.051 &  0.355 &  0.007 &  0.654 \\
      \hline
      $D ({\rm kpc})$ & 0.352 & 0.350 &  0.315 &  0.369 &  0.284 &  0.374 \\
      \hline 
\enddata
\tablecomments{Results obtained using our MN+PT-emcee sampling algorithm to fit a waveform model with two possibly different and overlapping uniform oval spots to \textit{NICER} energy channels 40--299 of a synthetic waveform generated assuming two different, non-overlapping uniform oval spots.}
\end{deluxetable*}

The results that we obtained by fitting this waveform model to the synthetic waveform data in \textit{NICER} energy channels 25--299 and 40--299 are shown in Table~\ref{tab:synth2ovalchan40}, which lists the median value of each model parameter computed using its 1D posterior probability distribution and the boundaries of the $\pm1\,\sigma$ and $\pm2\,\sigma$ credible intervals for each parameter.

\begin{figure*}[ht!]
\begin{center}
\vspace{-0.13truein}
  \resizebox{0.5\textwidth}{!}{\includegraphics{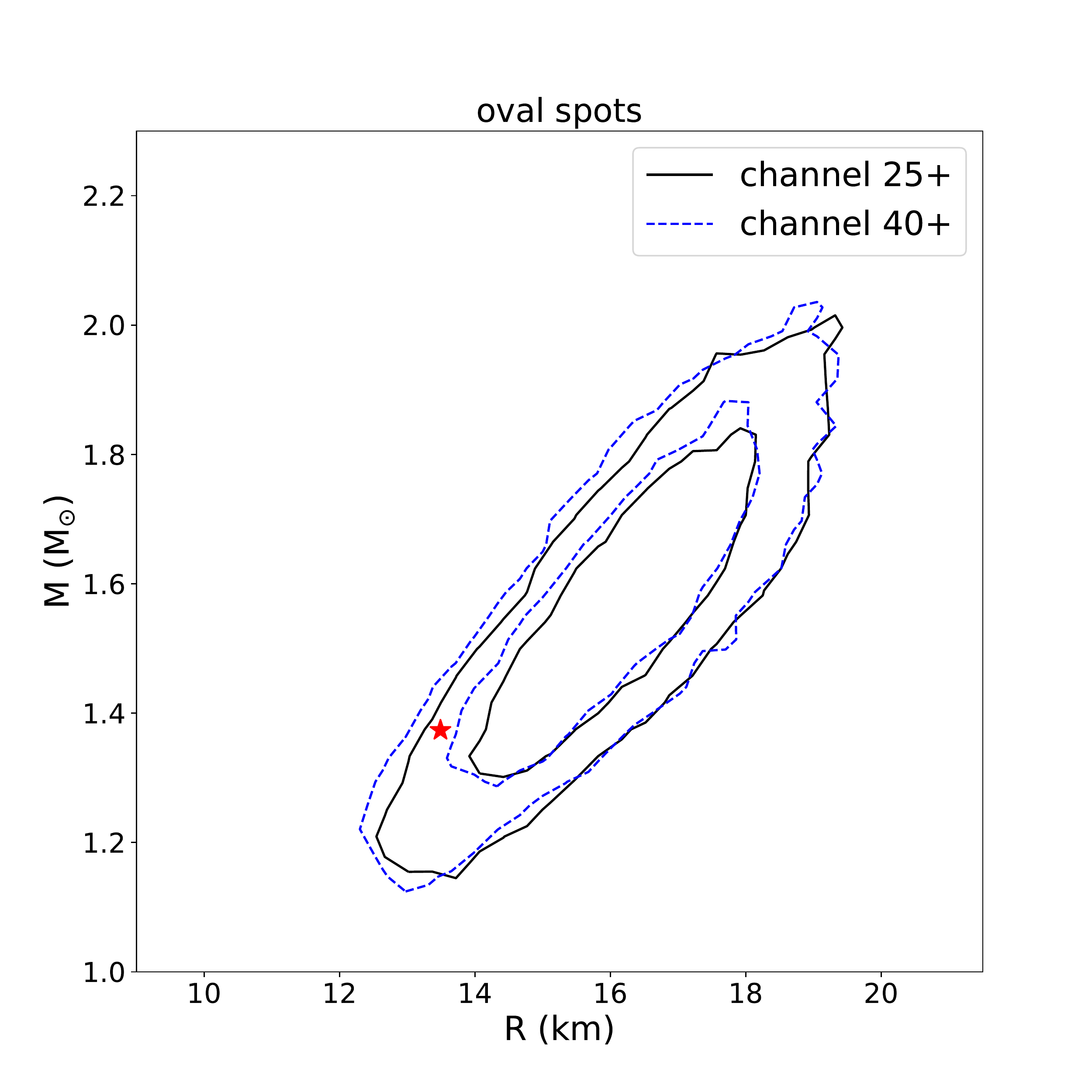}}
\vspace{-0.15truein}
   \caption{Plot of the 68\% (inner) and 95\% (outer) contours of the joint probability density distribution of $M$ and $R_e$ obtained by fitting a model waveform with two possibly different and overlapping uniform oval hot spots to a synthetic waveform that assumed two different, non-overlapping uniform oval spots. The solid black lines show the contours obtained when the synthetic waveform data in \textit{NICER} energy channels 25--299 were used, whereas the dashed blue lines show the contours obtained when the synthetic waveform data in energy channels 40--299 were used. Both joint probability distributions are consistent with the $M$ and $R_e$ values assumed when the synthetic data were generated, which are indicated by the star. These results show that using only the data in energy channels 40--299 neither biases the $M$ and $R_e$ estimates nor significantly reduces their precisions.}
\label{fig:synthovalmr}
\end{center}
\end{figure*}

We found that when the model waveform was fit to the synthetic waveform data in \textit{NICER} energy channels 25--299, the values of~10 of the~14 parameters assumed in generating the synthetic waveform are within the $\pm 1\,\sigma$ credible intervals for these parameters, the values assumed for three others are within the relevant $\pm 2\,\sigma$ credible intervals, and the value assumed for the remaining parameter is within the relevant $\pm 3\,\sigma$ credible interval. When this model waveform was fit only to the synthetic data in \textit{NICER} energy channels~40--299, the assumed values of~12 of the~14 parameters fall within their $\pm 1\,\sigma$ credible intervals and the assumed values of the other two parameters fall within their $\pm 2\,\sigma$ credible intervals. The waveform parameter values that were assumed when generating the synthetic waveform data are all consistent with the 1D posterior probability densities obtained by analyzing the data using either energy cut. These results are summarized in Table~\ref{tab:synth2oval}.

\begin{deluxetable*}{c|c|c|c}
    \tablecaption{Verification of Fits to Synthetic Data Generated by Two Oval Spots}
\tablewidth{0pt}
\tablehead{
      \colhead{Energy Channels} & \colhead{$\pm 1\sigma$ (68.3\%)} & \colhead{$\pm 2\sigma$ (95.4\%)} & \colhead{$\pm 3 \sigma$ (99.7\%)}
    \label{tab:synth2oval}
}
\startdata
      \hline
      $25-299$ & 10 & 13 & 14\\
      \hline
      $40-299$ & 12 & 14 & 14\\
      \hline 
\enddata
\tablecomments{The number of values of the waveform parameters assumed in generating the synthetic waveform that are within the indicated 1D credible intervals derived from our fit of the 14-parameter waveform model with two possibly different and overlapping oval spots to the synthetic waveform generated assuming two different, non-overlapping oval spots, using the data in the indicated energy channels.}
\end{deluxetable*}

Figure~\ref{fig:synthovalmr} shows that the joint probability density distribution for $M$ and $R_e$ obtained by fitting this model to the synthetic waveform data in \textit{NICER} energy channels 25--299, and by fitting it to the synthetic data in channels 40--299, are both consistent with the values of the stellar mass and radius assumed when the synthetic data was generated. Figure~\ref{fig:synthovalmr} also demonstrates that discarding the data in \textit{NICER} energy channels 25--39 does not significantly degrade the precision of the $M$ and $R_e$ estimates.

\begin{figure*}[ht!]
\gridline{\fig{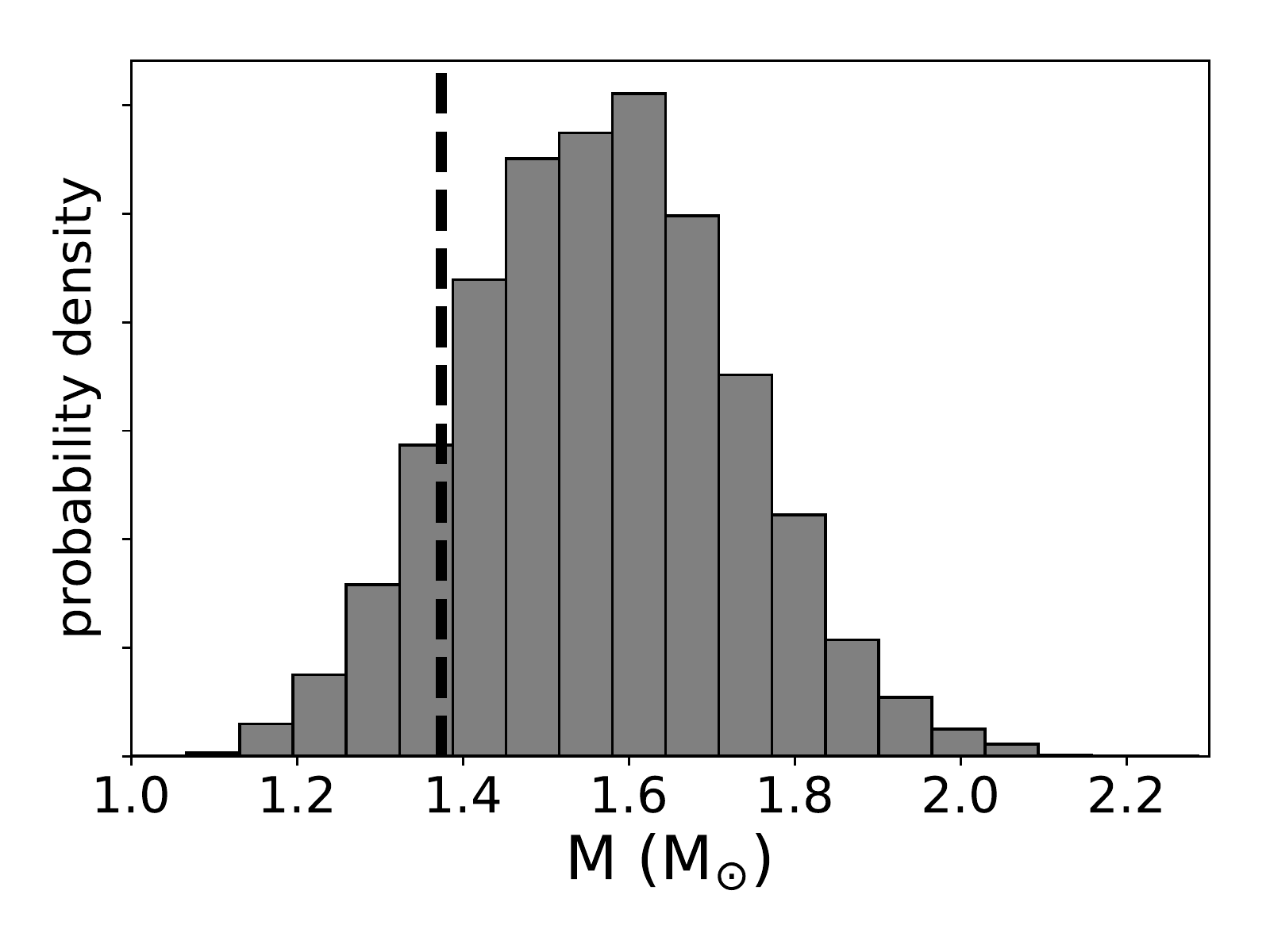}{0.3\textwidth}{(a)}
          \fig{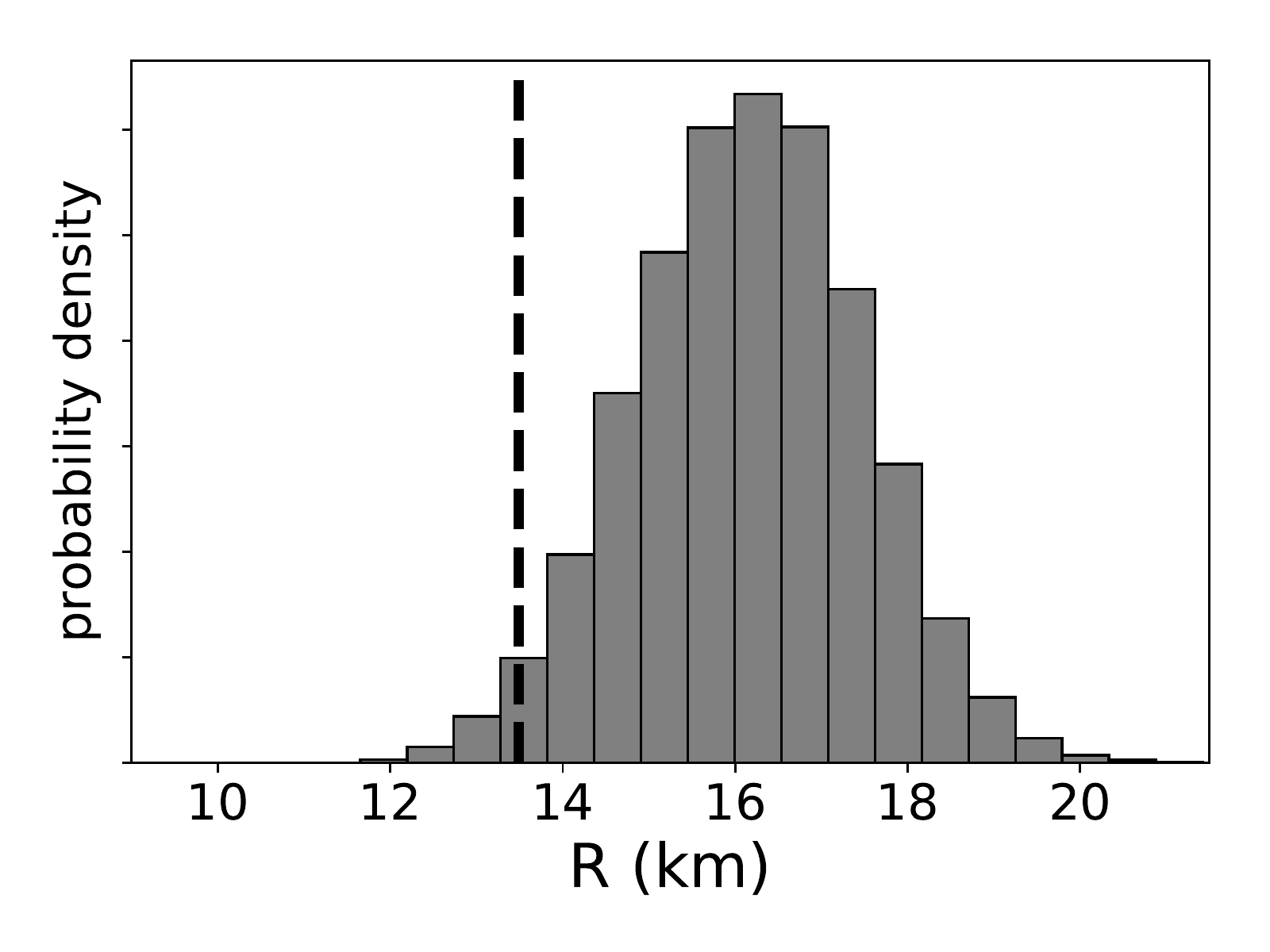}{0.3\textwidth}{(b)}
          \fig{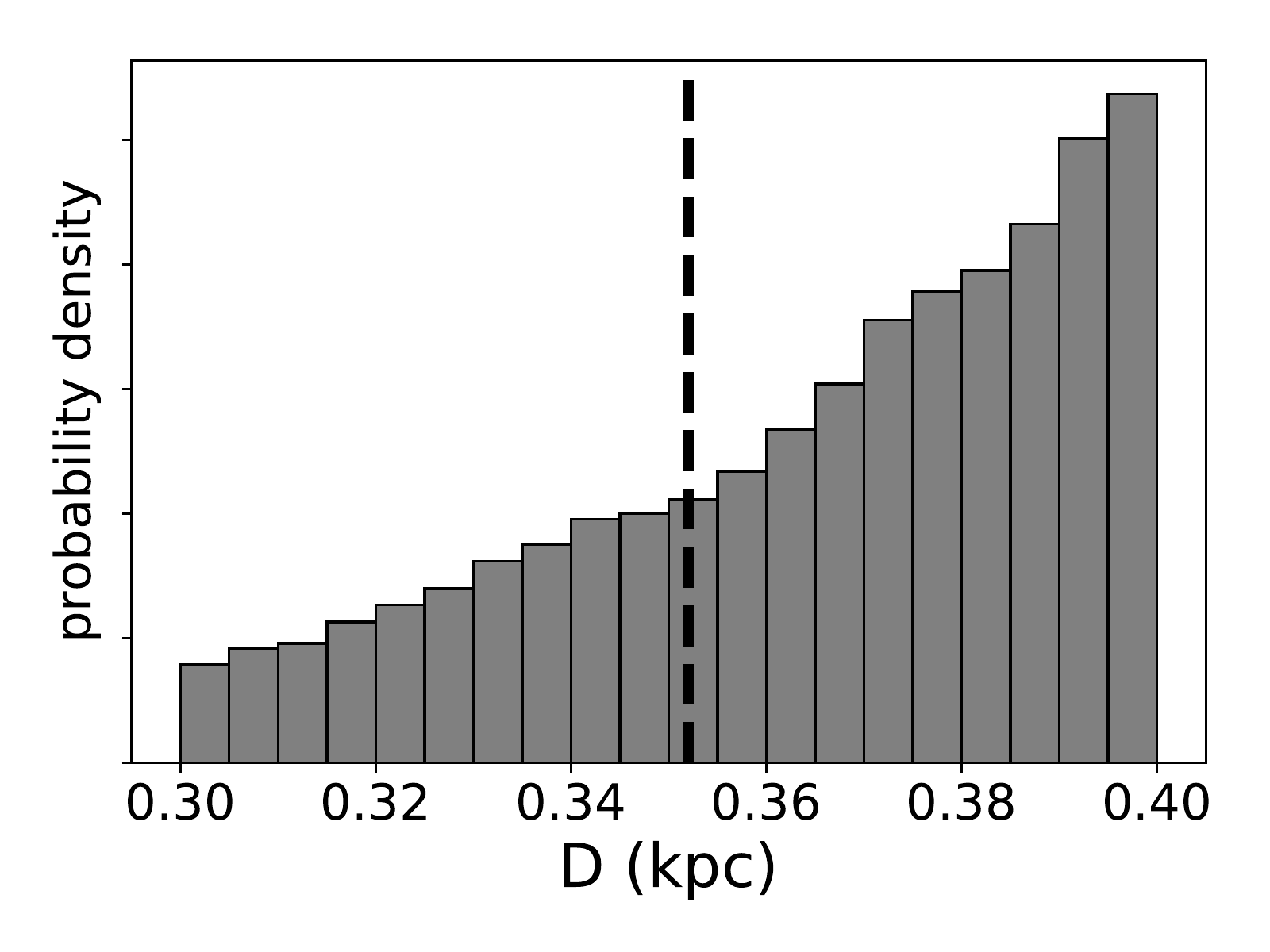}{0.3\textwidth}{(c)}}
\gridline{\fig{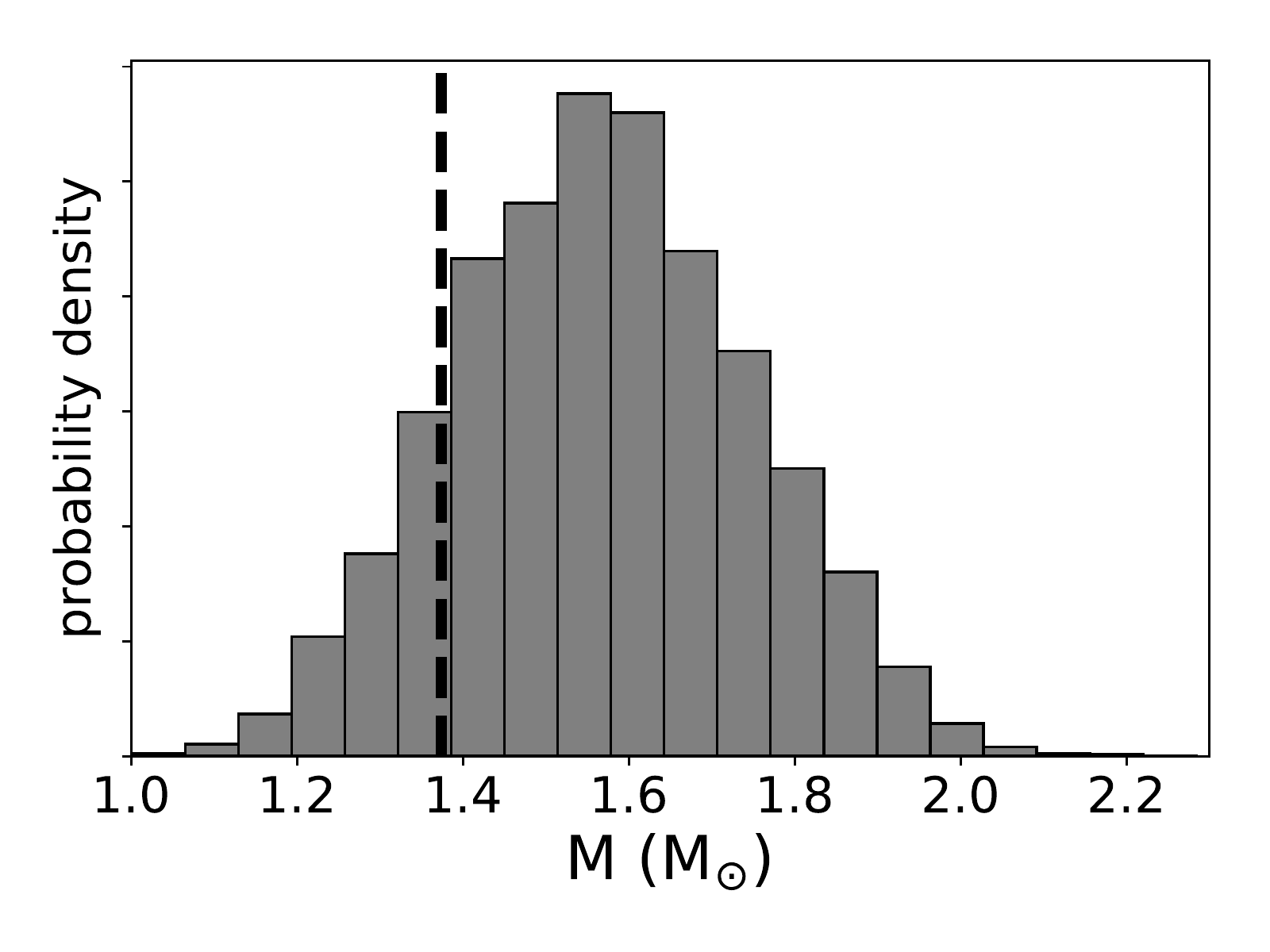}{0.3\textwidth}{(d)}
          \fig{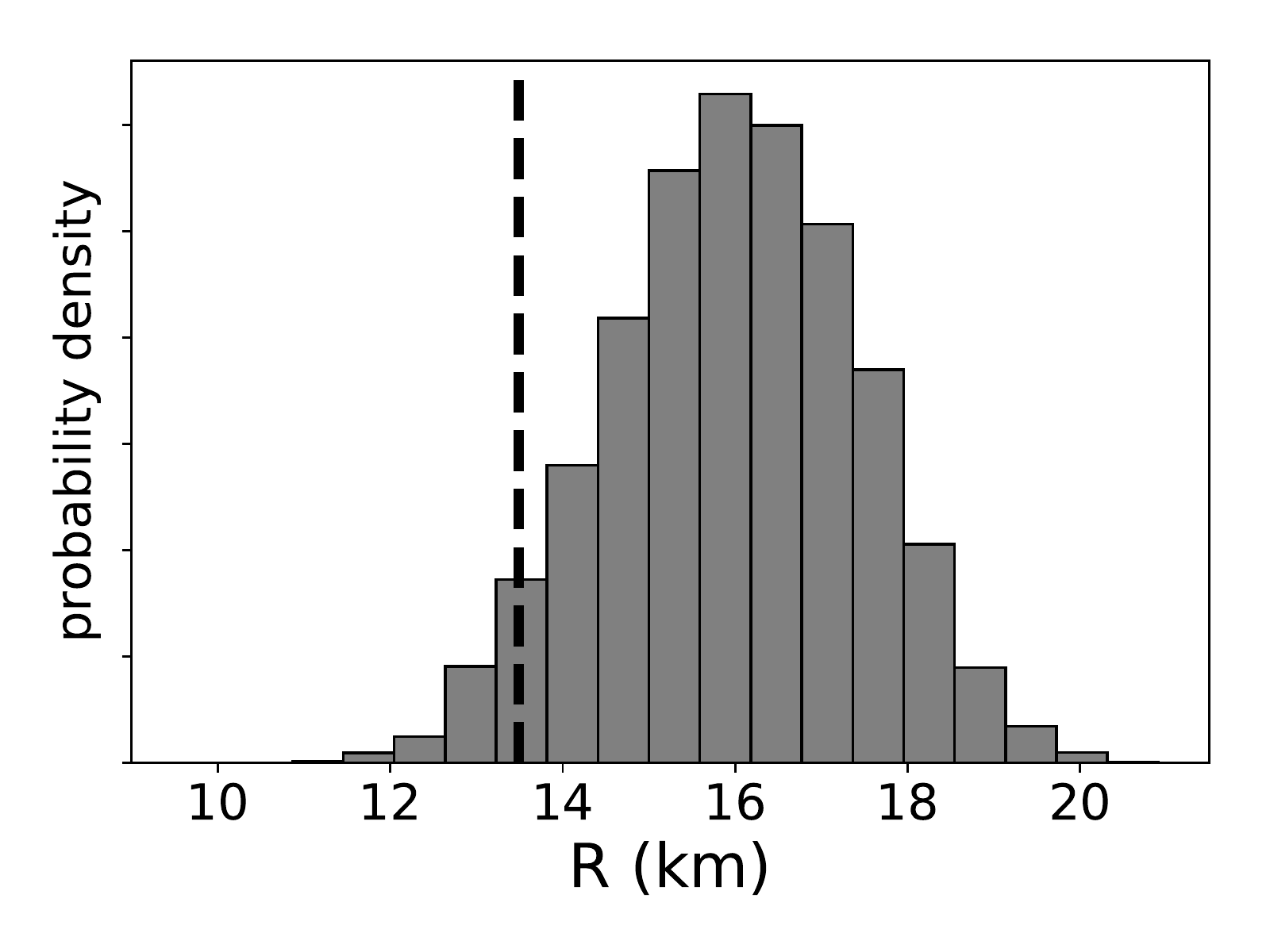}{0.3\textwidth}{(e)}
          \fig{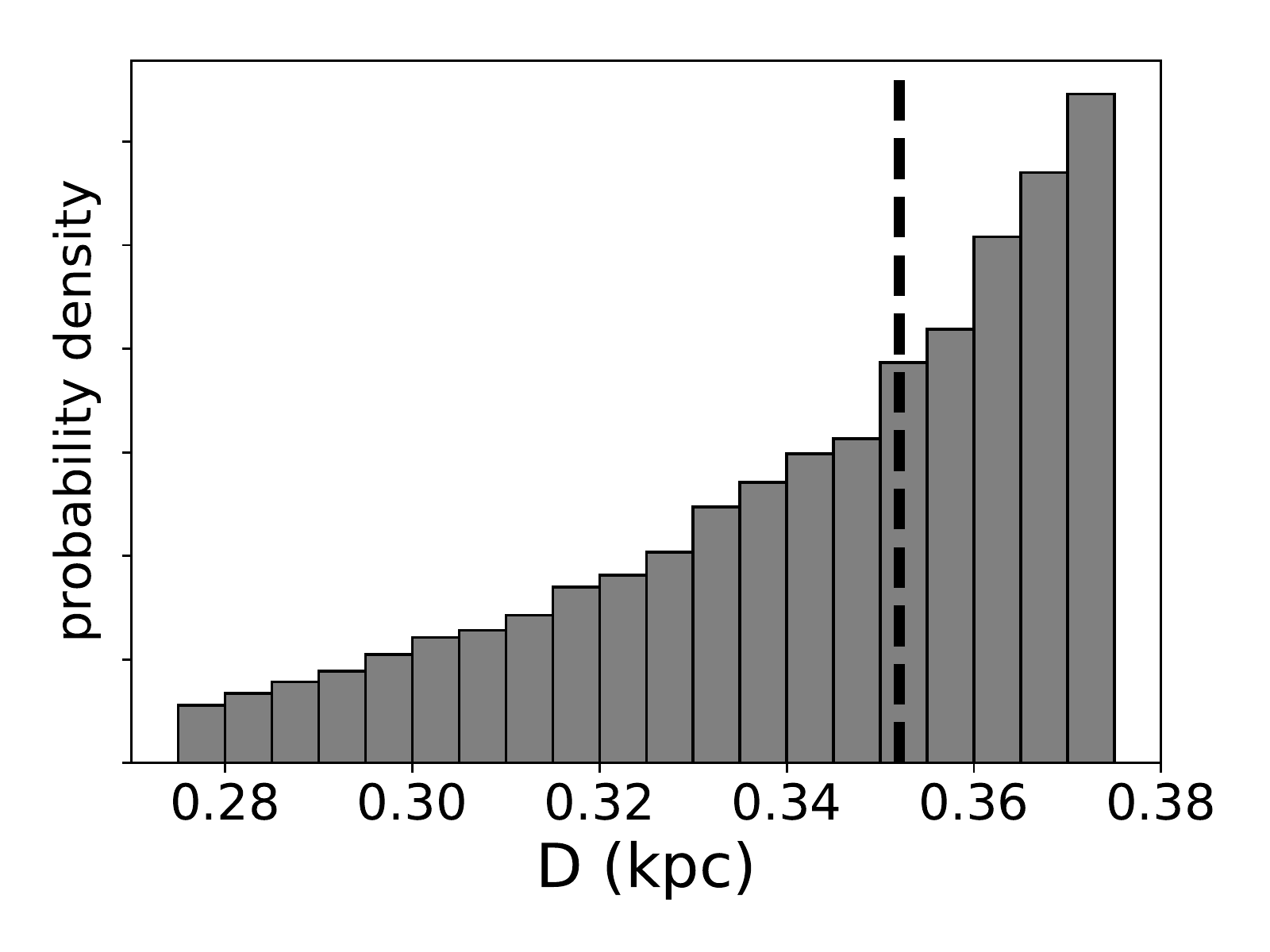}{0.3\textwidth}{(f)}}
\vspace{-0.1truein}
\caption{Posterior probability density distributions for the $M$, $R_e$, and $D$ estimates obtained by fitting the waveform model with two possibly different and overlapping uniform oval hot spots to synthetic waveform data generated using a model that assumed two different, non-overlapping uniform oval hot spots. 
\textit{Top row:} results obtained by fitting the data in \textit{NICER} energy channels 25--299. 
\textit{Bottom row:} results obtained by fitting the data in \textit{NICER} energy channels 40--299.  The vertical dashed line in each panel shows the value of that waveform parameter that was assumed when generating the synthetic waveform data.}
\label{fig:synthoval25vs40}
\end{figure*}

The panels in the top row of Figure~\ref{fig:synthoval25vs40} show the 1D probability distributions of $M$, $R_e$, and $D$ obtained by fitting this model to the synthetic waveform using the data in \textit{NICER} energy channels~25--299, whereas the panels in the bottom row show the results obtained using only the data in energy channels~40--299. We note several aspects of these distributions. 

First, these 1D probability distributions are consistent with each other and with the values of $M$, $R_e$, and $D$ that were assumed in generating the synthetic waveform. Second, these distributions are not significantly affected by using only the data in energy channels~40 and above. Finally, although the 1D probability distributions for $D$ obtained by fitting this model to these data are consistent with the value of $D$ assumed in generating the synthetic waveform (the cumulative probability near the assumed value of $D$ is substantial for both energy cuts and fits with wider priors show that the probability density eventually decreases with increasing distance), waveform analysis again does not determine $D$ very precisely. This further emphasizes the value of using the precise and accurate independent measurement of $D$ (see below) when analyzing the \textit{NICER} waveform data on PSR~J0030$+$0451.

\section{ANALYSIS OF THE \textit{NICER} PULSE WAVEFORM DATA}
\label{sec:fits-to-nicer-data}

In this section we describe our estimation of the radius and mass of  PSR~J0030$+$0451 using \textit{NICER} observations of its soft X-ray pulse waveform. We first describe our modeling of this waveform and detail the basis for our decision to use the data in \textit{NICER} energy channels 40--299. We then present our radius and mass estimates, and the evidence for the validity of the model that we used to make these estimates. Finally, we discuss our choices of waveform-phase resolution, the sizes and locations of the hot spots given by our modeling, the best-fit colatitude of the observer, and the combined, waveform- and phase-integrated X-ray spectrum of the hot spots given by our analysis. 

As we have previously described in Section~\ref{sec:parameter-estimation}, we explored a variety of algorithms for sampling the parameter space of waveform models but found the best results using the hybrid sampling algorithm that we refer to as MN+PT-emcee. Therefore, in this section we present only results obtained using this algorithm.

\subsection{Modeling the Waveform of PSR~J0030$+$0451}
\label{sec:modeling-the-J0030-waveform}

Our algorithm for developing a successful model of the heated regions on the surface of PSR~J0030$+$0451 was to proceed systematically from the simplest possible model to more complicated models, guided at each step by the differences between the waveform given by the best-fit simpler models and the waveform observed using \textit{NICER}, and ending when adding additional hot spots or complexity fails to increase significantly the evidence for the model. The sequence of models we considered in our quantitative modeling is presented in Table~\ref{tab:model-sequence}. We now discuss this sequence.

As mentioned in Section~\ref{sec:waveform-modeling}, visual examination of the PSR~J0030$+$0451 waveform shows evidence for two peaks. As a result, waveform models constructed using a single hot spot fail to reproduce the qualitative features of the observed waveform. We therefore began our quantitative modeling of the observed waveform using models with two uniform circular hot spots, allowing the spots to have different locations, areas, and temperatures. We also allowed the two circular spots to overlap partially or completely. The sampling algorithm therefore explored configurations in which the two circular spots (1)~are disjoint, with the same or different temperatures; (2)~partially overlap, potentially producing a somewhat oval-shaped uniform-temperature spot if the two component spots have the same temperature, or a spot with a complicated boundary and a complex temperature structure, if the two component spots have different temperatures; or (3)~fully overlap, producing (3a)~a single, uniform-temperature circular spot, if one of the spots completely covers the other, (3b)~a spot with a centered or off-center core-annulus structure in which the core is hotter or colder than the annulus, if the two spots have different temperatures, or (3c)~a circular spot partially surrounded by a hotter or colder crescent-shaped region. None of these models produced waveforms that adequately describe the observed waveform. Among other problems, the (relatively large) spot sizes required to reproduce the weak higher harmonic content of the observed waveform appeared to conflict with the (relatively small) spot areas required to reproduce the observed X-ray flux. We therefore considered more complicated heated regions.

In the second step of our exploration of waveform models, we considered models with three and four uniform circular hot spots. In all cases we allowed the spots to overlap in any way that improved the fit. This algorithm allowed the heated regions in the model to evolve toward a variety of different complex shapes as the fitting process proceeded (see Section~\ref{sec:waveform-modeling}). These explorations found no significant evidence for configurations with more than three circular hot spots on the stellar surface, for configurations in which three hot spots were near one another, or for overlapping or adjacent hot spots with significantly different temperatures, such as one would expect if there were spots on the stellar surface that had regions with different temperatures or a significant temperature gradient. We again found that the larger spot sizes required to reproduce the weak higher harmonic content of the waveform appeared to conflict with the smaller sizes required to reproduce the observed X-ray flux. This conflict led us to consider uniform-temperature oval hot spots as the next step in our sequence of trial models.

The first oval hot-spot model we considered examined heated regions that could be modeled using two uniform-temperature oval spots. We again allowed the two spots to have different properties and to overlap one another, partially or completely. The sampling algorithm could have collapsed this model to a model with two uniform-temperature circular spots, if that was preferred by the data, but it was not. The sampling algorithm therefore explored configurations in which the two oval spots (1)~are disjoint, with the same or different temperatures; (2)~partially overlap, potentially producing a uniform-temperature spot with a complicated boundary, if both spots have the same temperature, or a spot with a complicated boundary and a complex temperature structure, if the two spots have different temperatures; or (3)~fully overlap, producing (3a)~a single, uniform-temperature oval spot, if one of the spots completely covers the other, (3b)~a possibly oval spot with a potentially oval centered or off-center core-annulus structure in which the core is hotter or colder than the annulus, if the two spots have different temperatures, or (3c)~a possibly oval spot partially surrounded by a hotter or colder crescent-shaped region. 

Our analysis showed that a configuration with two non-overlapping, uniform oval spots with almost the same temperatures is preferred over any models with two overlapping spots, adequately describes the observed waveform, and is strongly favored over the best-fit model with two circular spots. Indeed, the log evidence for the best-fit model with two oval spots was 11.6 larger than the log evidence for the best-fit model with two circular spots. The fit with two uniform oval spots favors a configuration in which both spots are elongated in the east$-$west direction and both have almost the same temperature. The larger east$-$west extent of the best-fitting oval spots allows them to successfully reproduce the observed weak higher harmonic content of the waveform while the smaller north$-$south extent allows them to have the relatively small areas required to reproduce the observed X-ray flux.

To assess whether a model with two uniform-temperature oval spots is sufficient, we also considered configurations with three, possibly different, uniform-temperature oval spots. We found that a configuration with three different non-overlapping oval spots also describes the observed waveform adequately. This model is again strongly favored over the best-fit model with two uniform circular spots: the log evidence for the best-fit model with three oval spots was 13.3 greater than the log evidence for the best-fit model with two circular spots. We again found no evidence for more than one temperature in any of the heated regions. Two of the three spots in the best-fitting three-spot model are almost identical to the two spots in the best-fitting two-spot model. Like the two spots in the best-fitting two-spot model, the two larger spots in the best-fitting three-spot model are elongated in the east$-$west direction and have almost the same temperatures. The third spot in the best-fitting three-spot model has a much higher temperature but a much smaller area than either of the other two spots, is located close to the rotational pole on the far side of the star from the observer, and makes only a very small contribution to the waveform. The waveform produced by the best-fitting model with three oval spots is very similar to the waveform produced by the model with two oval spots.

The best-fitting models with two and three oval spots both provide acceptable fits to the phase-channel and bolometric waveform data in the $\chi^2$ sense, and are not statistically distinguishable from each other. Most importantly, the best-fit values of the stellar radius and mass given by the model with three oval spots are almost identical to, and are statistically indistinguishable from, the best-fit values of the stellar radius and mass given by the model with two oval spots. The evidence for the best-fit model with three oval spots is slightly, but not significantly, greater than the evidence for the model with two oval spots.

Based on these results, we conclude that the models with two and three oval spots both provide adequate descriptions of the observed waveform, that there is no evidence that a more complicated model is required to describe the observed waveform, and that the mass and radius estimates inferred using these two models are reliable. We slightly prefer, and therefore report here, the radius and mass estimates given by the model with three oval spots, only because the evidence for this model is slightly---although not significantly---greater than the evidence for the model with two oval spots.

These considerations and results are summarized in Table~\ref{tab:model-sequence}. We now provide the details. 

\begin{deluxetable*}{l|l}
    \tablecaption{Summary of pulse waveform models considered and the results.}
\tablewidth{0pt}
\tablehead{
     \colhead{Waveform Model} & \colhead{Results} 
    \label{tab:model-sequence}
}
\startdata
      \hline
      Two circular spots 
      & Strongly disfavored \\
      & The required areas of the spots appear to conflict with \\
      & ~~ the harmonic structure of the waveform \\
      \hline
      Three or four circular spots 
      & Disfavored \\
      & No evidence for overlapping spots \\
      & No evidence for configurations with more than three spots \\
      & No evidence for more than one temperature in any of the \\
      & ~~ heated regions \\
      & The required areas of the spots appear to conflict with \\
      & ~~ the harmonic structure of the waveform \\
      \hline
      Two oval spots &  Strongly favored over models with two circular spots \\
      & Two different non-overlapping oval spots adequately describe the \\
      & ~~ observed waveform data \\
      & The best-fit temperatures of the two spots are almost the same \\
      & One spot is slightly elongated in the east$-$west direction \\
      & The other spot is highly elongated in the east$-$west direction \\
      & No evidence that a more complicated configuration is needed \\
      & ~~ to describe the observed waveform \\
      \hline
      Three oval spots & Strongly favored over models with two circular spots \\ 
      & Three different non-overlapping oval spots adequately describe \\
      & ~~ the observed waveform data \\
      & Radius and mass estimates are statistically indistinguishable \\
      & ~~ from the estimates using the model with two oval hot spots \\
      & Two of the spots are almost identical to the two spots in the \\
      & ~~ best-fit model with two oval spots \\
      & The third spot makes almost no contribution to the waveform \\
      & Slightly but not significantly higher evidence than the model \\
      & ~~ with two oval spots \\ 
      & No evidence that a more complicated configuration is needed \\
      & ~~ to describe the observed waveform \\
      \hline
\enddata
\tablecomments{See text for details about the assumed prior probability distributions for each parameter.}
\end{deluxetable*}

\subsection{Choice of Energy Channels}
\label{sec:energy-channels}

\begin{figure*}[ht!]
\gridline{\fig{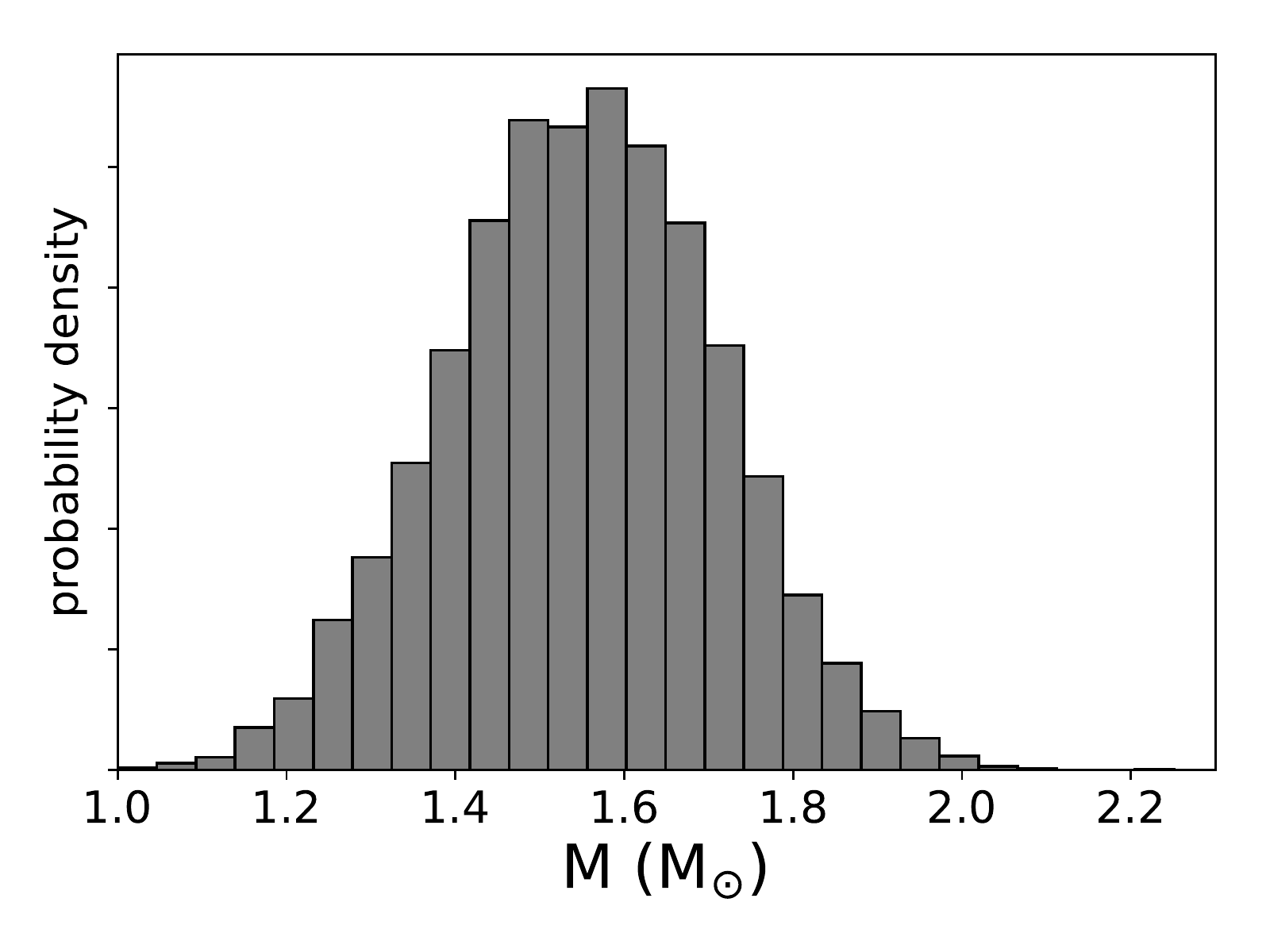}{0.3\textwidth}{(a)}
          \fig{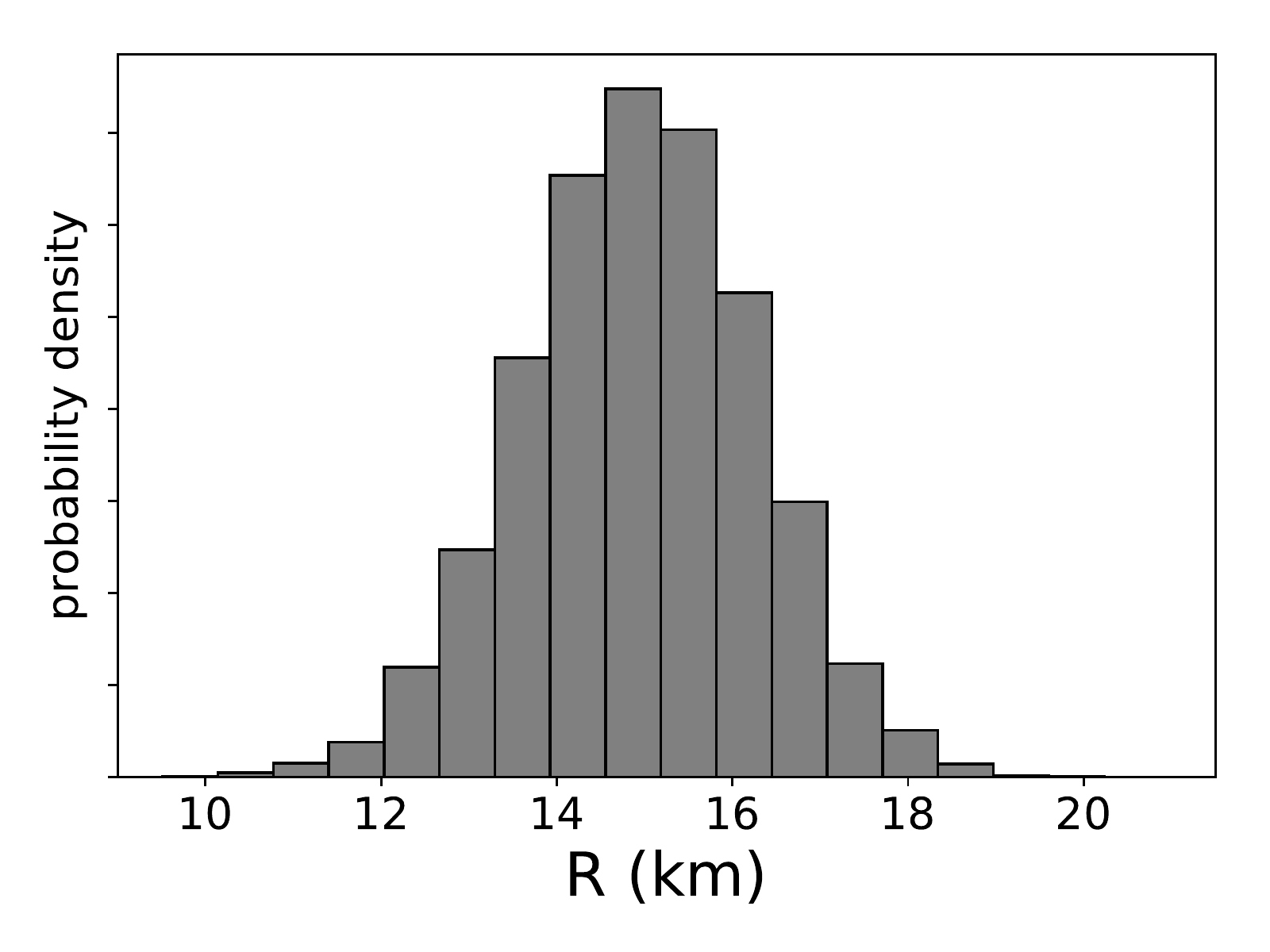}{0.3\textwidth}{(b)}
          \fig{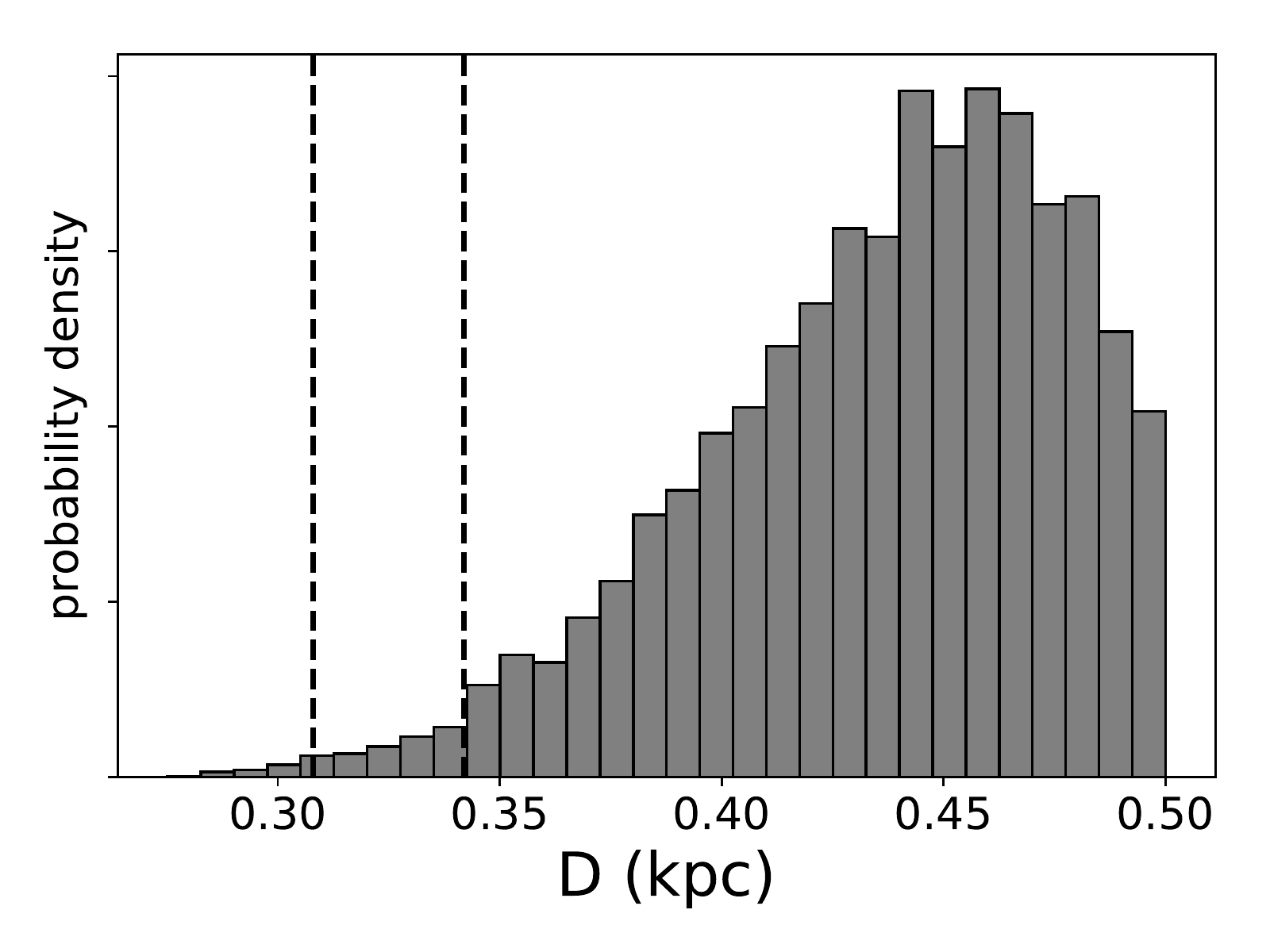}{0.3\textwidth}{(c)}}
\gridline{\fig{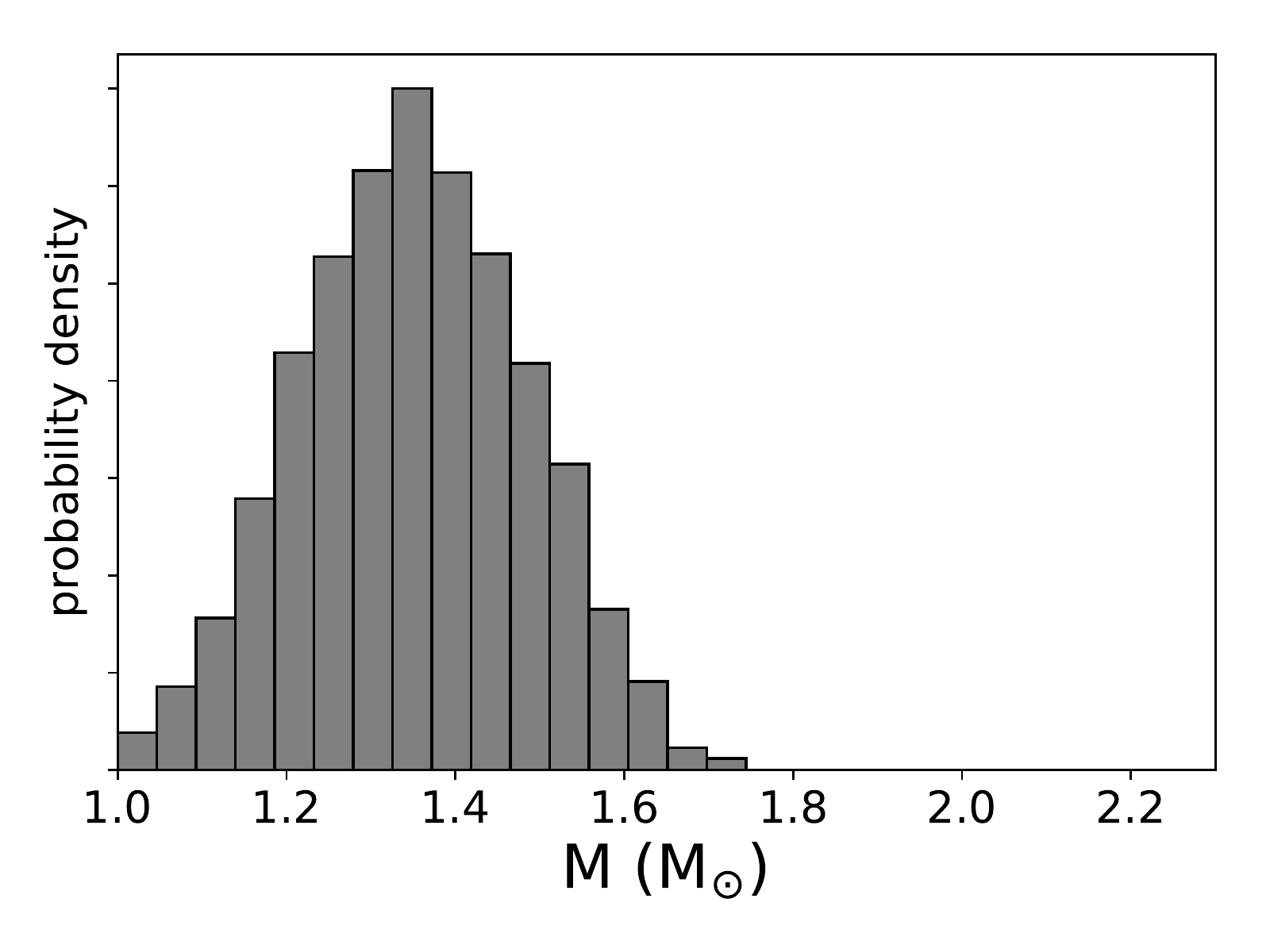}{0.3\textwidth}{(d)}
          \fig{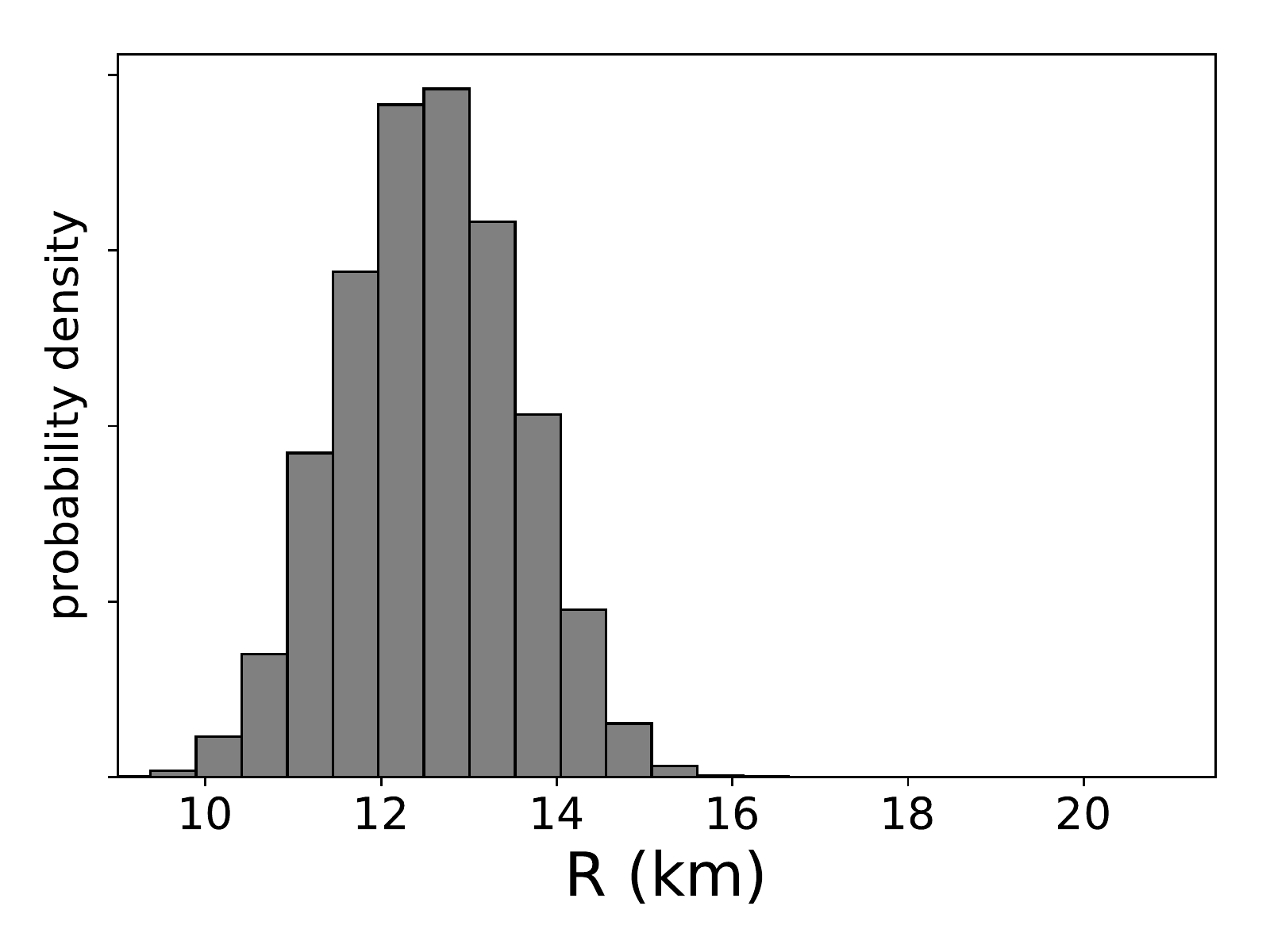}{0.3\textwidth}{(e)}
          \fig{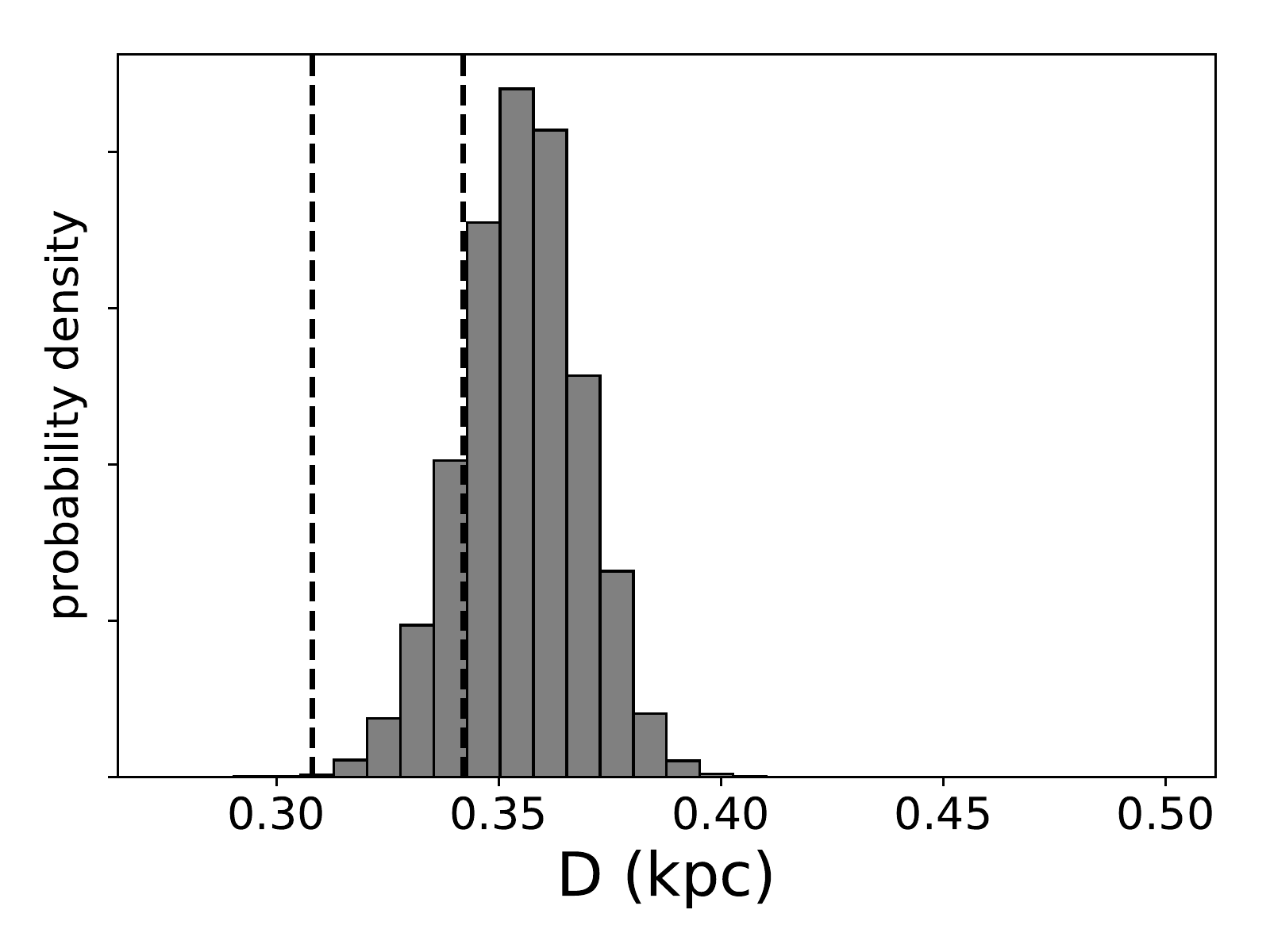}{0.3\textwidth}{(f)}}
\vspace{-0.1truein}
\caption{The 1D posterior probability distributions for $M$, $R_e$, and $D$ estimates obtained by fitting the waveform model with two possibly different uniform oval hot spots to two different sets of \textit{NICER} data on PSR~J0030$+$0451. Both fits assumed a distance prior flat between 0.275 and 0.5~kpc and zero outside this interval. The independently measured distance to PSR~J0030$+$0451 is $0.325 \pm 0.009$~kpc (see text). The vertical dashed lines in panels~(c) and~(f) indicate the uncertainty in $D$ obtained by combining the uncertainty in the measurement of $D$ and the uncertainty in $D$ that corresponds to the uncertainty in the effective area of \textit{NICER} (see text for details).
\textit{Top row:}  
posteriors obtained using the data in \textit{NICER} energy channels 25--299. Note that the distance posterior, which is plotted in panel~(c), peaks at $\sim\,0.45$~kpc and excludes the independently measured distance with high confidence. 
\textit{Bottom row:}
posteriors obtained using the data in \textit{NICER} energy channels 40--299. In contrast to the distance posterior obtained using the data in \textit{NICER} energy channels 25--299, the distance posterior obtained using this data set, which is plotted in panel~(f), peaks at a distance that is close to the independently measured distance.}
\label{fig:twooval25vs40rmd}
\end{figure*}

The lowest-energy \textit{NICER} data that was included in the standard pipeline data set that we analyzed was in energy channel~25 (see Section~\ref{sec:data-filtering}). As noted previously (see Section~\ref{sec:uncertainties-bias}), we found no detectable modulation of the X-ray flux in \textit{NICER} energy channels~300 and above. We therefore considered in our analyses only the data in channels~25--299.

The calibration errors in the data from energy channels~40--299 are relatively small. For example, the residuals between the Crab Nebula spectrum determined from \textit{NICER} measurements and the Crab Nebula spectrum determined from measurements made using other instruments are $\lesssim\,$5\% for channels~40--299 (see Sections~\ref{sec:calibration} and~\ref{sec:uncertainties-bias}; see also \citealt{2018ApJ...858L...5L}). In contrast, there are still substantial and poorly understood errors in the calibration of energy channels~25--39. We were therefore concerned that including data from channels 25--39 in our analysis would bias our estimates of the mass and radius of PSR~J0030$+$0451.

The distance to PSR~J0030$+$0451 has been accurately measured to be $0.325\pm 0.009$~kpc, independent of any waveform analyses (see \citealt{2018ApJ...859...47A}). We therefore decided to investigate the effect of including or not including the data in channels 25--39 on estimates of the distance made using the PSR~J0030$+$0451 waveform. We found that if we used the \textit{NICER} waveform data in channels 25--299, thereby including the data in channels 25--39, and assumed a wide, flat prior for the distance, all our analyses gave posterior probability distributions for the distance that peaked at distances substantially larger than the independently measured distance of 0.325~kpc and excluded 0.325~kpc with high confidence. This is illustrated by panel~(c) in Figure~\ref{fig:twooval25vs40rmd}, which shows the 1D posterior distributions for $M$, $R_e$, and $D$ obtained by fitting the waveform model with two possibly different and overlapping uniform oval hot spots to the \textit{NICER} data in channels 25--299. In contrast, when we used only the well-calibrated data in channels~40--299, we found posterior probability distributions for the distance that naturally peak close to the independently measured distance. This is illustrated by panel~(f) in Figure~\ref{fig:twooval25vs40rmd}, which shows the 1D posterior distributions for $M$, $R_e$, and $D$ obtained by fitting the waveform model with two possibly different and overlapping uniform oval hot spots to the \textit{NICER} data in channels 40--299.

These results suggested to us that including the poorly calibrated data in channels 25--39 in the analysis was likely to be biasing the distance estimate and could also be biasing the estimates of $M$ and $R_e$, about which we have no independent information. We therefore thought it was important to address the effects of the apparent calibration errors in channels~25--39, but we wanted to do this in a way that would not itself bias our results. Consequently, we decided that, rather than attempting to correct for the calibration errors in channels~25--39, which are poorly understood at present, the safest approach would be to discard the data in all these channels. This would prevent the calibration errors in these channels from biasing our $M$ and $R_e$ estimates, but it could in principle increase the uncertainties in our estimates of these parameters because the amount of data we used would be reduced.

Comparison of the results obtained by analyzing synthetic waveform data in \textit{NICER} energy channels 25--299 and 40--299 that were generated assuming two different, non-overlapping uniform circular spots (see Section~\ref{sec:analysis-2circ-synthetic-data}) shows that using only the data in the reduced energy range does not significantly bias the estimates of $M$ and $R_e$ or decrease the precisions of the estimates (compare panels~(a) and~(d) and panels~(b) and~(e) of Figure~\ref{fig:synthcirc25vs40})). (We note that for this synthetic waveform, the temperature assumed for one of the hot spots was substantially lower than any of the temperatures inferred for the hot spots of PSR~J0030$+$0451. Consequently, the cut at channel 40 discarded much more of the spectrum of this waveform than it does for the spectrum of PSR~J0030$+$0451 observed by \textit{NICER}.)

Similarly, comparison of the results obtained by analyzing synthetic waveform data in \textit{NICER} energy channels 25--299 and 40--299 that were generated assuming two different, non-overlapping uniform oval spots (see Section~\ref{sec:analysis-2oval-synthetic-data}) shows that using only the data in the reduced energy range does not significantly bias the estimates of $M$ and $R_e$ or decrease the precisions of the estimates (compare panels~(a) and~(d) and panels~(b) and~(e) of Figure~\ref{fig:synthoval25vs40}).

Finally, comparison of the results obtained by analyzing the \textit{NICER} data on the waveform of PSR~J0030$+$0451 in energy channels 25--299 and 40--299 shows that using only the data in the reduced energy range does not significantly decrease the precisions of the estimates of $M$ and $R_e$ (compare panels~(a) and~(d) and panels~(b) and~(e) of Figure~\ref{fig:twooval25vs40rmd}).

Taking into account all these results, we decided to use only the data in energy channels 40--299 to estimate the values of the parameters in our models of the PSR~J0030$+$0451 pulse waveform. Consequently, in the remainder of this Letter we present only results obtained using the data in these energy channels. Given that the independently measured distance of $0.325\pm 0.009$~kpc is very precise, we imposed a Gaussian prior on the distance that uses this measurement and an uncertainty that takes into account the uncertainty in this measurement and the global (in energy) uncertainty in the effective area of the \textit{NICER} instrument (see below).

\subsection{The Mass and Radius of PSR~J0030$+$0451}
\label{sec:estimating-J0030-mass&radius}

In this section, we present and discuss the results of our analysis of the \textit{NICER} data on PSR~J0030$+$0451, focusing on our estimates of its mass and radius. For the reasons described in Section~\ref{sec:energy-channels}, we present only the results we obtained by fitting the \textit{NICER} data in energy channels 40--299. When estimating the values of all the parameters except the distance, we assumed prior probability distributions constant within the intervals listed in Table~\ref{tab:wf-primary-parameters} and zero outside these intervals. When estimating the distance, we assumed a Gaussian prior. We chose the width of the Gaussian by combining linearly, rather than quadratically, the 0.009~kpc uncertainty in the measured distance to PSR~J0030$+$0451 (see \citealt{2018ApJ...859...47A}) and the estimated uncertainty in the overall normalization of the \textit{NICER} effective area, converted to an uncertainty in the distance. We assumed a 5\% uncertainty in the overall normalization of the \textit{NICER} effective area (see Section~\ref{sec:uncertainties-bias}). This translates into a distance uncertainty of 2.5\%. The distance to PSR~J0030$+$0451 has been independently measured to be 0.325~kpc \citep{2018ApJ...859...47A}. Consequently, a $1\sigma$ uncertainty of 2.5\% in the distance is $\approx 0.008$~kpc, yielding a total $1\sigma$ width for the Gaussian of 0.009~kpc $+$ 0.008~kpc $=0.017$~kpc.

We considered two waveform models, one with two possibly different and overlapping uniform-temperature oval spots, and one with three possibly different and overlapping uniform-temperature oval spots. The model with two oval spots has 14 primary parameters. These are listed in Table~\ref{tab:credible2}, along with their median values and the $\pm 1\,\sigma$ and $\pm 2\,\sigma$ boundaries of their 1D credible regions, computed using the 1D posterior probability density distribution we found for each parameter. The model with three oval spots has 19 primary parameters. These are listed in Table~\ref{tab:credible3}, along with their median values and the $\pm 1\,\sigma$ and $\pm 2\,\sigma$ boundaries of their 1D credible regions, computed using the 1D posterior probability density distribution we found for each parameter.\footnote{Samples from the full posterior probability distribution obtained by fitting the model with three uniform-temperature oval spots to the waveform data are available at https://zenodo.org/record/3473466.} Both fits favored models in which the spots do not overlap.

\begin{deluxetable*}{c|r|r|r|r|r|r}
    \tablecaption{Fits to \textit{NICER} Data with Two Oval Spots.}
\tablewidth{0pt}
\tablehead{
      \colhead{Parameter} & \colhead{Median} & \colhead{$-1\sigma$} & \colhead{$+1\sigma$} & \colhead{$-2\sigma$}  & \colhead{$+2\sigma$} & \colhead{Best Fit}
   \label{tab:credible2}
}
\startdata
      \hline
      $R_e ({\rm km})$ & 13.271 & 12.115 & 14.578 & 11.042 & 15.968 & 13.643 \\ 
      \hline
      $GM/(c^2R_e)$ & 0.160 & 0.152 & 0.169 & 0.143 & 0.176 & 0.161 \\
      \hline
      $M (M_\odot)$ & 1.442 & 1.282 & 1.619 & 1.141 & 1.802 & 1.488\\
      \hline
      $\theta_{\rm c1} ({\rm rad})$ & 2.251 & 2.165 & 2.334 & 2.084 & 2.419 & 2.232 \\ 
      \hline
      $\Delta\theta_1 ({\rm rad})$ & 0.035 & 0.030 & 0.041 & 0.026 & 0.047 & 0.031 \\
      \hline
      $f_1$ & 5.347 & 3.950 & 6.981 & 3.136 & 8.362 & 6.024 \\
      \hline
      $kT_{\rm eff,1} ({\rm keV})$ & 0.117 & 0.114 & 0.120 & 0.111 & 0.122 & 0.119 \\
      \hline
      $\theta_{\rm c2} ({\rm rad})$ & 2.417 & 2.333 & 2.495 & 2.245 & 2.560 & 2.394\\
      \hline
      $\Delta\theta_2 ({\rm rad})$ & 0.033 & 0.028 & 0.039 & 0.025 & 0.044 & 0.029 \\
      \hline
      $f_2$ & 15.490 & 12.317 & 18.396 & 10.020 & 19.748 & 17.744 \\
      \hline
      $kT_{\rm eff,2} ({\rm keV})$ & 0.117 & 0.115 & 0.119 & 0.112 & 0.122 & 0.119 \\
      \hline
      $\Delta\phi_2 ({\rm cycles})$ & 0.459 & 0.457 & 0.461 & 0.455 & 0.463 & 0.458\\
      \hline
      $\theta_{\rm obs} ({\rm rad})$ & 0.848 & 0.751 & 0.951 & 0.660 & 1.043 & 0.827\\
      \hline
      $N_{\rm H} (10^{20}~{\rm cm}^{-2})$ & 0.266 & 0.072 & 0.593 & 0.011 & 1.08 & 0.047\\
      \hline
      $D ({\rm kpc})$ & 0.327 & 0.309 & 0.345 & 0.293 & 0.361 & 0.332 \\
      \hline
\enddata
\tablecomments{1D credible regions, and best fit, obtained by fitting the model with two possibly different uniform oval spots to channels 40--299 of the \textit{NICER} data on PSR~J0030$+$0451.}
\end{deluxetable*}

\begin{deluxetable*}{c|r|r|r|r|r|r}
    \tablecaption{Fits to \textit{NICER} Data with Three Oval Spots.}
\tablewidth{0pt}
\tablehead{
      \colhead{Parameter} & \colhead{Median} & \colhead{$-1\sigma$} & \colhead{$+1\sigma$} & \colhead{$-2\sigma$}  & \colhead{$+2\sigma$} & \colhead{Best Fit}
   \label{tab:credible3}
}
\startdata
	  \hline
	  $R_e$ (km)  &  13.019 & 11.959 & 14.255 & 10.938 & 15.500 & 13.466 \\
	  \hline
      $GM/(c^2R_e)$ & 0.163 &  0.154 &  0.171 &  0.144 &  0.179 & 0.156 \\
      \hline
      $M (M_\odot)$ & 1.443 &  1.299  & 1.594 &  1.164 &  1.745 & 1.423 \\
      \hline
      $\theta_{\rm c1} ({\rm rad})$ & 2.270 &  2.179  & 2.357 &  2.093 &  2.442 & 2.330 \\
      \hline
      $\Delta\theta_1 ({\rm rad})$ & 0.036 &  0.031  & 0.040 &  0.028 &  0.045 & 0.032 \\
      \hline
      $f_1$ & 5.352 &  4.364  & 6.502 &  3.568 &  7.664 & 5.335 \\
      \hline
      $kT_{\rm eff,1} ({\rm keV})$ & 0.117 &  0.114  & 0.120 &  0.110 &  0.122 & 0.113 \\
      \hline
      $\theta_{\rm c2} ({\rm rad})$ & 2.417 &  2.341  & 2.486 &  2.252 &  2.540 & 2.446 \\
      \hline
      $\Delta\theta_2 ({\rm rad})$ & 0.033 &  0.029  & 0.038 &  0.025 &  0.043 & 0.029 \\
      \hline
      $f_2$ & 15.769 & 13.017 & 18.498 & 10.923 & 19.79 & 16.588 \\
      \hline
      $kT_{\rm eff,2} ({\rm keV})$ & 0.115 &  0.112  & 0.118 &  0.107 &  0.121 & 0.105 \\
      \hline
      $\Delta\phi_2 ({\rm cycles})$ & 0.460 &  0.458  & 0.463 &  0.456 &  0.466 & 0.463 \\
      \hline
      $\theta_{\rm c3} ({\rm rad})$ & 2.988  & 2.865  & 3.063 &  2.691 &  3.094 & 3.056 \\
      \hline
      $\Delta\theta_3 ({\rm rad})$ & 0.056  & 0.021  & 0.103 &  0.004 &  0.168 & 0.087 \\
      \hline
      $f_3$ &1.215 &  0.493  & 2.096 &  0.161 &  3.511 & 1.253 \\
      \hline
      $kT_{\rm eff,3} ({\rm keV})$ & 0.239 &  0.143  & 0.354 &  0.029 &  0.470 & 0.209 \\
      \hline
      $\Delta\phi_3 ({\rm cycles})$ & 0.420 &  0.378  & 0.534 &  0.071 &  0.932 & 0.427 \\
      \hline
      $\theta_{\rm obs} ({\rm rad})$ & 0.878 &  0.769  & 0.973 &  0.675 &  1.051 & 1.012 \\
      \hline
      $N_{\rm H} (10^{20}~{\rm cm}^{-2})$ & 0.244 & 0.082  & 0.441 &  0.011 &  0.723 & 0.187  \\
      \hline
      $D ({\rm kpc})$ & 0.327 &  0.307  & 0.347 &  0.290 &  0.365 & 0.317 \\
      \hline
\enddata
\tablecomments{1D credible regions, and best fit, obtained by fitting the model with three possibly different uniform oval spots to channels 40--299 of the \textit{NICER} data on PSR~J0030$+$0451.}
\end{deluxetable*}

Table~\ref{tab:summarycomp} compares the median values of $R_e$ and $M$ and the boundaries of the $\pm 1\,\sigma$ credible regions computed from the posterior probability density distributions obtained by fitting the two models described here.

\begin{deluxetable*}{c|r|r|r|r|r|r}
    \tablecaption{Comparison between Two-oval and Three-oval fits to \textit{NICER} data.}
\tablewidth{0pt}
\tablehead{
     \colhead{Fit} & \colhead{$-1\sigma,\ R$} & \colhead{Median $R_e$} & \colhead{$+1\sigma,\ R$} & \colhead{$-1\sigma,\ M$}  & \colhead{Median $M$} & \colhead{$+1\sigma,\ M$}
   \label{tab:summarycomp}
}
\startdata
      \hline
      {Two ovals} & 12.115 & 13.271 & 14.578 & 1.282 & 1.442 & 1.619 \\
      \hline
      {Three ovals} & 11.959 & 13.019 & 14.255 & 1.299 & 1.443 & 1.594 \\
      \hline
\enddata
\tablecomments{Comparison of the radius (in km) and mass (in solar masses) estimates given by the fits of the models with two and three oval spots to the \textit{NICER} data on PSR~J0030$+$0451.}
\end{deluxetable*}

\begin{figure*}[ht!]
\gridline{\fig{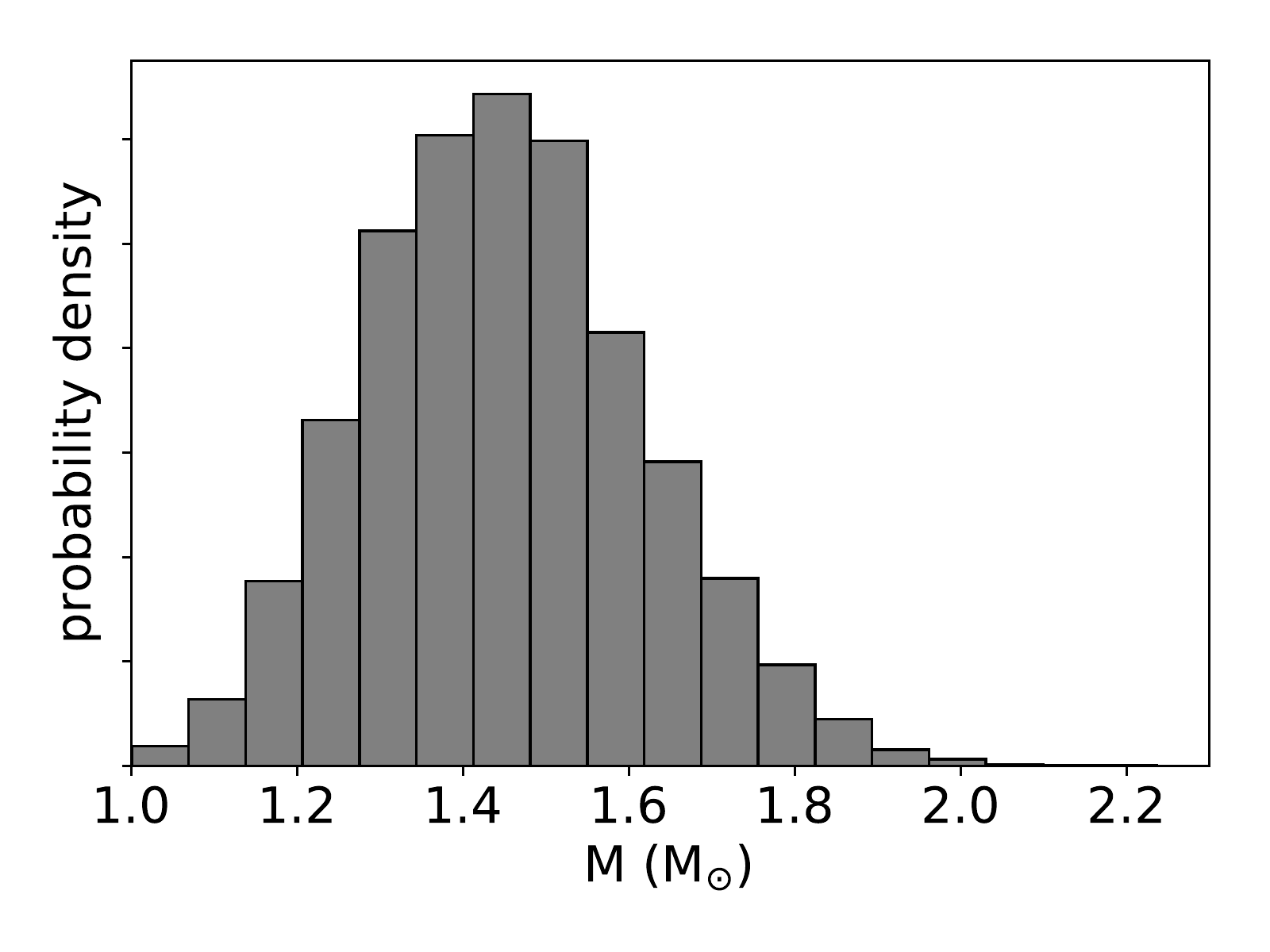}{0.3\textwidth}{(a)}
          \fig{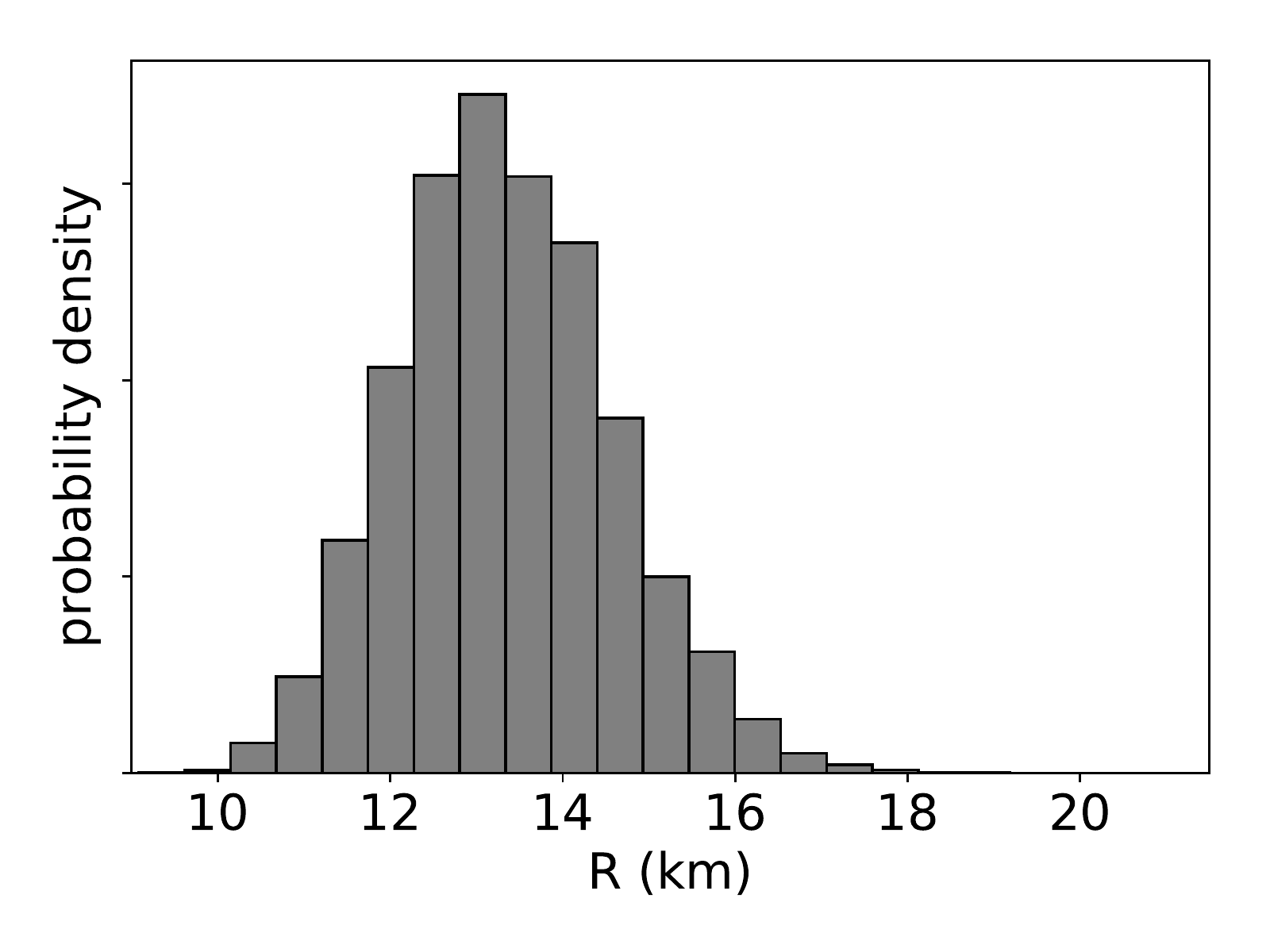}{0.3\textwidth}{(b)}
          \fig{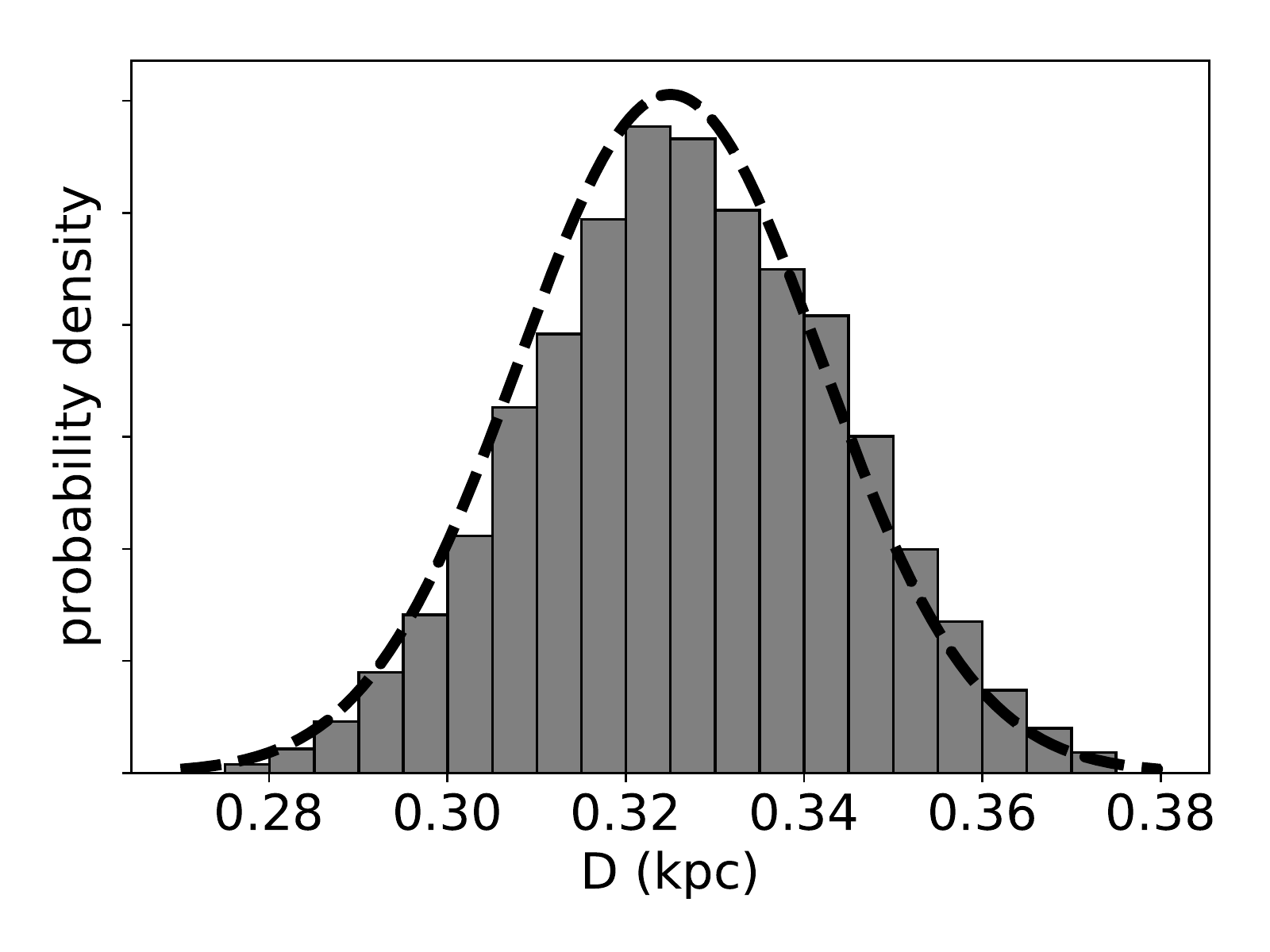}{0.3\textwidth}{(c)}}
\gridline{\fig{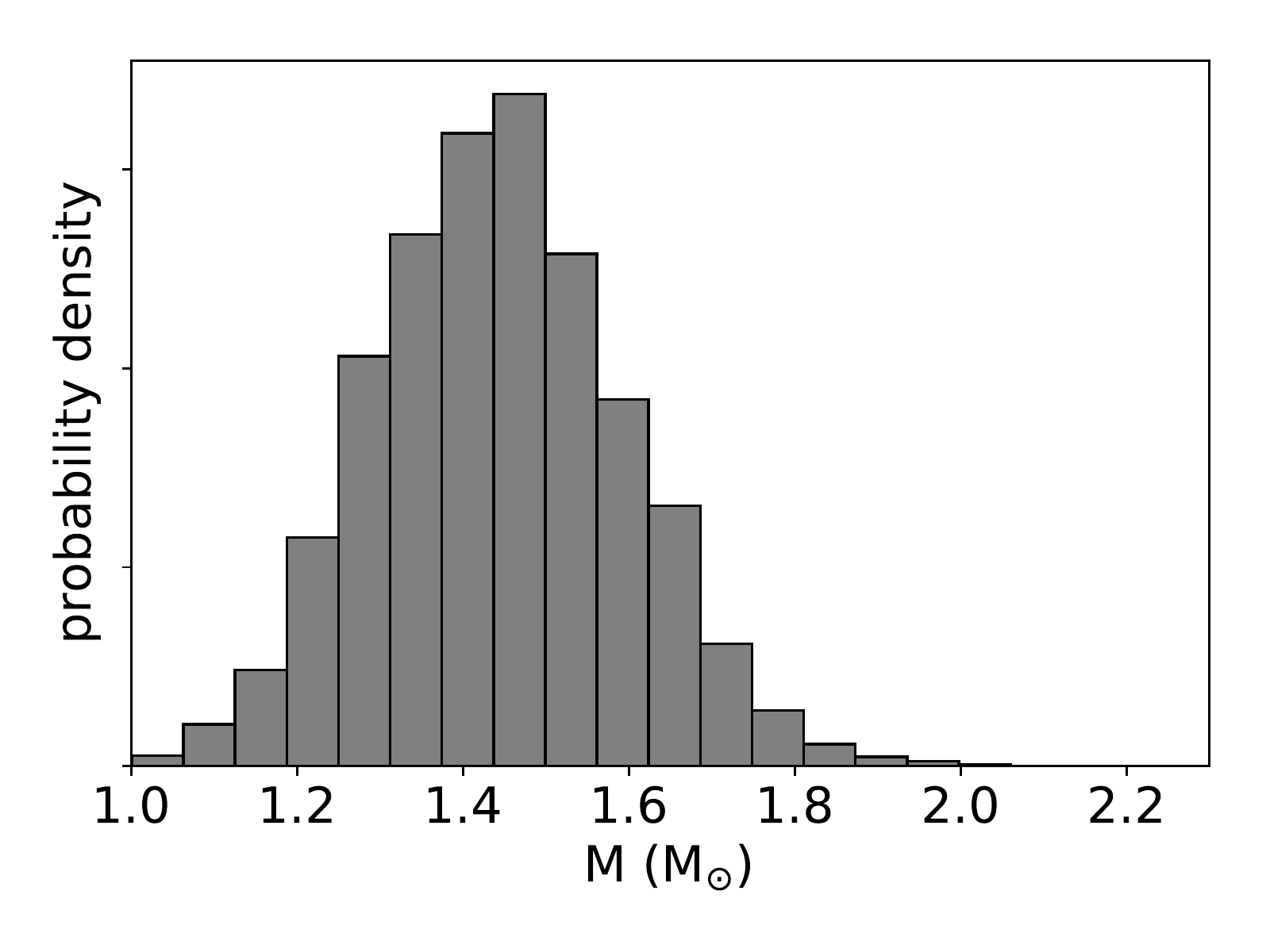}{0.3\textwidth}{(d)}
          \fig{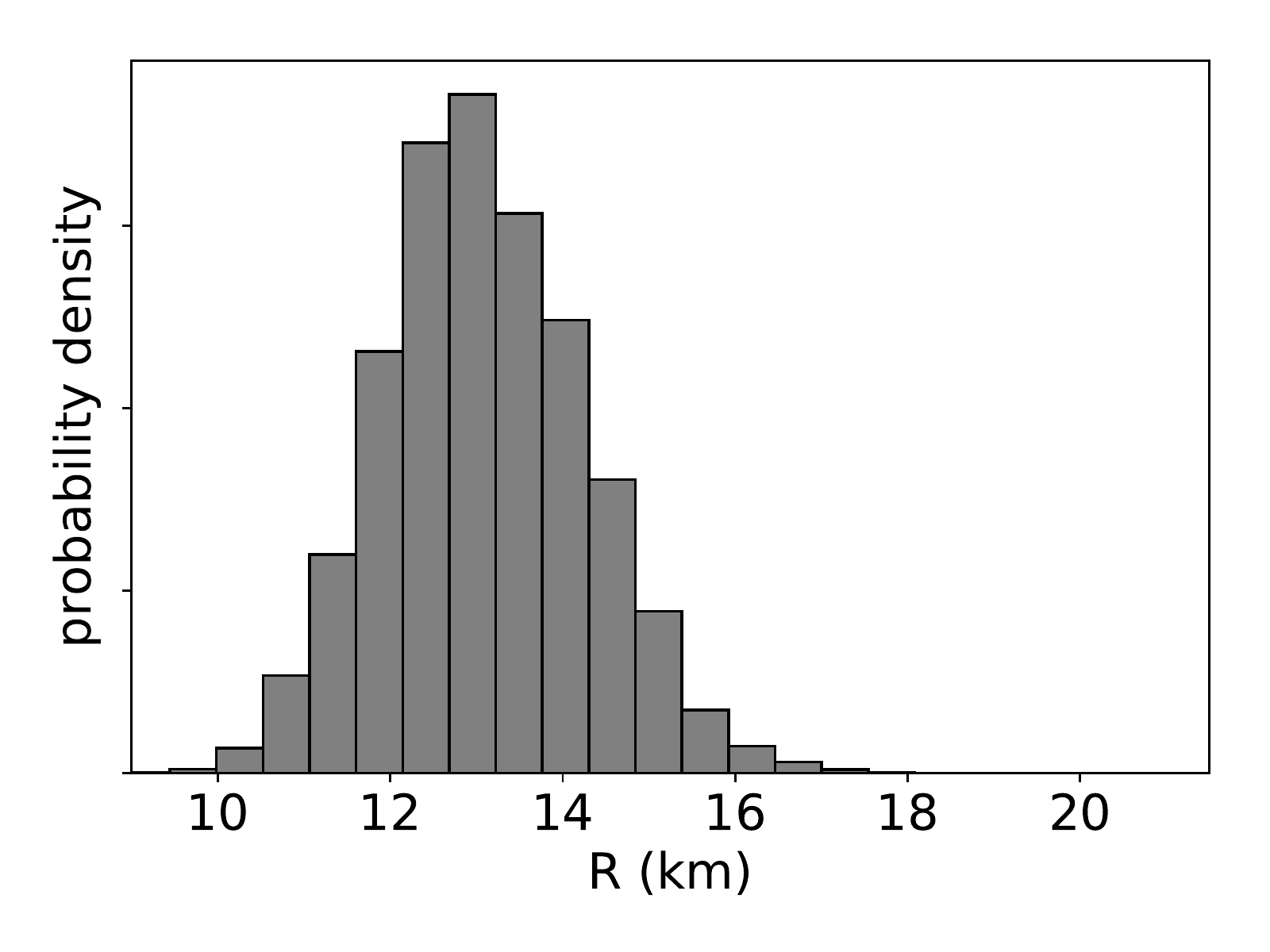}{0.3\textwidth}{(e)}
          \fig{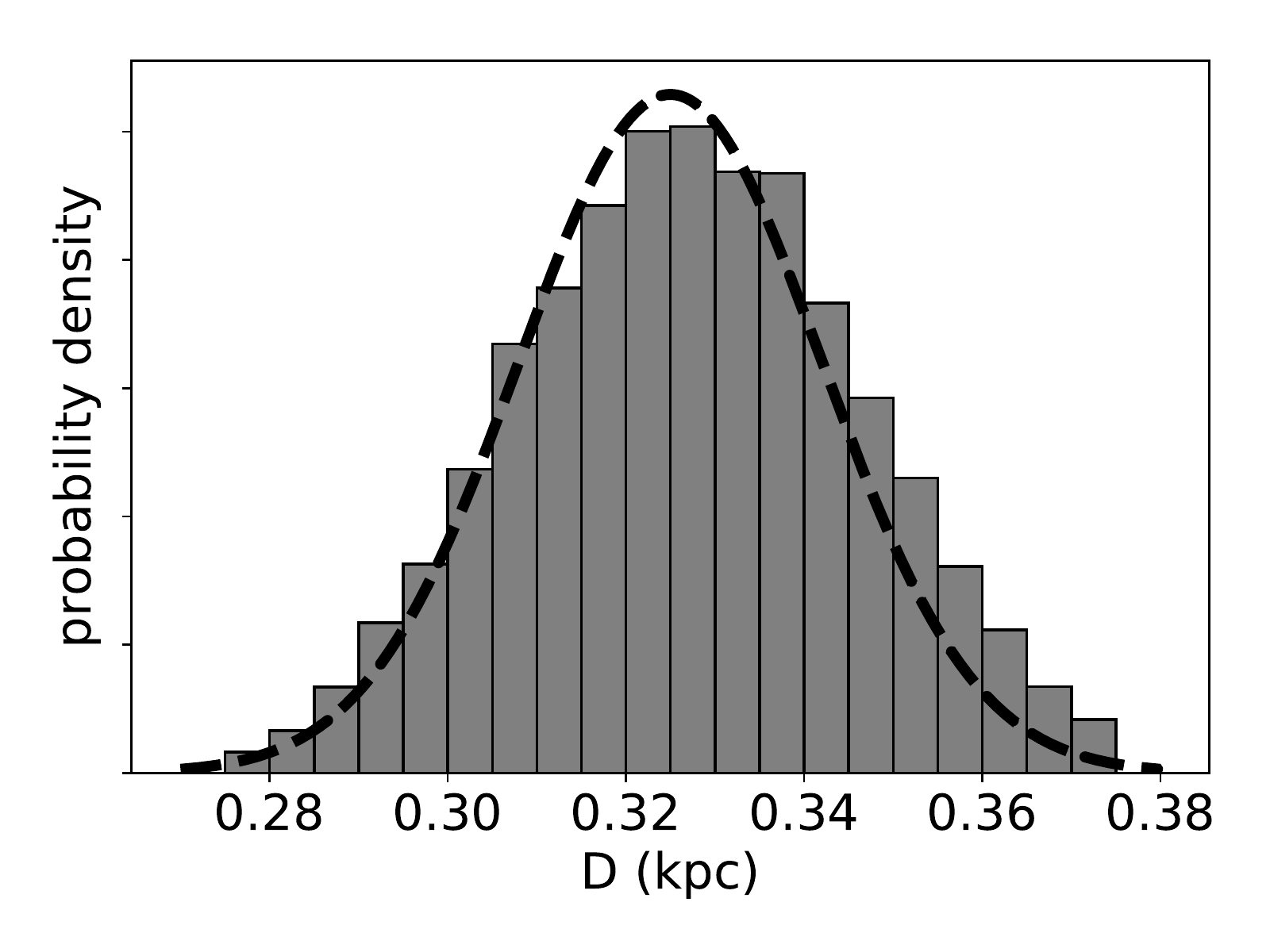}{0.3\textwidth}{(f)}}
\vspace{-0.1truein}
\caption{Comparison of the 1D posterior probability density distributions of the stellar mass (panels~(a) and~(d)), stellar radius (panels~(b) and~(e)), and the distance to the pulsar (panels~(c) and~(f)) given by the best fits of the waveform model with two (top row of panels) and three (bottom row of panels) oval spots. Both models were fit to the \textit{NICER} data in energy channels 40--299, assigned to 32 phase bins. The dashed lines in panels~(c) and~(f) show the Gaussian prior that was used for the distance estimate (see text). The agreement of the distributions given by the two models is excellent.}
\label{fig:1dcomparison}
\end{figure*}

Figure~\ref{fig:1dcomparison} compares the -D posterior probability density distributions for $M$, $R_e$, and $D$. Note that the waveform data modifies the posterior probability distributions for the distance only slightly compared to the Gaussian prior we assumed based on the independently measured distance to PSR~J0030$+$0451 and the uncertainty in the \textit{NICER} effective area (see Section~\ref{sec:energy-channels}), showing that the distance estimates obtained by fitting both models to the waveform data are consistent with the measured distance.

\begin{figure*}[ht!]
\vspace{-0.2truein}
\gridline{\fig{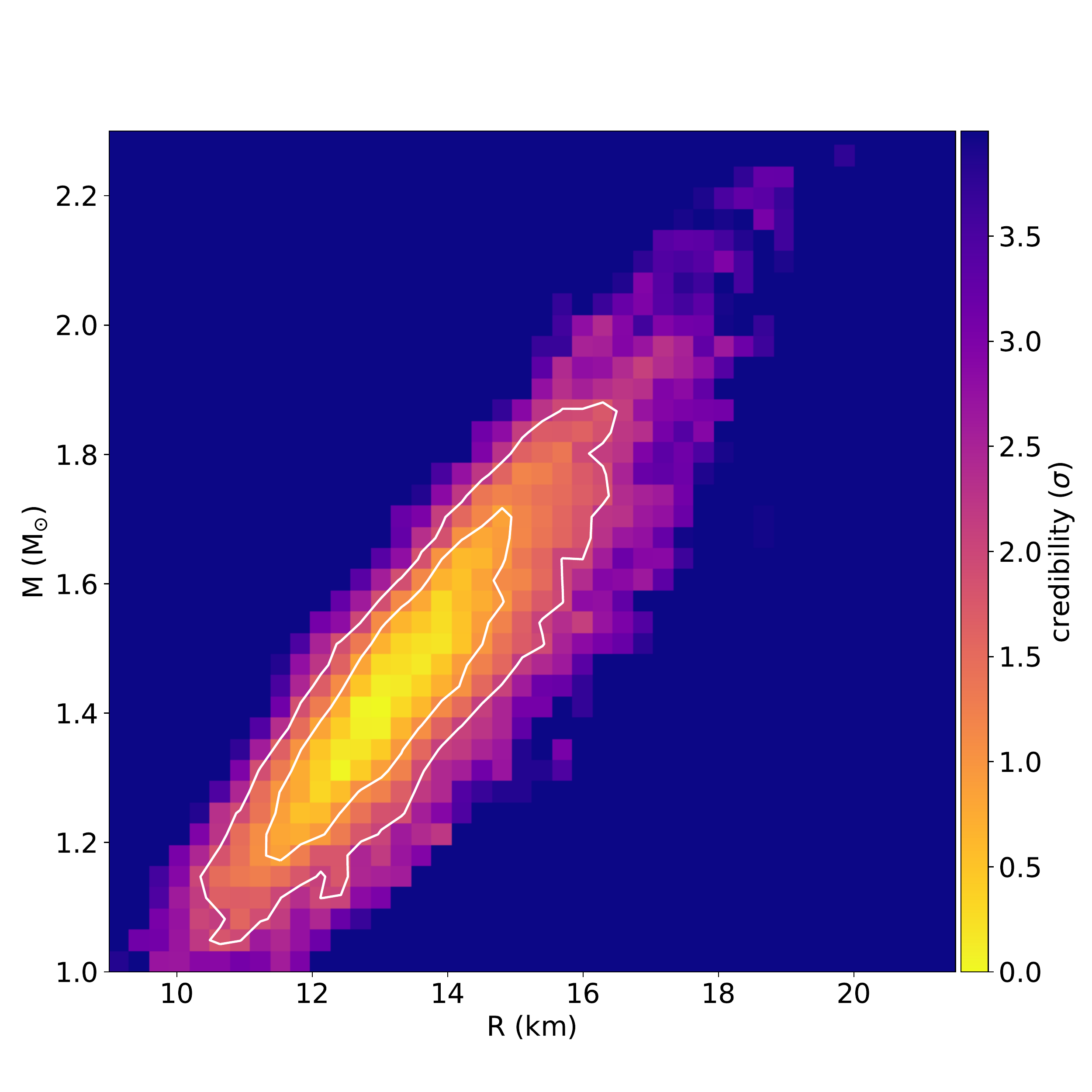}{0.48\textwidth}{(a)}
          \fig{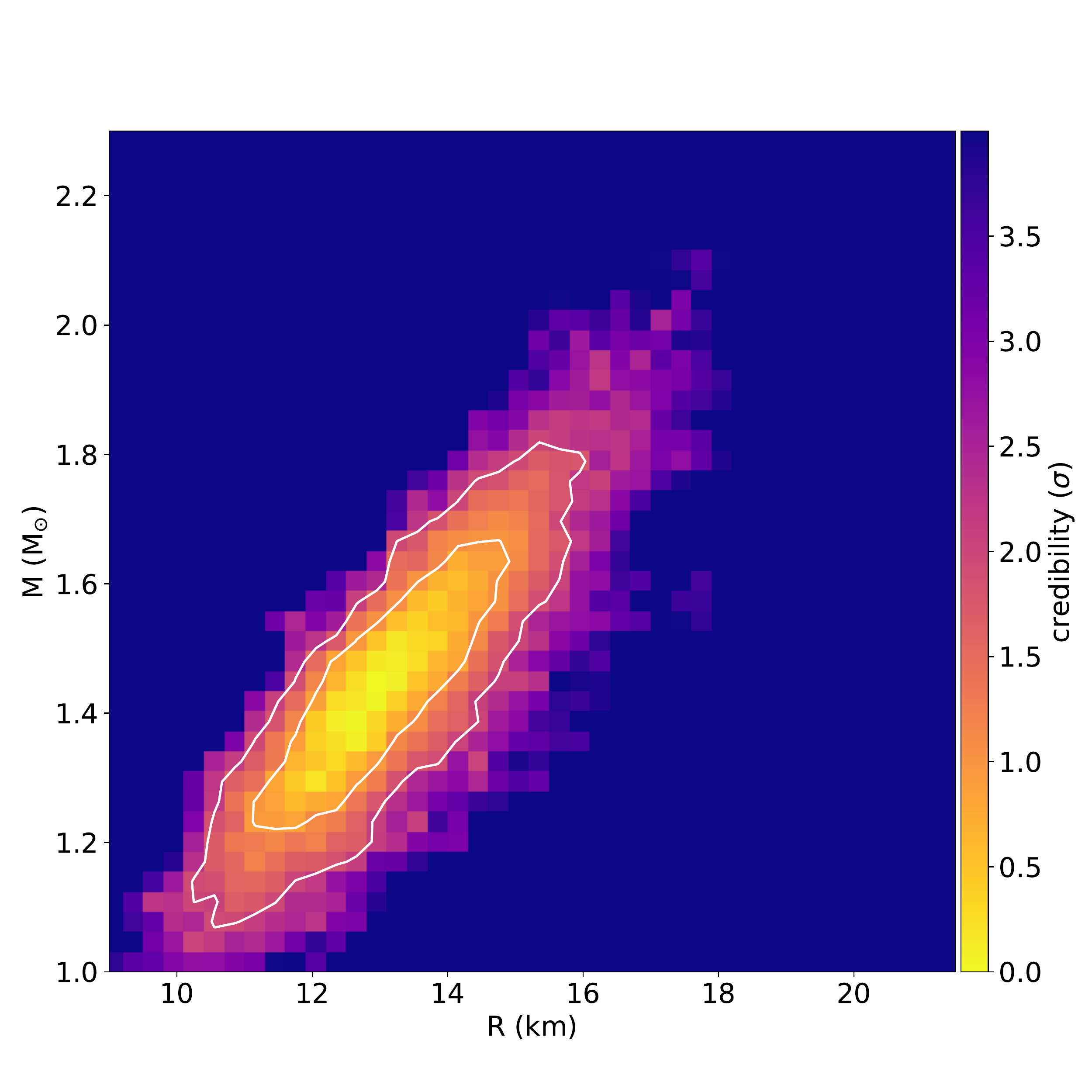}{0.48\textwidth}{(b)}}
\caption{Comparison of the joint posterior probability density distributions for $M$ and $R_e$ given by the best fits of the waveform model with two (panel~(a)) and three (panel~(b)) uniform-temperature oval spots. The inner contour shown in each panel contains 68.3\% of the posterior probability, whereas the outer contour contains 95.4\%. The color indicates the credibility in standard deviations of each point in the posterior probability density distribution. Again, the agreement of the distributions given by the two models is excellent.}
\label{fig:2dcomparison}
\end{figure*}

Figure~\ref{fig:2dcomparison} compares the joint posterior probability density distributions for $R_e$ and $M$ that were obtained by fitting the two models to the \textit{NICER} waveform data.

\begin{figure*}[ht!]
\gridline{\fig{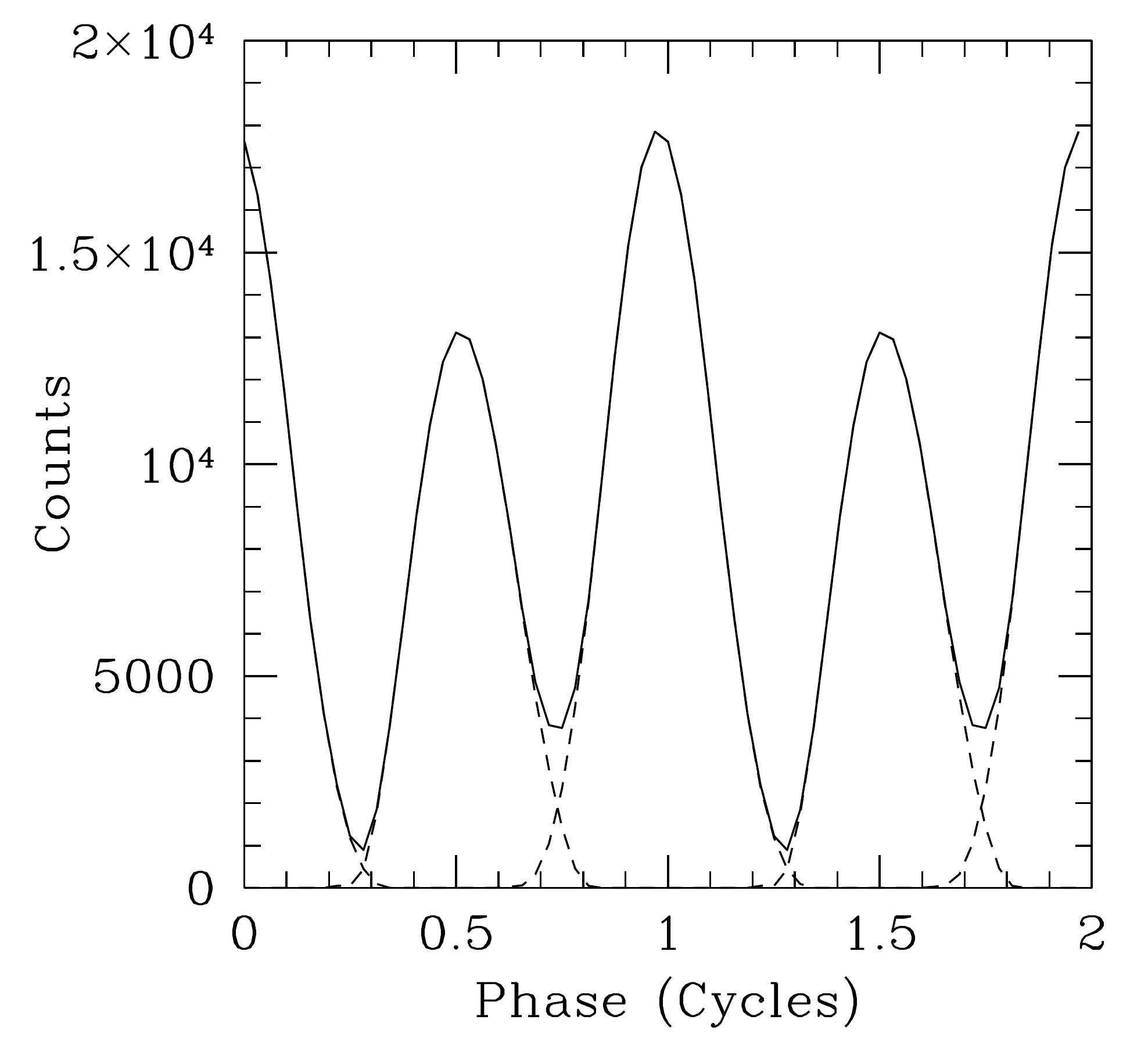}{0.45\textwidth}{(a)}
          \fig{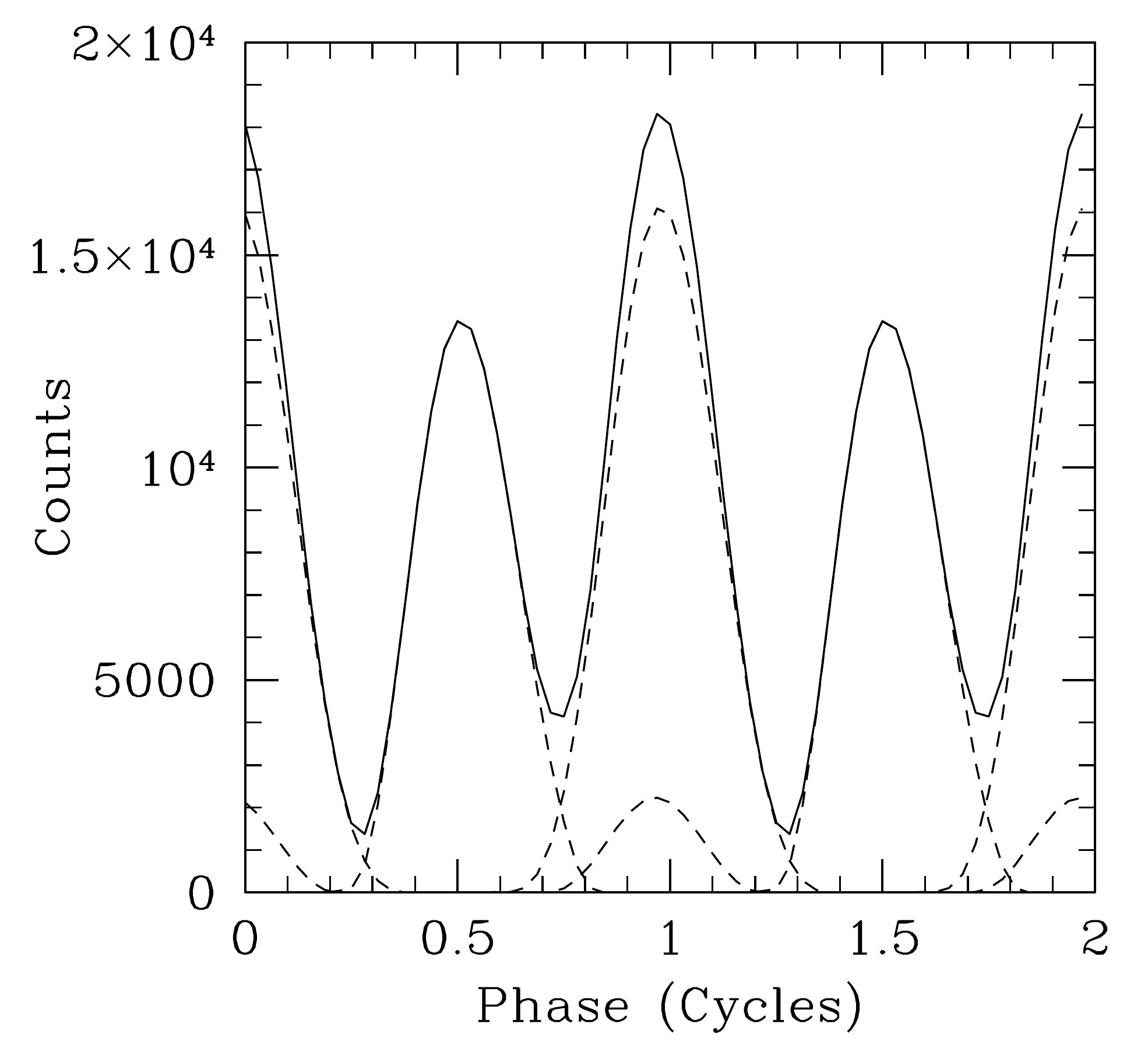}{0.45\textwidth}{(b)}}
\caption{Comparison of the bolometric waveforms given by the best-fit waveform models with two (panel~(a)) and three (panel~(b)) oval spots. The solid curves show the full waveforms; the dashed curves show the contributions to the full waveform made by the individual hot spots. The components that generate the full waveforms are very similar for the two models.}
\label{fig:waveform-comparison}
\end{figure*}

Figure~\ref{fig:waveform-comparison} compares the bolometric waveforms given by the best-fit waveform models with two and three oval spots. The waveform components are very similar for the two models, and the full waveforms are almost identical.

All these comparisons show that the models with two and three uniform-temperature oval spots give mutually consistent credible regions for all the parameters they have in common, most notably for the mass and radius of the pulsar.  The log evidence for the best fit of the model with three oval spots is higher than the log evidence for the best fit of the model with two oval spots, but only by 1.7, which is less than the estimated uncertainty in the log evidence. Thus, both models are comparably good. This consistency suggests that, at least for the types of waveform models considered here, no further complexity is needed to adequately describe the observed waveform. The 68\% credible intervals for $M$ and $R_e$ given by the fit of the model with three oval spots to the \textit{NICER} data from PSR~J0030$+$0451 are 1.30--1.59$~M_\odot$ and  11.96--14.26~km.

Our results for mass and radius of PSR~J0030$+$0451 may be compared with the recent results of \citet{2018MNRAS.476..421S}, who examined the evidence from neutron stars that are transient X-ray sources and conclude that the radius of a $1.4\,M_\odot$ neutron star is most likely between 10 and 14~km, which is consistent with our estimates of the mass and radius of PSR~J0030$+$0451.

Our results may also be compared with those of \citet{2017A&A...608A..31N}, who estimated that the mass of the (neutron star) X-ray burst source 4U~1702$-$429 is between 1.6 and $2.2\,M_\odot$ and that its circumferential radius is between 12.0 and 12.8~km (with 68\% credibility). Although this estimate of the mass of 4U~1702$-$429 is higher than our estimate of the mass of PSR~J0030$+$0451, for most neutron star EoS currently under consideration, the radius of a neutron star depends only weakly on its mass in this mass range, in which case this estimate of the radius of 4U~1702$-$429 is consistent with our estimate of the radius of PSR~J0030$+$0451. 

The full posterior probability density distributions for the parameters in the fit of the model with two oval spots to the \textit{NICER} waveform data are shown in Figure~\ref{fig:full-posteriors-for-two-oval-spots} in Appendix~\ref{sec:two-oval-spot-corner-plot}. The full posterior probability density distributions for the parameters in the fit of the model with three oval spots are shown in Figure~\ref{fig:full-posteriors-for-three-oval-spots} in Appendix~\ref{sec:three-oval-spot-corner-plot}. Among other things, these ``corner'' plots show that there is a correlation between estimates of the stellar radius and the distance, but that this correlation is fairly weak. Thus, any errors in the data that bias the estimate of the distance may not significantly bias the estimate of the radius. There are indications that our sampling of the posterior distribution of the stretching parameter for the second oval spot did not explore a wide enough range, but this probably did not significantly affect our results, because there is no correlation between the value of this parameter and the values of the mass, radius, or distance parameters.

\subsection{Adequacy of the Models}

We performed $\chi^2$ analyses (see Equation~\ref{chi-squared}) to determine whether the models we have used to estimate the radius and mass of PSR~J0030$+$0451 adequately describe the \textit{NICER} data. Specifically, we compared our best-fit energy-resolved model waveforms with the energy-resolved waveform data (i.e., the pulse-phase--energy-channel data) collected by \textit{NICER}, and compared the energy-integrated (bolometric) waveforms given by our best-fit energy-resolved waveform models with the bolometric waveforms observed by \textit{NICER}, using the values of $\chi^2$ given by these comparisons. (In principle, we could also compare the pulse-phase-integrated spectra given by our models with the pulse-phase-integrated spectrum observed by \textit{NICER} by computing the relevant values of $\chi^2$, but our procedure for modeling the observed unmodulated counts---see Section~\ref{sec:fitting-procedure}---guarantees a nearly perfect description of the photon energy spectrum, so this comparison would be uninformative.)

Assigning the \textit{NICER} data to 32~phase bins and 260~energy channels, the best-fit energy-resolved waveform model with two oval spots gives a $\chi^2$ of 8204.68 when compared with this data set, which consists of the number of counts in each of $32 \times 260=8320$ phase-energy bins. As discussed in Section~\ref{sec:waveform-models}, the model waveform with two oval spots has 14 primary parameters, 260 ancillary parameters related to the non-star emission, and one additional parameter that describes its overall phase, yielding a total of $8320-260-14-1=8045$ degrees of freedom. The resulting $\chi^2/{\rm degrees\ of\ freedom}$ (dof) is therefore 8204.68/8045. If this model is correct, the probability of finding a value of $\chi^2/{\rm dof}$ this large or larger is 0.104. Thus, according to the $\chi^2$ test this model provides an acceptable description of this data set.

\begin{figure*}[t!]
\begin{center}
  \resizebox{0.9\textwidth}{!}{\includegraphics{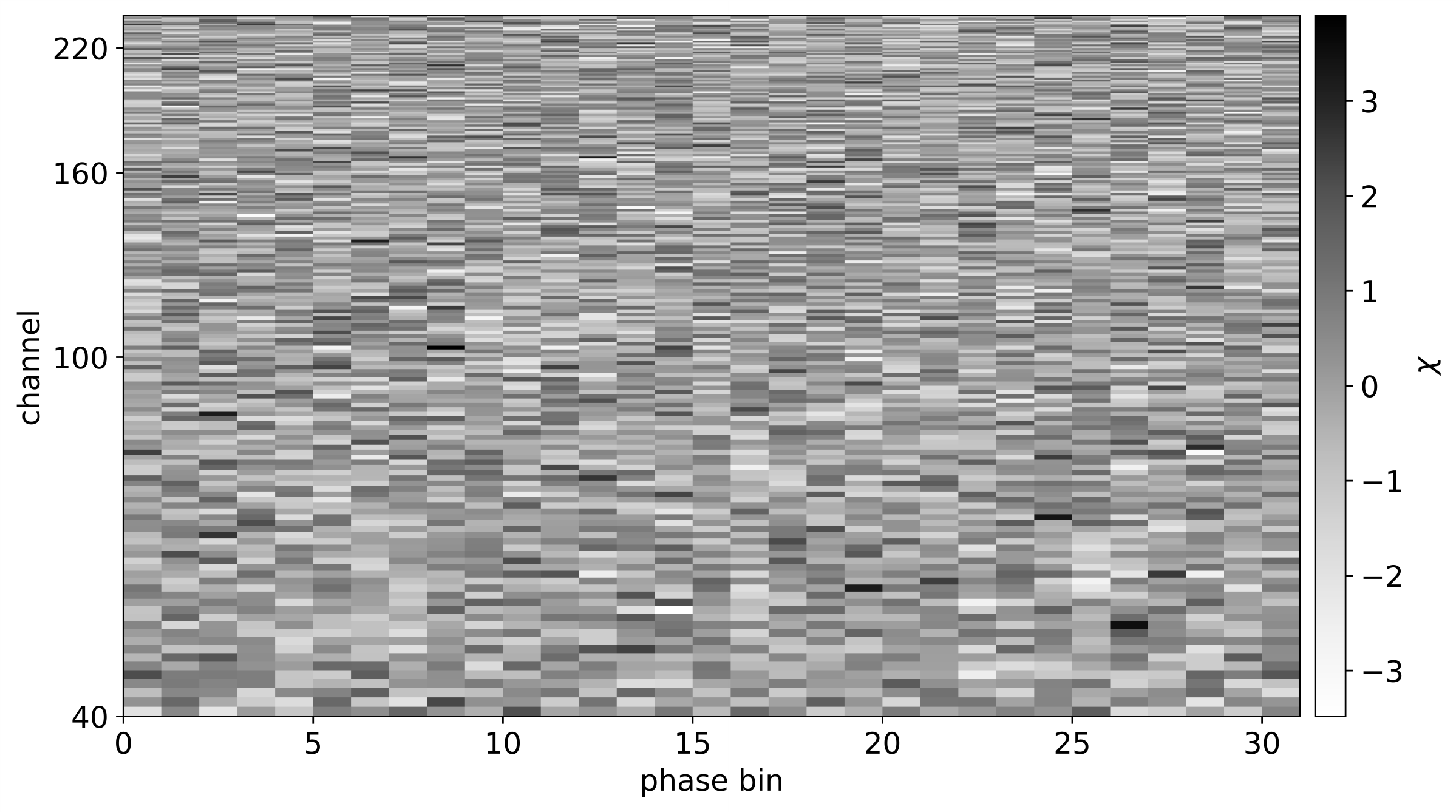}}
\caption{Value of $\chi$ in each of the 8320 phase-energy bins (32~phase bins and 260~energy bins), for the best fit of our energy-resolved waveform model with three oval spots to the \textit{NICER} data grouped in 32~phase bins. The energy channel numbers are plotted on a logarithmic scale. No patterns in the values of $\chi$ are evident as a function of phase or energy, which is what one would expect for a good fit.}
\label{fig:chi-plot}
\end{center}
\end{figure*}

Again assigning the \textit{NICER} data to 32~phase bins and 260~energy channels, the best-fit energy-resolved waveform model with three oval spots gives a $\chi^2$ of 8188.99 when compared with the \textit{NICER} data binned in this way. This model has 19 primary parameters, yielding a total of 8040 degrees of freedom. The resulting $\chi^2/{\rm dof}$ is therefore 8188.99/8040. If this model is correct, the probability of finding a $\chi^2/{\rm dof}$ this large or larger by chance is 0.120, so the fit of this model to this data set is also acceptable, according to the $\chi^2$ test. In Figure~\ref{fig:chi-plot}, we show the value of $\chi$ in each of the 8320 phase-energy bins, for this fit. No patterns in the values of $\chi$ are evident as a function of phase or energy, which is what one would expect for a good fit.

When the bolometric waveforms predicted by the models with two oval spots and three oval spots that best fit the \textit{NICER} phase-channel data with 32 phases are compared with the 32-phase bolometric waveform constructed using the \textit{NICER} data, the values of $\chi^2/{\rm dof}$ are considerably larger. When the bolometric waveform given by the model with two oval spots is compared with the 32-phase bolometric waveform constructed using the \textit{NICER} data, the value of $\chi^2$ is 40.6. There are $32-14-1-1=16$ dof in this 32-phase bolometric data. The $\chi^2/{\rm dof}$ is therefore 40.6/16, which has a probability $6.3\times 10^{-4}$ of occurring by chance. This probability is low enough that it indicates that this model provides a description of this data set that is at least incomplete. When the bolometric waveform given by the model with three oval spots is compared with the 32-phase bolometric waveform constructed using the \textit{NICER} data, the value of $\chi^2$ is 41.7. There are $32-19-1-1=11$ dof in this 32-phase bolometric data. The $\chi^2/{\rm dof}$ is therefore 41.7/11, which has a probability $2\times 10^{-5}$ of occurring by chance. This probability again indicates that this model provides a description of this data set that is at least incomplete.

Noting these low probabilities, we performed exploratory fits specifically designed to minimize the value of $\chi^2$ obtained when the 32-phase bolometric waveforms predicted by these two energy-resolved waveform models are compared with the 32-phase bolometric waveform constructed from the \textit{NICER} data. The probabilities of the models given by these fits were not significantly larger than the values given by the best fits of these models to the full phase-channel data.

Next, we constructed 64-phase bolometric waveforms using the \textit{NICER} data and compared these with the 64-phase bolometric waveforms predicted by the best fit of the 32-phase energy-resolved waveform model with two oval spots to the 32-phase energy-resolved \textit{NICER} waveform data, re-fitting the phase shift and the number of phase-independent background counts in each energy channel. The resulting minimum value of $\chi^2/{\rm dof}$ is 58.64/48, which has a chance probability of 0.140, indicating that this model provides an acceptable description of the 64-phase bolometric waveform data. When the predictions of this model for the 64-phase energy-resolved waveform are compared with 64-phase energy-resolved \textit{NICER} data, the resulting $\chi^2/{\rm dof}$ is 16363.9/16365, which has a probability of 0.501, much higher than the probability we found when we divided the data into only 32 phase bins, and indicating that this energy-resolved waveform model also provides an acceptable description of the energy-resolved \textit{NICER} waveform data with 64 phase bins.

\begin{figure*}[t!]
\begin{center}
\vspace*{-1.0truein}
  \resizebox{0.9\textwidth}{!}{\includegraphics{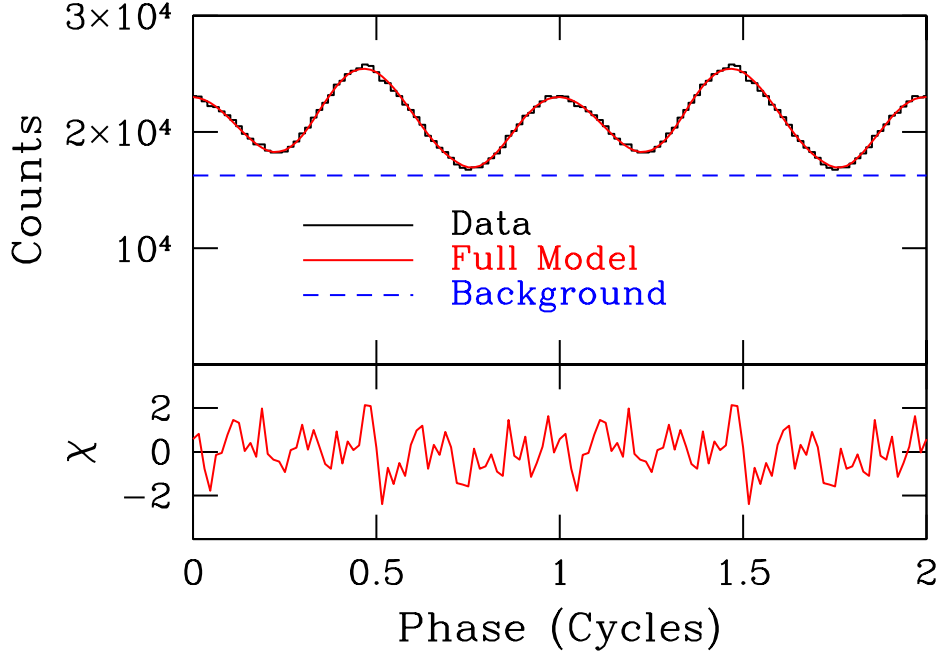}}
\vspace{-3.6truein}
   \caption{
Top: comparison of the 64-phase bolometric waveform constructed using the \textit{NICER} data on PSR~J0030$+$0451 with the 64-phase bolometric waveform model given by the 32-phase energy-resolved waveform model with three oval spots that best fits the 32-phase energy-resolved waveform data.
The dashed blue line shows the fitted unmodulated background that was added to the counts produced by the three hot spots as part of the fitting procedure (see Section~\ref{sec:fitting-procedure}).
The zero of time of the model bolometric waveform was adjusted to minimize the value of $\chi^2$; this adjustment was +0.49 cycles.
Bottom: the resulting value of $\chi$ as a function of phase. The $\chi^2/{\rm dof}$ is 59.6/43, which has a probability of 0.0473.
    }
\label{fig:bolo-waveform+residuals}
\end{center}
\end{figure*}

Finally, we compared the 64-phase bolometric waveforms constructed using the \textit{NICER} data with the 64-phase bolometric waveforms predicted by the energy-resolved waveform model with three oval spots that best fits the 32-phase energy-resolved \textit{NICER} waveform data (again re-fitting the phase shift and the number of phase-independent counts in each energy channel). The results are shown in Figure~\ref{fig:bolo-waveform+residuals}. The minimum value of the bolometric $\chi^2/{\rm dof}$ is 59.6/43, which has a probability of 0.0473, indicating that this 64-phase model also provides an acceptable description of the 64-phase bolometric waveform data. When the predictions of this best-fit model for the energy-resolved waveform with 64 phase bins are compared with the 64-bin energy-resolved \textit{NICER} waveform data, the resulting $\chi^2/{\rm dof}$ is 16347.5/16360, which has a probability of 0.526, again much higher than the probability we found when we divided the data into only 32 phase bins, and indicating that this energy-resolved waveform model provides an acceptable description of the energy-resolved \textit{NICER} waveform data with 64 phase bins.

These results indicate that our best-fit models with two and three oval spots provide good descriptions of the \textit{NICER} waveform data at high phase resolutions, and that the radius and mass estimates inferred from them are therefore credible. Why the bolometric waveforms given by the models that best fit the 32- and 64-phase energy-resolved \textit{NICER} waveform data differ from the 32-phase bolometric waveforms constructed from the \textit{NICER} data but agree with 64-phase bolometric waveforms is unclear, but is most likely due to temporal fluctuations in the \textit{NICER} data that are not yet understood. This question deserves further study.

\newpage

\subsection{Other Aspects of the Models}

\subsubsection{Locations of the Hot Spots}
\label{sec:spot-locations}

\begin{figure*}[ht!]
\gridline{\fig{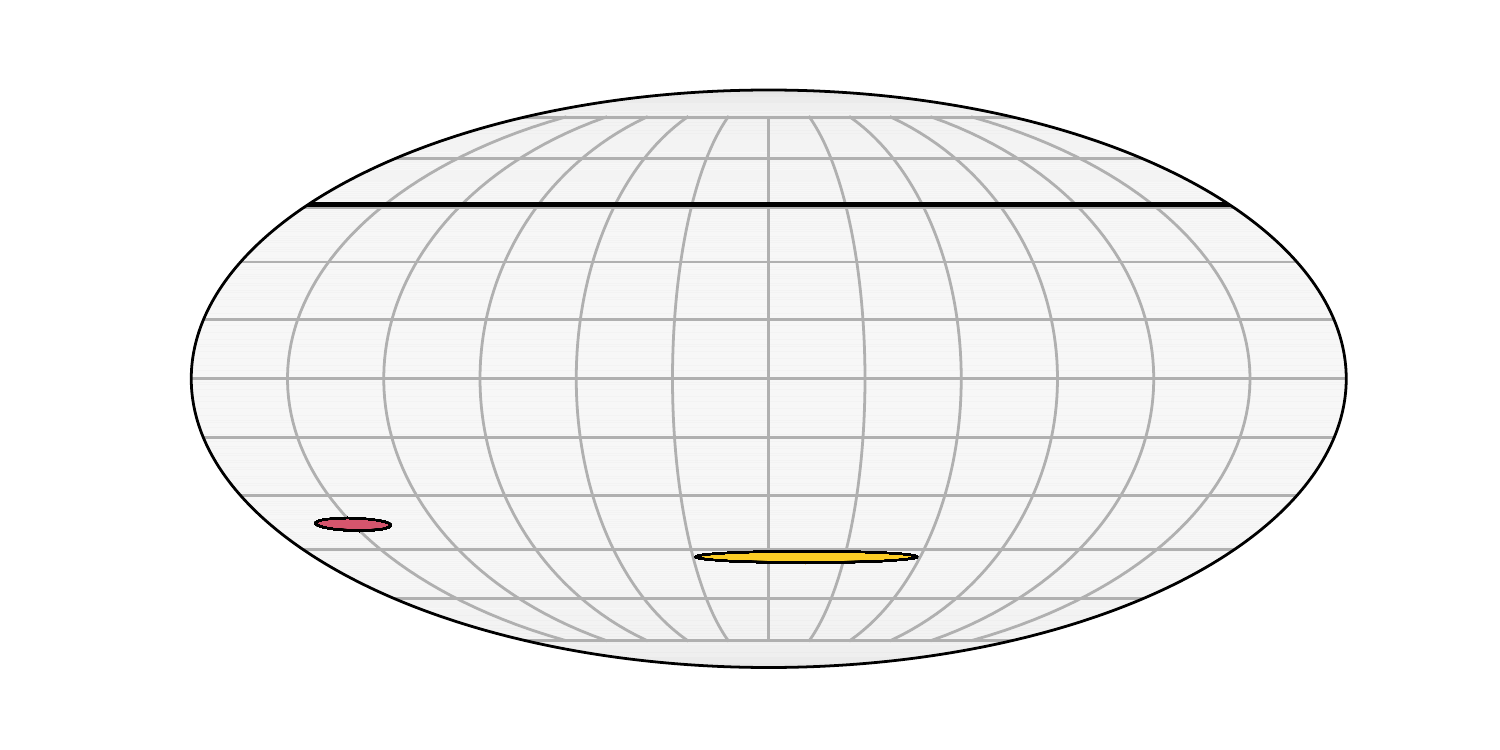}{0.6\textwidth}{(a)}
          \fig{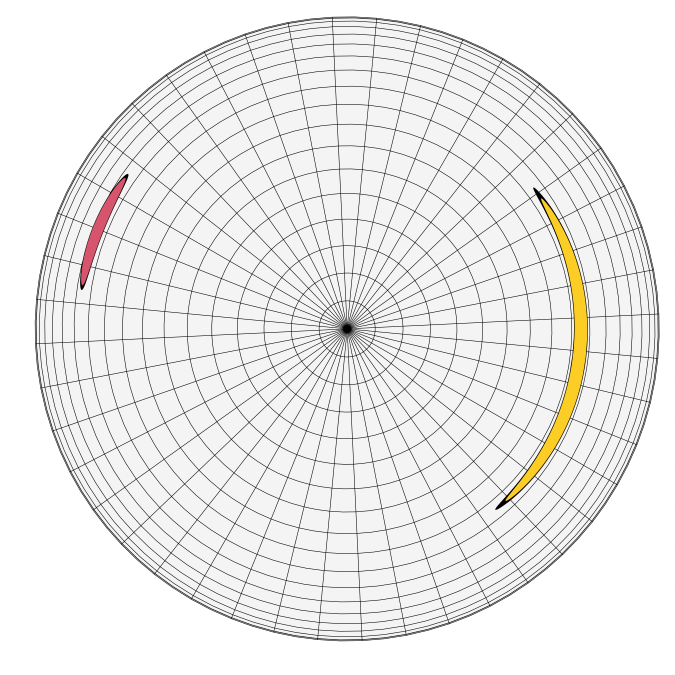}{0.3\textwidth}{(b)}}
\gridline{\fig{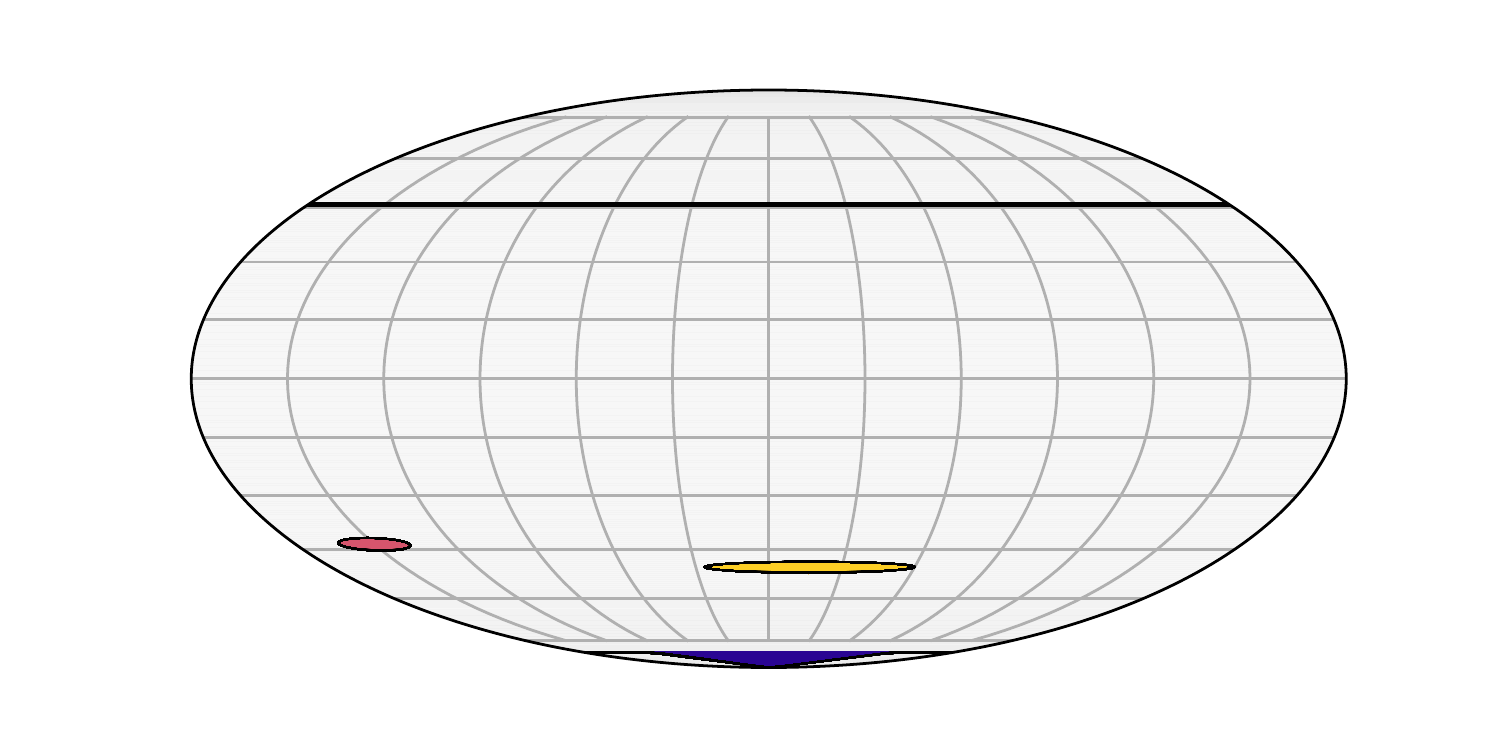}{0.6\textwidth}{(c)}
          \fig{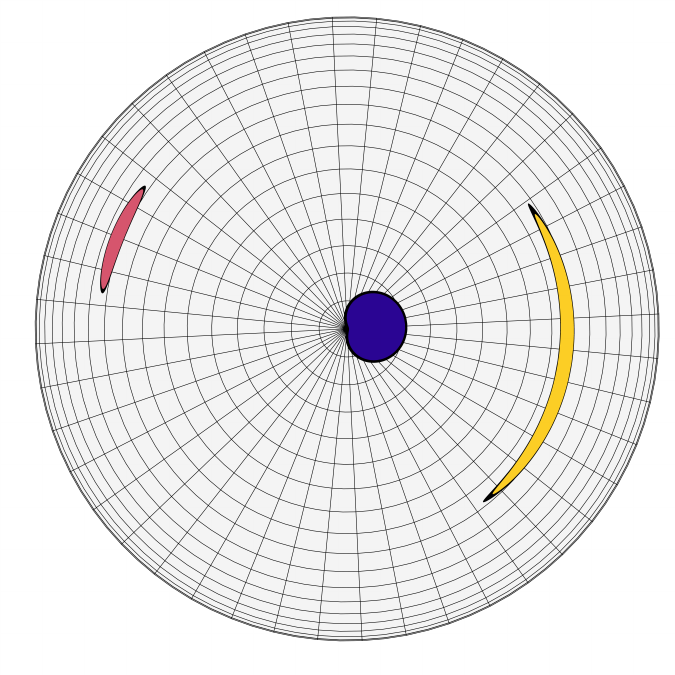}{0.3\textwidth}{(d)}}
\caption{Locations, shapes, and sizes of the hot spots in the best-fit waveform models with two oval spots (panels~(a) and~(b)) (see Table~\ref{tab:credible2}) and three oval spots (panels~(c) and~(d)) (see Table~\ref{tab:credible3}). Panels~(a) and~(c) show equal-area projections, centered on the rotational equator. Panels~(b) and~(d) are views from the south pole. The cooler main spot is indicated by yellow, the hotter main spot is indicated by red, and in the three-spot model, the hottest spot is indicated by blue. For both fits, the horizontal line shows the inferred colatitude of the observer. Clearly, both spots in the model with two oval spots and the two main spots in the model with three oval spots are very similar in location, size, and shape; the third spot in the three-spot model has a very small area and makes only a minor contribution to the waveform.}
\label{fig:spotlocations}
\end{figure*}

Figure~\ref{fig:spotlocations} compares the best-fit locations and shapes of the hot spots obtained by fitting the models with two and three oval spots to the \textit{NICER} waveform data.
Both hot spots in the model with two oval spots and all three hot spots in the model with three oval hot spots are located in the southern rotational hemisphere, well away from the sightline to the observer (recall our convention that the sightline to the observer defines the northern hemisphere). Fits that force any of the spots to be in the same hemisphere as the sightline to the observer are strongly disfavored.  As we noted earlier, when we discussed the phase-channel $\chi^2$ values for the fits of both models to the \textit{NICER} waveform data, these spot locations provide formally excellent fits.

The fundamental reason spot locations in the southern hemisphere are favored appears to be that the modulation fraction of the waveform observed by \textit{NICER} is high and its harmonic content is substantial. Both waveform properties favor spot locations in the southern hemisphere, because when the spots are visible for only a small fraction of a rotational cycle, both the modulation amplitude of the waveform and the strengths of the higher harmonics of the spin frequency are larger. Spots located in the northern hemisphere are visible for most of each rotational cycle and therefore produce waveforms with smaller modulation amplitudes and weaker high harmonics. In contrast, radio observations as well as joint fitting of radio and $\gamma$-ray profiles suggest geometries in which one of the two radio pulses comes from the northern hemisphere \citep{2014ApJS..213....6J}. However, these studies assume a star-centered dipolar field, whereas the \textit{NICER} results suggest that the magnetic field of PSR~J0030$+$0451 is not a centered dipole (see also \citealt{2019MNRAS.490.1774L}). Although the locations of the soft, thermal X-ray emission and the nonthermal radio emission need not be the same, this difference should be investigated further.

\subsubsection{Colatitude of the Observer}
\label{sec:observer-colatitude}

One recent analysis of $\gamma$-ray observations of PSR~J0030$+$0451 suggests that the colatitude $\theta_{\rm obs}$ of the observer lies between 0.942 and 1.222 rad (the boundaries of the $\pm 1\sigma$ interval; see \citealt{2019MNRAS.483.1796C}). The lower end of this interval is consistent with the $\pm 1\sigma$ range of the probability density distributions for $\theta_{\rm obs}$ that we infer from the fit of our model with two oval spots and the fit of our model with three oval spots, which is encouraging.

\subsubsection{Spectrum of the Emission from the Hot Spots}
\label{sec:hot-spot-spectra}

\begin{figure*}[ht!]
\begin{center}
\vspace{-0.8truein}
  \resizebox{0.8\textwidth}{!}{\includegraphics{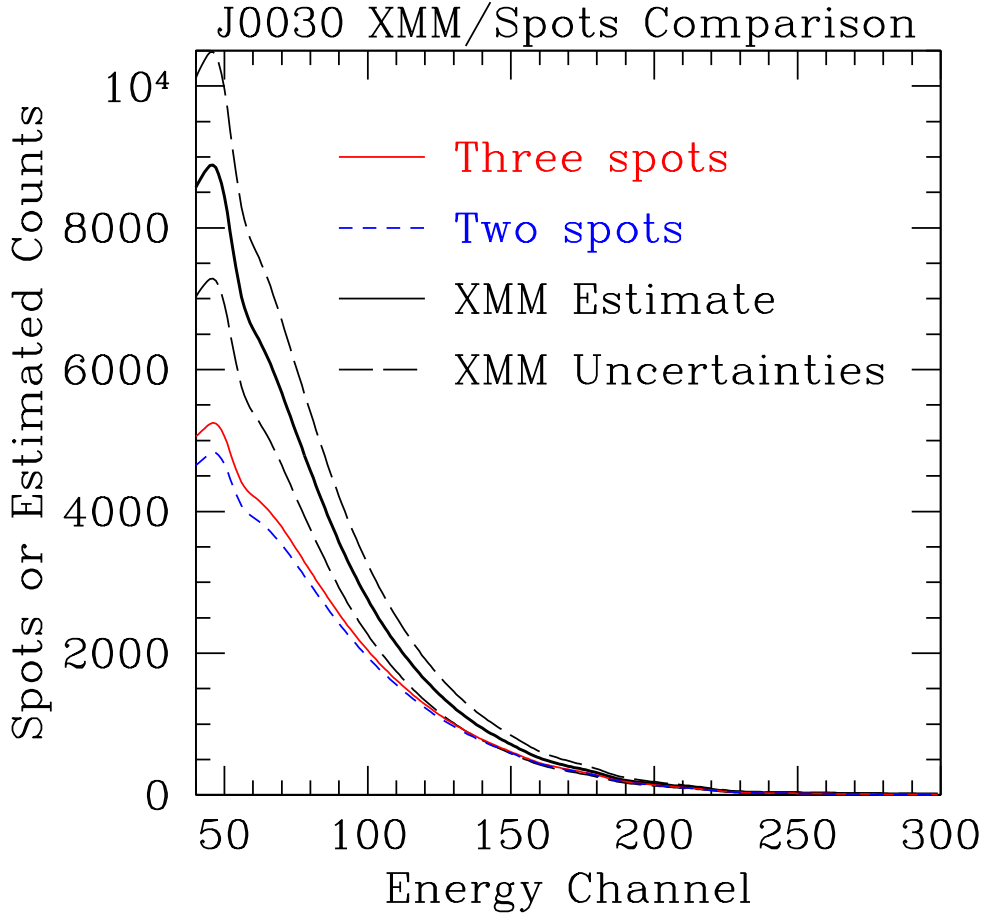}}
\vspace{-1.7truein}
   \caption{Comparison of the predicted total count spectrum that should be observed from PSR~J0030$+$0451 by \textit{NICER} (solid black line), based on the \textit{XMM-Newton} observations of PSR~J0030$+$0451, with the total count spectrum expected from all the hot spots in the best-fit model with two (dashed blue line) and three (solid red line) oval hot spots. The dashed black lines indicate the $\pm 1\,\sigma$ uncertainty in the count spectrum predicted from the \textit{XMM-Newton} observations. They are a linear combination of the statistical uncertainty in the \textit{XMM-Newton} observations and an estimate of the systematic error in the \textit{NICER} calibration. The total emission predicted in the lower-energy channels using the emission observed by \textit{XMM-Newton} is greater than the total emission expected from all the hot spots in the best-fit hot spot models. This indicates that the emission provided by the hot spots in these models does not account for all of the emission.}
\label{fig:spectcompare}
\end{center}
\end{figure*}

An independent estimate of the total X-ray emission from PSR~J0030$+$0451 can be made using observations made with \textit{XMM-Newton} \citep{2009ApJ...703.1557B}. Although the \textit{XMM-Newton} EPIC~MOS1/2 observations have substantially fewer source counts, they have a much lower background in the point-source spatial extraction region compared to the non-imaging \textit{NICER} data, which means that to first order all of the counts detected by \textit{XMM-Newton} come from the star rather than from unassociated sources. If the phase-integrated data from \textit{XMM-Newton} are fit using a two-temperature non-magnetic hydrogen atmosphere model (as in \citealt{2009ApJ...703.1557B}) and the predictions of the model are folded through the \textit{NICER} response matrix, we obtain an estimate for the total number of \textit{NICER} counts from the star and the spectral shape we expect to see in \textit{NICER}. This predicted spectral flux can be compared with the total spectral flux expected from all the hot spots in our best-fit models of the pulse waveform emission. Figure~\ref{fig:spectcompare} shows this comparison for our models with two oval spots and three oval spots. In both cases, the combined spectral flux expected from the hot spots at low photon energies falls short of the spectral flux predicted by the \textit{XMM-Newton} observations.  

A first thought would be that the missing emission might be unpulsed thermal emission from a substantial fraction of the stellar surface. There are, however, serious difficulties with such an interpretation. First, the emission would have to be almost exactly axisymmetric around the stellar rotation axis in order to avoid generating detectable pulsed emission. This forces one to consider emission patterns that are highly tuned: filled circular emitting regions around one or both rotation poles, annular emitting regions centered around one or both rotation axes, or some combination of these would be required to avoid producing detectable flux modulation. Second, if the emission is thermal, the total area of the axisymmetric emitting region(s) would have to be a very small fraction of the stellar surface.

We illustrate these difficulties by an example. In order to make up the observed deficit in the flux at low energies and not produce detectable modulation, a circular spot centered on the north rotational pole (the pole nearer the observer), would have to have an angular radius of just 0.075~rad, assuming thermal emission at the best-fit effective temperature of $kT_{\rm eff} \approx 0.075$~keV. Such a spot would also have to be very nearly circular and centered on the north rotational pole: a deviation of more than $\sim\,0.01$~rad would cause a flux modulation that would be inconsistent with the observed waveform. Because of its smaller projected area, a circular spot centered on the south rotational pole would have to have an angular radius $\sim\,0.7$~rad, about 10 times larger than a spot at the north rotational pole, but would have to be even more precisely circular and centered than a spot at the north rotational pole: a deviation of more than $\sim\,0.005$~rad would cause a flux modulation inconsistent with the observed waveform.

Other thermal emission geometries with the same projected area are theoretically possible (e.g., one or more very thin axisymmetric annuli), but these seem to us even more fine-tuned. The spectrum could also be produced by processes in which the peak in the spectrum required to make up the deficit does not reflect the temperature of the emitting plasma, but is instead created by a steep decrease of the optical depth of the emitting region with increasing photon energy. High-harmonic cyclotron emission in a region where the optical depth to cyclotron absorption decreases with increasing photon energy is one such possibility \citep{1999A&AT...18..447P}, but usually requires a population of higher-energy electrons, in which case the optically thick (Rayleigh-Jeans) portion of the spectrum would have to have a very small area, in order to avoid producing too much emission.

It does appear worthwhile to explore other possibilities. For example, the spot pattern is clearly not that of a centered dipole. If this means that the total magnetic field at the surface of the star has significant multipolar components, and is therefore much stronger than the nominal dipole component of $\sim 2\times 10^8$~G, then perhaps cyclotron emission could contribute to the missing emission.

\section{IMPLICATIONS FOR THE DENSE MATTER EoS}
\label{sec:implications-for-EoS}

The joint posterior probability distribution we have obtained for the mass and radius of PSR~J0030$+$0451 provides additional constraints on the EoS of cold catalyzed matter, through the effect of the EoS on the stellar structure. The EoS can be expressed as a relation between the pressure $p$ and the total mass-energy density $\rho$, or a similar relation between $p$ and a different thermodynamic variable.

\begin{figure*}[t!]
\begin{center}
\vspace{-1.0truein}
  \resizebox{0.8\textwidth}{!}{\includegraphics{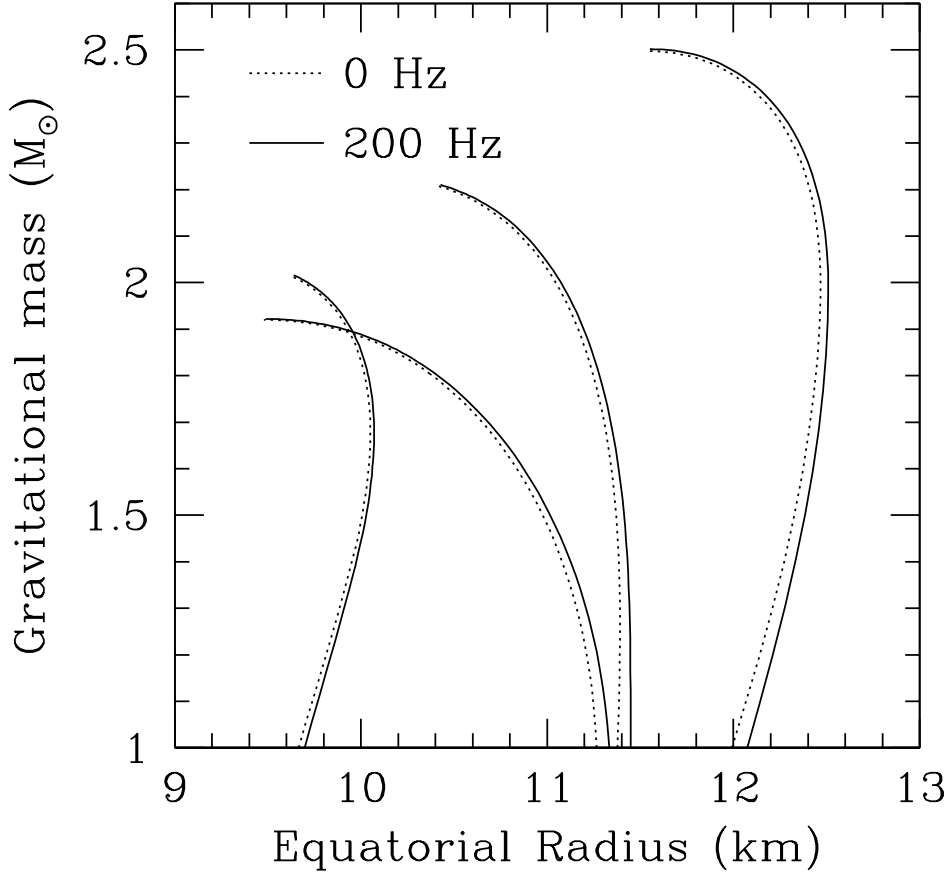}}
\vspace{-1.7truein}
   \caption{Comparison of the mass-radius relations for a wide variety of EoS computed exactly for a stellar spin rate of 200~Hz (solid curves) and approximately using the TOV equation (dotted curves). 
The EoS are (in order of increasing radius at $1.5~M_\odot$)
HLPS1 \citep{2013ApJ...773...11H}, 
BBB2 (the higher mass model that uses the Paris two-body interaction potential from \citealt{1997A&A...328..274B}), 
APR (Model A18+$\delta(v)$+UIX* from \citealt{1998PhRvC..58.1804A}), and 
HLPS2 \citep{2013ApJ...773...11H}.  
For the $\sim\,$$1.4\,M_\odot$ estimated mass of PSR~J0030$+$0451, the exact radii for a spin rate of 200~Hz differ by less than 1\% from the approximate radii given by the TOV equation. This difference is far smaller than the uncertainty in the estimate of the radius of PSR~J0030$+$0451 we have obtained from analyzing its soft X-ray waveform.}
\label{fig:validity-of-TOV}
\end{center}
\end{figure*}

Here we use the Tolman$-$Oppenheimer$-$Volkoff (TOV) structure equation \citep{1939PhRv...55..364T,1939PhRv...55..374O}
\begin{equation}
{dp\over{dr}}=-{(\rho c^2+p)(M(<r)+4\pi r^3p/c^2)\over{r(c^2r/G-2M(<r))}} \; ,
\label{eq:TOV}
\end{equation}
where $r$ is the circumferential radius and $M(<r)$ is the gravitational mass inside $r$. The TOV equation assumes that the star is cold enough that its EoS can be treated as barotropic, which is expected to be an excellent approximation for PSR~J0030$+$0451. The TOV equation also assumes that the star is nonrotating and spherically symmetric. Figure~\ref{fig:validity-of-TOV} shows that this is an excellent approximation for PSR~J0030$+$0451, which has a rotation frequency of 205~Hz. For a given central mass-energy density, the compactness ratio $GM/(R_ec^2)$ is even less sensitive to the rotation rate than either the mass or the radius considered separately.

Given $p(\rho)$ and a central density, the TOV equation can be integrated to yield the gravitational mass $M$ and circumferential radius $R_e$ of a star.

There are numerous papers that apply various Bayesian techniques to the problem of constraining a parameterization of $p(\rho)$ using various astronomical measurements (e.g., \citealt{2015PhRvD..92b3012A}; \citealt{2015PhRvD..91d3002L}; \citealt{2016EPJA...52...69A}; \citealt{2018MNRAS.478.1093R}; \citealt{2019MNRAS.485.5363G}).  Here we use the approach of \citet{2019arXiv190408907M}, who emphasize using full posterior probability distributions rather than strict bounds on the mass or other quantities.  They also emphasize that mass-radius measurements such as the ones we report in Section~\ref{sec:estimating-J0030-mass&radius} need to be incorporated in the constraints on the EoS by fully marginalizing over the central density, rather than (for example) simply picking the point in a mass-radius curve from an EoS that has the highest likelihood from the mass-radius measurement.  

There is at present no agreement on the EoS of neutron star matter at the densities expected near the centers of the most massive stars, or even on how best to incorporate the information that is currently available from nuclear and particle physics. A variety of parameterized EoS models are therefore currently being considered (for recent examples, see \citealt{2018ApJ...860..149T}; \citealt{2019ApJ...885...42B}; and \citealt{2019MNRAS.485.5363G}). For this reason, we prefer here to use generic parameterizations to illustrate the new constraints on EoS models that are provided by the new measurements of the mass and radius of PSR~J0030$+$0451 reported above. We caution, though, that as has been emphasized by \citet{2019MNRAS.485.5363G} and others, the constraints on the EoS derived from a given set of observations depend somewhat on the parameterization chosen for the EoS. Here we present results for the following two parameterizations:

\begin{enumerate}

\item The spectral parameterization of \citet{2010PhRvD..82j3011L} (see also \citealt{2018PhRvD..97l3019L}).  In this approach, the adiabatic index $\Gamma(p)=[(\rho+p)/p](dp/d\rho)$ is represented using
\begin{equation}
\Gamma(p)=\exp\left(\sum_k\gamma_kx^k\right)\; ,
\end{equation}
where $x\equiv\log(p/p_0)$ and $p_0$ is the pressure at $\rho_s/2$, where $\rho_s$ is the density of nuclear saturation, which we take to be $\approx 2.7\times 10^{14}~{\rm g~cm}^{-3}$. Following \citet{2018PhRvD..98f3004C} and \citet{2018PhRvL.121p1101A}, we carry out the expansion up to $x^3$. The intervals in $x$ that we use and the corresponding uniform priors are $\gamma_0\in[0.2,2]$, $\gamma_1\in[-1.6,1.7]$, $\gamma_2\in[-0.6,0.6]$, and $\gamma_3\in[-0.02,0.02]$.

\item A piecewise polytrope with transition densities that are also parameters.  Here the adiabatic index from $\rho_s/2$ to $\rho_2\in[3/4,5/4]\rho_s$ is $\Gamma_1\in[2,3]$; from $\rho_2$ to $\rho_3\in[3/2,5/2]\rho_s$ is $\Gamma_2\in[0,5]$; from $\rho_3$ to $\rho_4\in[3,5]\rho_s$ is $\Gamma_3\in[0,5]$; from $\rho_4$ to $\rho_5\in[6,10]\rho_s$ is $\Gamma_4\in[0,5]$; and from $\rho_5$ to $\infty$ is $\Gamma_5\in[0,5]$ (all priors are uniform in the listed range).  The relatively restricted range of $\Gamma_1$ is based on the study of \citet{2013ApJ...773...11H}.  Note also that the large number of parameters in this parameterization is intended to give flexibility to our description of the equation of state; we are not attempting to produce a concise fit to realistic microphysics. Unlike the spectral parameterization, the piecewise polytrope can represent phase transitions, in regions where the adiabatic index is close to zero.

\end{enumerate}

\noindent   For both parameterizations, we use the EoS of \citet{2001A&A...380..151D} up to $\rho = \rho_s/2$.  We also use as a prior the requirement that all candidate EoS be able to support stars with masses of at least $1.4~M_\odot$. Unlike in \citet{2019arXiv190408907M}, we do not incorporate information about the nuclear symmetry energy, but otherwise we follow their procedure.

For a given EoS, we do not consider densities above the density at which the adiabatic sound speed for that particular EoS exceeds the speed of light ($dp/d\rho>c^2$).  We generate 150,000 random equations of state from each parameterization and then weight them, first, based only on the priors, then using our mass and radius measurement for PSR~J0030$+$0451, and finally also including the high measured masses of PSR~J1614$-$2230 \citep{2018ApJ...859...47A}, PSR~J0348$+$0432 \citep{2013Sci...340..448A}, and PSR~J0740$+$662 \citep{2019arXiv190406759T}, and the tidal deformability constraints from the GW170817 binary neutron star coalescence event \citep{2018PhRvL.121p1101A,2018PhRvL.121i1102D}. 

In order to perform our analysis, we need a way to turn discrete samples of the $(M,R)$ posterior into an estimate of the continuous posterior probability distribution.  We use the standard approach of kernel density estimation (see \citealt{rosenblatt1956,parzen1962,silverman1986} for details).  In kernel density estimation, each point in the sample is replaced by an extended probability distribution (that is, a kernel), and then the probability density at any $(M,R)$ is proportional to the sum of the kernels associated with every point, evaluated at $(M,R)$.  One must make a choice for both the shape of the kernel and its width (called its bandwidth in this context).  For our kernels we use a Gaussian form, and a bandwidth matrix ${\bf H}$ based on the covariance matrix.  More specifically, our estimate of the probability density at a
given point ${\bf x}=(M,R)$ given ${\bf H}$ and samples at the points
${\bf X}_1,{\bf X}_2,\ldots,{\bf X}_n$ is
\begin{equation} {\hat f}_{\bf H}({\bf x})={1\over n}\sum_{i=1}^n {1\over{{\rm det}({\bf H})}}{\cal K}\left\{{\bf H}^{-1}({\bf x}-{\bf X}_i)\right\}\;.
\end{equation}
Here ${\cal K}({\bf\nu})$, where ${\bf\nu}$ is a column vector, is
\begin{equation}
{\cal K}({\bf\nu})=\exp(-{\bf\nu}^T{\bf\nu}/2)\; .
\end{equation}

For ${\bf H}$ we employ the covariance matrix $\Sigma$.  To calculate $\Sigma$, assume that each sample point in $(M,R)$ is a 2D vector $(M_i,R_i)$.  Let $\mu_M$ be the mean of $M$ over the sample, let $\mu_R$ similarly be the mean of $R$ over the sample, and let $E[(Y_i-\mu_i)(Y_j-\mu_j)]$ be the mean of $(Y_i-\mu_i)(Y_j-\mu_j)$ over the sample where $i$ and $j$ are $M$ or $R$. Then the $ij$ element of $\Sigma$ is
\begin{equation}
\Sigma_{ij}=E[(Y_i-\mu_i)(Y_j-\mu_j)]\; .
\end{equation}
The standard rule-of-thumb from \citet{silverman1986} is that the bandwidth matrix is
\begin{equation}
{\bf H}=n^{-1/(d+4)}{\bf\Sigma}^{1/2}
\end{equation}

\begin{figure*}[ht!]
\begin{center}
\vspace*{-1.5truein}
  \resizebox{1.06\textwidth}{!}{\includegraphics{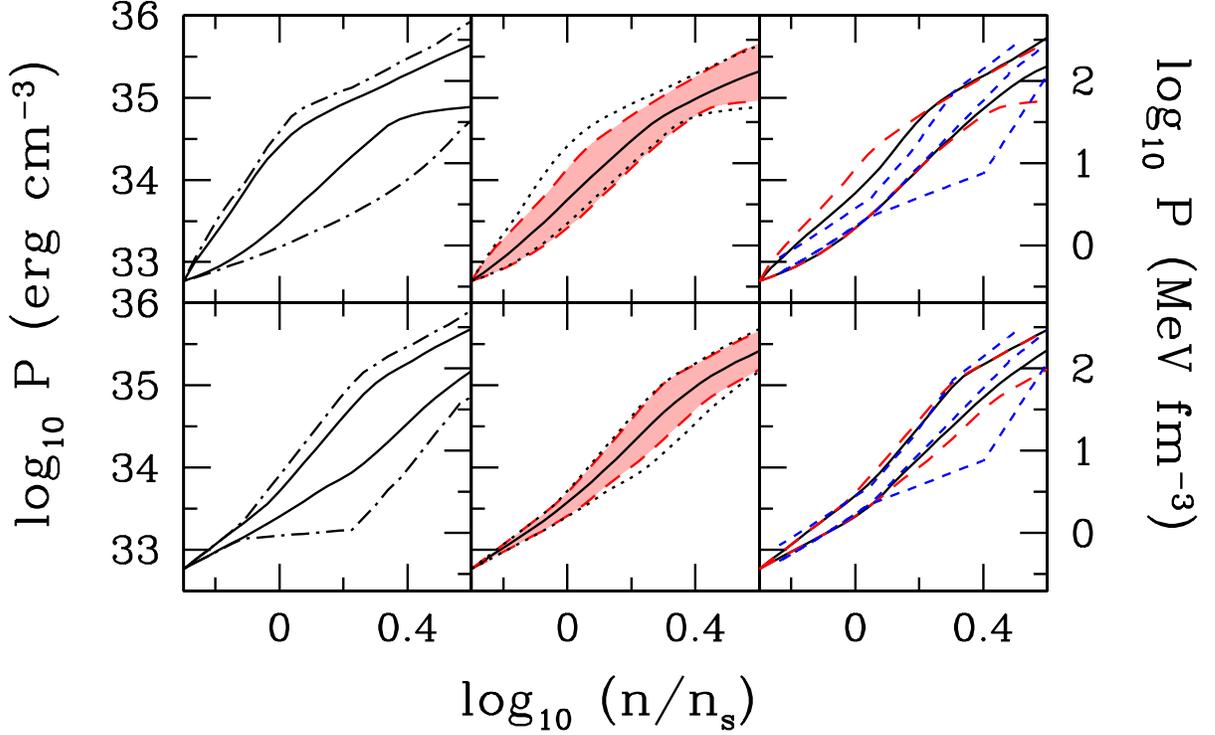}}
\vspace{-4.4truein}
   \caption{Constraints on the EoS (pressure vs.\ equivalent number density) of cold, catalyzed matter with and without taking into account the \textit{NICER} measurements of the mass and radius of PSR~J0030$+$0451 described earlier in this Letter, for the two parameterizations of the EoS described in the text. We define the equivalent number density as the rest mass density divided by the mass of a neutron.
   Top row: constraints on the spectral EoS described in the text. 
   Top left: the vertical separation of the black dashed-dotted lines shows the full range of the assumed prior for the pressure at each equivalent number density, for the spectral EoS; the black solid lines show the 5\% (bottom) and 95\% (top) credibility contours for the pressure at each number density, when only the priors are used. 
   Top middle: the black dotted lines show the 5\% and 95\% contours in the credible region when only the prior on the pressure is used, whereas the red shaded region, which is bounded by the red dashed lines, shows the range of pressures between the 5\% and 95\% contours when the priors are augmented by the mass and radius estimates that we report here; the solid black line shows the median pressure within the credible region at each density. 
   Top right: the red dashed lines are the same as in the middle panel, whereas the blue dashed lines show the soft, intermediate, and hard EoS described in \citet{2013ApJ...773...11H}; the solid black lines show the 5\% and 95\% contours when the information provided by the measured masses of PSR~J1614$-$2230 \citep{2018ApJ...859...47A}, PSR~J0348$+$0432 \citep{2013Sci...340..448A}, and PSR~J0740$+$662 \citep{2019arXiv190406759T}, and the tidal deformability estimate from the GW170817 event is added to the information provided by the mass and radius estimates reported here. 
   Bottom row: the same as in the top row, but for the piecewise-polytropic EoS described in the text. 
   These results show that the measurements of the mass and radius of PSR~J0030$+$0451 made using \textit{NICER} definitely tighten the constraints on the EoS of the cold, catalyzed matter in the interiors of neutron stars.}
\label{fig:EoS}
\end{center}
\end{figure*}

\noindent where $d$ is the number of dimensions in the data ($d=2$ in our case) and $n$ is the number of samples from the posterior.  This bandwidth is a compromise: if the width is too great then information is lost (an infinitely broad kernel would lead to a constant probability density) whereas if the width is too narrow then the probability density is choppy, with many peaks near the points that happened to be sampled.

We find, however, that our measurement of $M$ and $R$ for PSR J0030+0451 is precise enough that the standard bandwidth produces an inappropriately broadened posterior.  For example, using the standard bandwidth we find a smoothed posterior in $M/R$ that is $\sim 50\%$ broader than the unsmoothed posterior.  We therefore multiply the standard bandwidth by 0.1, so that for $n$ samples and $d=2$ dimensions, ${\bf H}=0.1n^{-1/6}{\bf\Sigma}^{1/2}$.
This modification yields a smoothed posterior that retains the information without introducing choppiness.

As Figure~\ref{fig:EoS} shows, the measurements of the mass and radius of PSR~J0030$+$0451 made using \textit{NICER} have definitely tightened the constraints on the EoS of the cold, catalyzed matter in the interiors of neutron stars.

\section{CONCLUSIONS}
\label{sec:conclusions}

Our analysis of the \textit{NICER} pulse waveform data on PSR~J0030$+$0451 and the independent analysis by \citet{2019ApJ...887L..21R} are the first analyses of this type on data with enough counts to provide a precise and reliable measurement of the mass and radius of a neutron star.  The radius and mass estimates given by our analysis are $R_e = 13.02^{+1.24}_{-1.06}$~km and $M = 1.44^{+0.15}_{-0.14}~M_\odot$ (68\%). These estimates imply an allowed range of high-density equations of state that is consistent with previous measurements of neutron star masses  (see, e.g., \citealt{2013Sci...340..448A,2018ApJ...859...47A,2019arXiv190406759T}), with the tidal deformability of the neutron stars in GW170817 (see, e.g., \citealt{2018PhRvL.121p1101A}; \citealt{2018PhRvL.121i1102D}), and with general nuclear physics considerations (see, e.g., the discussion in \citealt{2018PhRvL.121i1102D}). The \textit{NICER} measurements have significantly tightened constraints on the EoS. The consistency of our results with those of \citet{2019ApJ...887L..21R} adds confidence that the systematic errors in these results are not large, but further pulse waveform modeling and analysis of additional \textit{NICER} data will be helpful.  

In summary, our results and those of \citet{2019ApJ...887L..21R} provide a new constraint on the radius and mass of PSR~J0030$+$0451, and through them, on the properties of the dense matter in its core (see also \citealt{2019ApJ...887..22R}). This work, and future analyses of \textit{NICER} data on other sources, represents an important contribution to the ongoing transition to an era of unprecedented precision in our knowledge of neutron star interiors.

\acknowledgements

This work was supported by NASA through the \textit{NICER} mission and the Astrophysics Explorers Program. 
The authors acknowledge the University of Maryland supercomputing resources\break
(http://hpcc.umd.edu) that were made available for conducting the research reported in this Letter. M.C.M. is grateful for the hospitality of the Kavli Institute for Theoretical Physics at the University of California, Santa Barbara, where part of this Letter was written. This Letter was therefore supported in part by the National Science Foundation under grant No.\ NSF PHY-1748958.  M.C.M. was also supported by a Visiting Researcher position at Perimeter Institute for Theoretical Physics in the last stages of this project.  W.C.G.H. appreciates the use of computer facilities at the Kavli Institute for Particle Astrophysics and Cosmology.  J.M.L. acknowledges support from NASA through grant 80NSSC17K0554 and the U.S. DOE from grant DE-FG02-87ER40317.  R.M.L. acknowledges the support of NASA through Hubble Fellowship Program grant HST-HF2-51440.001.
The authors acknowledge the use of NASA's Astrophysics Data System (ADS) Bibliographic Services and the arXiv. 

\facility{\textit{NICER} (\citealt{2016SPIE.9905E..1HG})}

\software{emcee (\citealt{2013PASP..125..306F}), MultiNest (\citealt{2009MNRAS.398.1601F}), Python and NumPy (\citealt{2007CSE.....9c..10O}), Matplotlib (\citealt{2007CSE.....9...90H}), Cython (\citealt{2011CSE....13b..31B})}, SuperMongo (https://www.astro.princeton.edu/$\sim$rhl/sm/sm.html)

\clearpage
\appendix

\section{Posterior Distributions for the Model With Two Oval Spots}
\label{sec:two-oval-spot-corner-plot}
\restartappendixnumbering

\renewcommand\thefigure{\thesection.\arabic{figure}}
\setcounter{figure}{0}

\begin{figure*}[ht!]
\begin{center}
\vspace*{-0truein}
  \resizebox{0.90\textwidth}{!}{\includegraphics{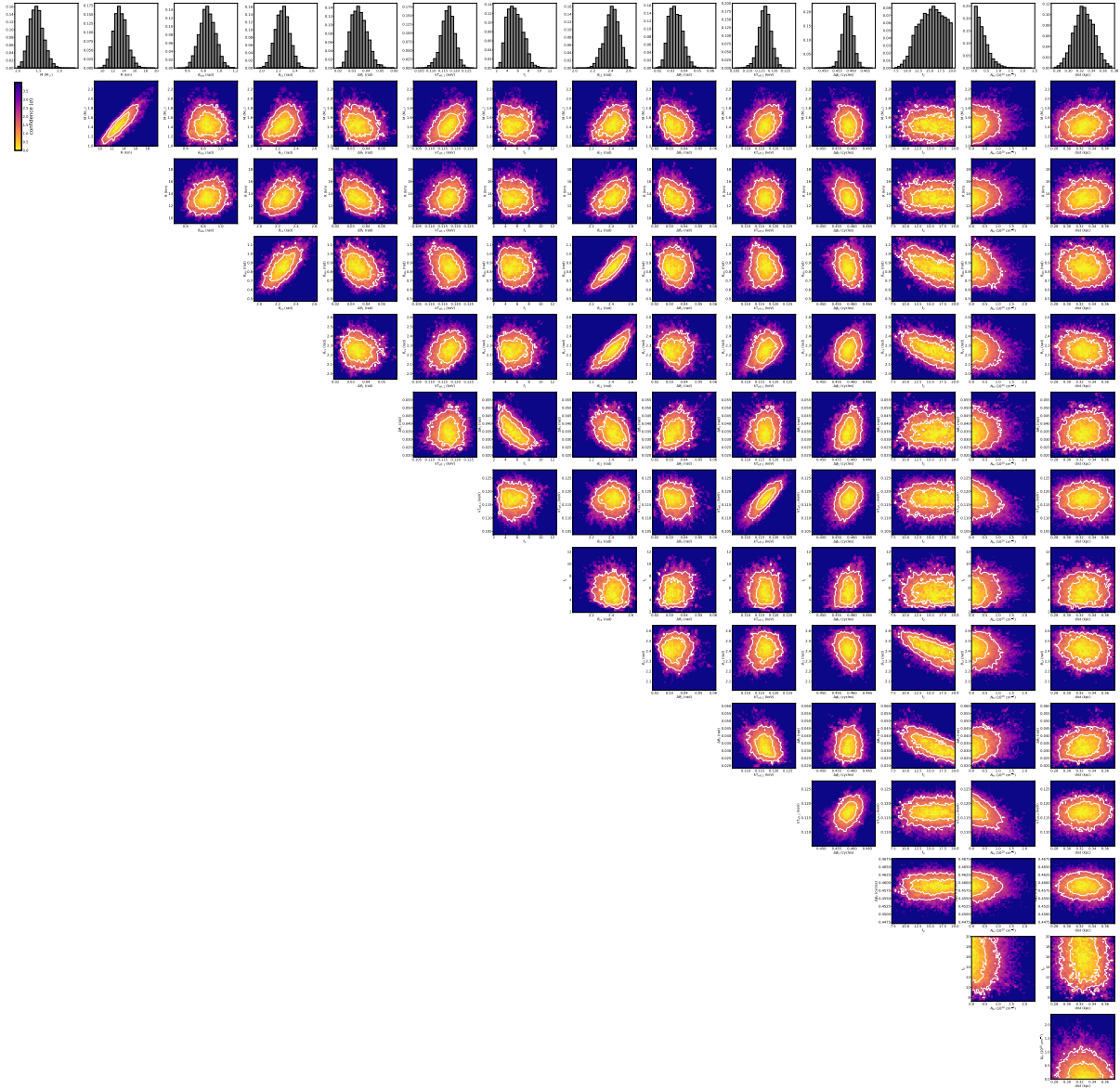}}
\vspace{-0.0truein}
   \caption{Posterior probability density distributions for the parameters in the best-fitting model pulse waveform produced by two, uniform, oval hot spots}
\label{fig:full-posteriors-for-two-oval-spots}
\end{center}
\end{figure*}

Figure \ref{fig:full-posteriors-for-two-oval-spots} shows the posterior probability density distributions for each of the parameters in the model pulse waveform produced by two, uniform, oval hot spots that best fits the NICER pulse waveform data on J0030+0451.

\clearpage

\section{Posterior Distributions for the Model With Three Oval Spots}
\label{sec:three-oval-spot-corner-plot}
\restartappendixnumbering

Figure \ref{fig:full-posteriors-for-three-oval-spots} shows the posterior probability density distributions for each of the parameters in the model pulse waveform produced by three, uniform, oval hot spots that best fits the NICER pulse waveform data on J0030+0451.

\renewcommand\thefigure{\thesection.\arabic{figure}}
\setcounter{figure}{0}

\begin{figure*}[ht!]
\begin{center}
\vspace*{-0truein}
  \resizebox{0.90\textwidth}{!}{\includegraphics{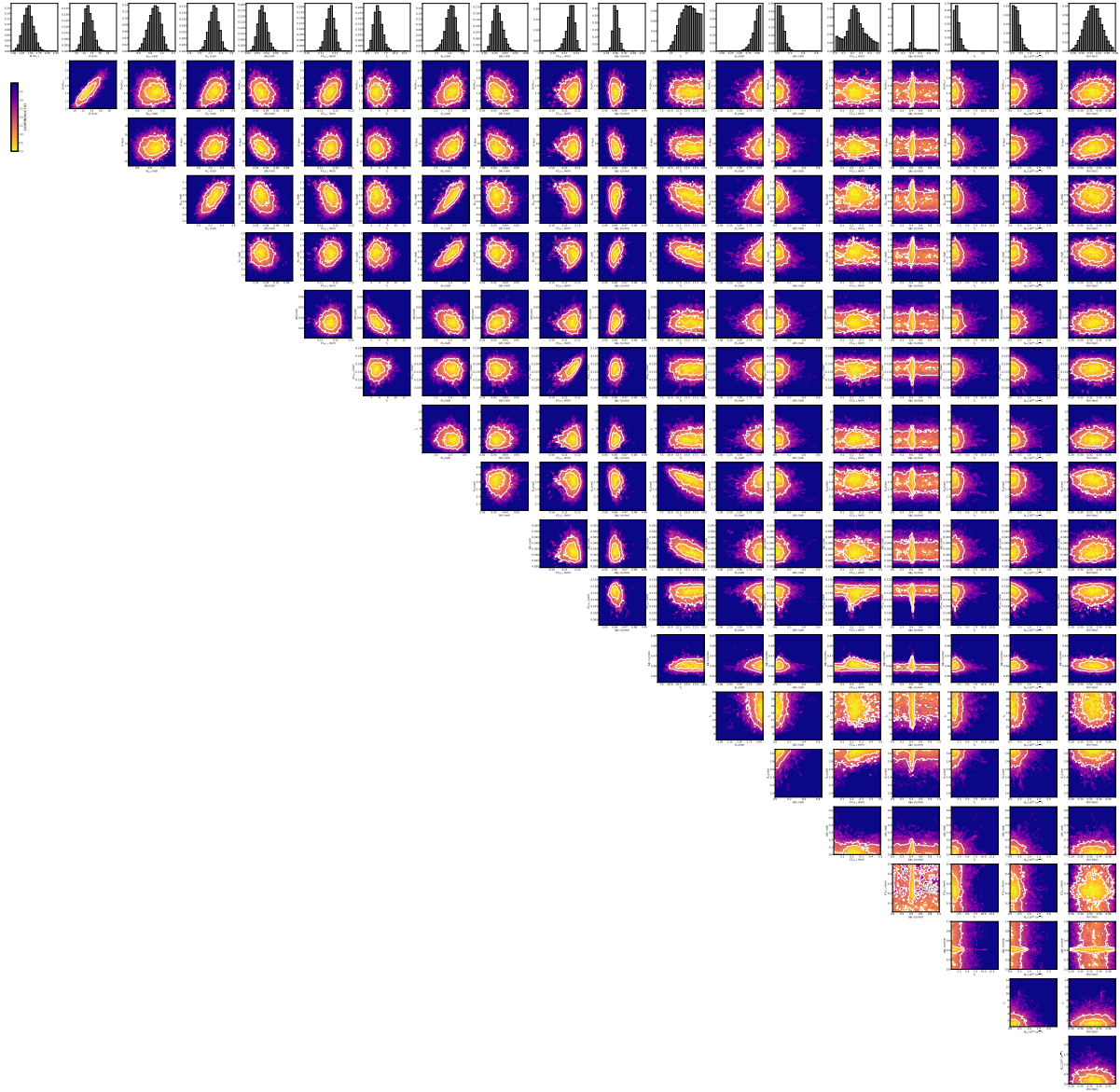}}
\vspace{-0.0truein}
   \caption{Posterior probability density distributions for the parameters in the best-fitting model pulse waveform produced by three, uniform, oval hot spots}
\label{fig:full-posteriors-for-three-oval-spots}
\end{center}
\end{figure*}
\clearpage

\bibliography{bibfile}

\end{document}